\documentclass[10pt,epsfig]{article} 
\usepackage{enumerate}
\usepackage{float}
\usepackage{subfig}
\usepackage{color}
\usepackage{slashed}
\usepackage{amssymb, amsmath,mathrsfs}
\usepackage{graphicx}
\usepackage{multirow}
\usepackage{cite}
\usepackage{chngcntr}
\usepackage[colorlinks=true,
            linkcolor=red,
            urlcolor=blue,
            citecolor=blue]{hyperref}
\usepackage{dcolumn}
\usepackage{bm}
\usepackage{color}
\long\def\rpl#1!!#2!!{\textcolor{red}{#1} \textcolor{blue}{#2}}
\def\baselinestretch{1.3}
\newcommand{\beq}{\begin {equation}}  
\newcommand{\eeq}{\end   {equation}} 
\newcommand{\bea}{\begin {eqnarray}} 
\newcommand{\eea}{\end   {eqnarray}}  
\newcommand{\baa}{\begin {array}   } 
\newcommand{\eaa}{\end   {array}   }     
\newcommand{\bit}{\begin {itemize} }
\newcommand{\eit}{\end   {itemize} }
\newcommand{\be }{\begin {equation}} 
\newcommand{\ee }{\end   {equation}}
\newcommand{\nn }{\nonumber        }

\allowdisplaybreaks
\begin{document}


%
\begin{flushright}
{
ACFI-T16-22 }
\end{flushright}

\begin{center}


{\Large \textbf {A Tale of Two Twin Higgses: Addressing Little Hierarchy in Natural 2HDM Framework}}\\[10mm]


Jiang-Hao Yu$^{\dagger}$\footnote{jhyu@physics.umass.edu}  \\
$^{\dagger}${\em Amherst Center for Fundamental Interactions,  Department of Physics, \\
University of Massachusetts Amherst, Amherst, MA 01002, USA}\\[5mm] 

\end{center}


\begin{abstract} 

In original twin Higgs model,	vacuum misalignment between electroweak and new physics scales 
is realized by adding explicit $\mathbb{Z}_2$ breaking term. 
Introducing additional twin Higgs could accommodate spontaneous $\mathbb{Z}_2$ breaking,
which explains the origin of this misalignment.
We introduce a class of two twin Higgs doublet models with most general scalar potential, and discuss general conditions which trigger electroweak and $\mathbb{Z}_2$ symmetry breaking.  
Various scenarios on realising the vacuum misalignment are systematically discussed in a natural two Higgs double model framework: explicit $\mathbb{Z}_2$ breaking, radiative $\mathbb{Z}_2$ breaking, tadpole-induced $\mathbb{Z}_2$ breaking, and quartic-induced $\mathbb{Z}_2$ breaking.
We investigate the Higgs mass spectra and the Higgs phenomenology in these scenarios. 

\end{abstract}

\newpage
\setcounter{footnote}{0}

\def\baselinestretch{1.5}
\counterwithin{equation}{section}


\section{Introduction}
\label{sec:intro}

The discovery of a 125 GeV Higgs boson at the LHC~\cite{Aad:2012tfa, Chatrchyan:2012xdj} 
is a great triumph of the Standard Model (SM) of particle physics. 
Although it confirms the Higgs mechanism, it sharpens existing naturalness problem.  
Naturalness tells us that 
the weak scale should be insensitive to quantum effects from physics at very higher scale. 
However, in SM, the large, quadratically divergent radiative corrections to the Higgs mass parameter 
destabilize the electroweak scale. 
From theoretical point of view, the SM should be well-behaved up to Planck scale. 
The existing hierarchy between the Planck and weak scales requires that 
the quantum corrections to the Higgs mass parameter should cancel against the Higgs bare mass
to obtain the observed 125 GeV Higgs boson mass. 
The large cancellation indicates existence of fine-tuning between the tree-level Higgs mass parameter and loop-level Higgs mass corrections. 
This is the well-known hierarchy problem~\cite{Barbieri:2000gf}.

The dynamical solution to the naturalness problem is 
to introduce a new symmetry which protects the Higgs mass against large radiative corrections. 
Under this direction are weak scale supersymmetry~\cite{Martin:1997ns}, and composite Higgs~\cite{Kaplan:1983fs, Agashe:2004rs, ArkaniHamed:2002qy}, etc. 
These new physics (NP) models introduce symmetry partners of the SM fields
that cancel the quadratically divergent corrections to the Higgs boson mass. 
Because the dominant quantum correction to the Higgs mass
involves in the SM top quark in the self-energy loop, 
the top quark partner is typically most relevant new particle to the quadratic cancellation. 
The new symmetry not only relates the top partner with the SM top quark, but also 
relates the Higgs coupling of the top partner to the one of the top quark. 
This enforces quadratic cancellation between the top quark and top partner contributions. 
Since the top partners typically carry SM color charge, 
the search limits of these top partners at the LHC have 
reached 700$\sim$ 800 GeV. 
This already leads to around $10$\% level of tuning between the weak scale and NP scale. 
This is known as the little hierarchy problem.


One way to avoid the little hierarchy problem is the neutral naturalness~\cite{Chacko:2005pe, Burdman:2006tz, Cai:2008au}, 
that symmetry partners are not charged under the SM gauge groups. 
This lowers the NP cutoff scale, and thus softens the little hierarchy problem. 
The twin Higgs model~\cite{Chacko:2005pe} 
[see also Refs.~\cite{Chacko:2005un, Craig:2014aea, Burdman:2014zta, Craig:2015pha, Barbieri:2005ri} and \cite{Batra:2008jy, Barbieri:2015lqa, Low:2015nqa}] 
introduces the mirror copy of the SM, the twin sector, 
which is neutral under the SM gauge group. 
The Higgs sector respects the approximate global $U(4)$ symmetry, which is
broken spontaneously to $U(3)$ at NP scale $f$.
The $U(4)$ symmetry is broken at the loop level via radiative corrections from the gauge and Yukawa interactions.
Thus the Higgs boson is the pseudo Goldstone Boson (PGB) of the symmetry breaking. 
Imposing a discrete $\mathbb{Z}_2$ symmetry between SM and twin sectors ensures that 
radiative corrections to the Higgs mass squared are still $U(4)$ symmetric. 
Thus there is no quadratically divergent radiative corrections to the Higgs mass terms. 
At the same time, the $\mathbb{Z}_2$ symmetry needs to be broken {\it at electroweak scale}.
Otherwise, the $\mathbb{Z}_2$ symmetry induces symmetric VEVs at NP scale.   
It is necessary to realize have vacuum misalignment $v \ll f$ (and thus some level of little hierarchy) to separate 
the electroweak and NP scales.  
This implies a moderate amount of tuning (approximately $\frac{2 v^2}{f^2}$).

If the Higgs boson is the PGB, the Higgs field should respect the shift symmetry. 
The shift symmetry is approximately broken by radiative corrections. 
Considering radiative corrections only, the typical Higgs potential~\cite{Bellazzini:2014yua} could be parametrized as
\bea
	V(h) \simeq a f^4 \sin^2\frac{h}{f} + b f^4 \sin^4\frac{h}{f}. 
\eea
Here $a$ and $b$ denote radiative corrections, with the form  
\bea
	a \simeq b \simeq \frac{g_*^2 }{(4\pi)^2} \log\frac{m_*^2}{f^2},
\eea
where $g_*$ denotes the typical SM couplings, such as top Yukawa coupling, and $m_*$ represents the 
top partner mass. 
If there is no other contribution than the radiative corrections, the Higgs VEV can be obtained  as 
\bea
	\langle h \rangle = \sqrt{a/b} f \simeq f. 
\eea
To realize the vacuum misalignment, additional contributions need to be added to $b$ or subtracted to $a$ and 
have $a/b \simeq v^2/f^2$. 
In the littlest Higgs model~\cite{ArkaniHamed:2002qy}, additional hard quartic terms are added to $b$ by hand to enhance the $b$. 
Instead, one could introduce soft term to $a$ to reduce $a$. 
In the original twin Higgs model, the $\mathbb{Z}_2$ symmetry is broken explicitly by introducing soft or hard $\mathbb{Z}_2$ breaking terms
in the scalar potential. 
The soft mass term  is added only to visible or twin sector to reduce $a$. 
To soften the tuning between $v$ and $f$, the Higgs sector is extended to incorporate 
two twin Higgses. 
Refs.~\cite{Chacko:2005vw, Goh:2006wj}  introduce two twin Higgses, and several choices of the  soft mass terms are introduced to breaking the $\mathbb{Z}_2$ symmetry
and reduces level of fine tuning.
In the  supersymmetric realization of the twin Higgs model~\cite{Chang:2006ra, Craig:2013fga}, two twin Higgses are also naturally introduced. 
In these literatures, the soft $\mathbb{Z}_2$ symmetry breaking term is introduced by hand, and its origin is 
unknown.  
Actually two twin Higgs setup provides more variants of $\mathbb{Z}_2$ symmetry breaking.

The spontaneous $\mathbb{Z}_2$ breaking mechanism provide a complete description of the 
electroweak symmetry breaking and vacuum misalignment.
The two twin Higgses are necessary to obtain such spontaneous breaking mechanism, without introducing the explicit $\mathbb{Z}_2$ breaking term. 
%
Refs.~\cite{Beauchesne:2015lva, Harnik:2016koz} discussed the tadpole induced spontaneous  $\mathbb{Z}_2$ breaking by introducing the bilinear term between two twin Higgses.
Without bilinear term, the VEVs of the first Higgs preserve $\mathbb{Z}_2$ while the other breaks it spontaneously. 
The bilinear Higgs mass term could transmit the $\mathbb{Z}_2$ breaking from the broken one to the unbroken one. 
It serves as the effective tadpole induced symmetry breaking and induces the vacuum misalignment naturally.   
Ref.~\cite{Yu:2016bku} realized that the spontaneous  $\mathbb{Z}_2$ breaking could be realized even without tree-level bilinear term, the 
``radiative  $\mathbb{Z}_2$ breaking". 
In this scenario, both the symmetry breaking and  $\mathbb{Z}_2$ breaking are obtained by opposite but comparable radiative corrections from the gauge and Yukawa arrangements. 
It seems that it is very hard to  realize such radiative $\mathbb{Z}_2$ breaking, because
typically the gauge corrections is much smaller than the Yukawa corrections, and thus the cancellation in the Higgs mass squared term  is not  adequate. 
But the gauge corrections could be enhanced by adjusting the VEVs of the two twin Higgses to be hierarchical. 
Through this way, the purely radiative corrections could induce  spontaneous  $\mathbb{Z}_2$ breaking.

In this work, we consider the general conditions which trigger the electroweak symmetry and the $\mathbb{Z}_2$ breaking. 
Both the tadpole-induced and radiative $\mathbb{Z}_2$ breaking scenarios could be deduced from the general conditions. 
We find that there is another novel spontaneous $\mathbb{Z}_2$ breaking mechanism. 
Instead of introducing the bilinear term in two twin Higgs potential, 
the quartic terms $\lambda_{4,5}$ could play the role of  breaking  $\mathbb{Z}_2$ symmetry spontaneously.
This is the ``quartic induced $\mathbb{Z}_2$ breaking". 
Similar to radiative symmetry breaking, the tree-level quartic terms $\lambda_{4,5}$ contribute to 
cancellation of the Higgs mass squared term. 
At the same time, similar to the tadpole-induced scenario, 
turning on $\lambda_{4,5}$ gradually transits the VEV of one Higgs to another one of another Higgs. 
Thus it provides another natural way to realize vacuum misalignment.

To systematically classify various $\mathbb{Z}_2$ breaking scenarios, 
we investigate the most general scalar potential in the two twin Higgs doublet framework.
Integrating out the twin particles, the visible Higgs sector contains the 
2HDM scalar potential. 
Depending on the breaking pattern, 
the scalars in visible sector could be partially Goldstone Bosons or 
complete Goldstone bosons.  
Through the 2HDM framework, physics behind the spontaneous $\mathbb{Z}_2$ breaking scenarios could be explained. 
The above radiative, tadpole induced, and quartic induced symmetry breaking mechanisms 
are also classified and considered in a unified framework for the two twin Higgses. 
The collider phenomenology of the two twin Higgs models is quite similar to the one of 2HDM. 
Only when we see the signatures of the twin Higgses, we will be able to distinguish these two models.

The paper is organized as follows. 
In Section 2 we briefly review the original twin Higgs and the vacuum misalignment in this model.
In Section 3 we introduce the most general scalar potential and its radiative corrections in the two twin Higgs model. 
Then we investigate the conditions for symmetry breaking and vacuum misalignment in Section 4.
Subsequently in section 5 we classify various $\mathbb{Z}_2$ symmetry breaking
scenarios in a natural two Higgs doublet framework.
Section 6 discuss the Higgs phenomenology in each scenario. 
Finally we conclude this paper. 
In Appendix A and B, we list the calculation details of the two twin Higgs models.

%

\section{Original Twin Higgs and Vacuum Misalignment}
\label{sec:twin}

We first briefly review the twin Higgs model~\cite{Chacko:2005pe, Craig:2014aea, Burdman:2014zta, Craig:2015pha} and how the vacuum misalignment is realized in this model.
The original twin Higgs model consists of a mirror copy of the SM content, called the twin sector. 
We use the labels A and B to denote the SM and twin sector respectively. 
The twin sector is related to the SM sector by a $\mathbb{Z}_2$ exchange symmetry: $A \leftrightarrow B$.
The Higgs sector consists of the SM Higgs doublet $H_A$ and the twin Higgs doublet $H_B$.
Due to the $\mathbb{Z}_2$ symmetry,  
the Higgs potential preserves an approximate global symmetry $U(4)$:
\bea
	V_{\rm tree} = - \mu^2 (H_A^2 + H_B^2) + \lambda (H_A^2 + H_B^2)^2 = - \mu^2 H^2 + \lambda H^2,
\eea
with the $U(4)$ invariant field $H \equiv \left(\begin{array}{c} H_A \\ H_B \end{array}\right)$.
If the $\mu^2$ is positive, the global $U(4)$ symmetry is spontaneously broken down to $U(3)$ 
and there are seven Goldstone Bosons modes. 
Assuming the VEV $\langle H \rangle = f $ lies along $H_B$, three Goldstone bosons are eaten by the 
twin gauge bosons, and the $H_A$ remains massless. 
Assuming the radial model is heavy, the field $H$ can be parametrized~\footnote{
Different notations on  field definition and VEVs are used in literatures~\cite{Chacko:2005pe}. 
Here we define the field and take notation on field VEVs $\langle H_B \rangle = f$ and $\langle H_A \rangle = v = 174$ GeV. 
Using the same field definition, 
another notation on field VEVs $\langle H_B \rangle = f$ and $\langle H_A \rangle = v/\sqrt2 = 174$ GeV are also used in literature~\cite{Burdman:2014zta}. 
Finally some literature~\cite{Cheng:2015buv} uses the following field definition
\bea
	H \equiv \exp\left(\frac{i}{f}\Pi\right)\left(\begin{array}{c} 0 \\ 0 \\ 0 \\\hline \frac{f}{\sqrt2} \end{array}\right),
	\quad
	\Pi=\left(\begin{array}{ccc|c}
	0&0&0&h_1\\
	0&0&0&h_2\\
	0&0&0&h_3\\ \hline
	h_1^{\ast}&h_2^{\ast}&h_3^{\ast}&h_0
	\end{array} \right),
\eea
Note that the normalization of the $h_i$ is different, with ${\rm Re} h_i + i {\rm Im} h_i$ to have correct field normalization.
Under this notation, the VEVs are $\langle H_B \rangle = f/\sqrt2$ and $\langle H_A \rangle = v/\sqrt2 = 174$ GeV.
} 
nonlinearly  as
\bea
	H \equiv \left(\begin{array}{c} H_A \\ H_B \end{array}\right)
	= \exp\left(\frac{i}{f}\Pi\right)\left(\begin{array}{c} 0 \\ 0 \\ 0 \\\hline f \end{array}\right), \quad 
	\Pi=\left(\begin{array}{ccc|c}
	0&0&0&h_1\\
	0&0&0&h_2\\
	0&0&0&h_3\\ \hline
	h_1^{\ast}&h_2^{\ast}&h_3^{\ast}&h_0
	\end{array} \right),
\eea
with $h_i = \frac{{\rm Re} h_i + i {\rm Im} h_i}{\sqrt2}$ to have correct field normalization.
Expanding out the exponential and taking the unitary gauge we obtain the explicit form
\bea
H=
\left(\begin{array}{c}
f\frac{i \bm{h}}{\sqrt{\bm{h}^{\dag}\bm{h}}}\sin\left( \frac{\sqrt{\bm{h}^{\dag}\bm{h}}}{f} \right)\\
0\\
f \cos\left( \frac{\sqrt{\bm{h}^{\dag}\bm{h}}}{f} \right)
\end{array} \right)
\simeq 
\left(\begin{array}{c}
i \bm{h}\\
0\\
f - \frac{1}{2  f}\bm{h}^{\dag}\bm{h}
\end{array} \right),
\eea
where the field $\bm{h}$ denotes the SM Higgs doublet 
$\bm{h} = \left(\begin{array}{c}  h^+  \\  h^0  \end{array}\right)$.

The global symmetry  $U(4)$ is explicitly broken once the SM and its mirror gauge group ${\textrm{SM}}_A \times {\textrm{SM}}_B$ are gauged, and 
the Yukawa interactions are introduced.
Both the gauge and Yukawa interactions give rise to radiative corrections to the quadratic part of the scalar potential.
The leading correction to the potential induced by gauging the  ${\textrm{SM}}_A \times {\textrm{SM}}_B$ is
\bea
	\Delta V \supset \frac{9\Lambda^2}{64\pi^2} \left( g_A^2 H_A^\dagger H_A + g_B^2 H_B^\dagger H_B\right) \xrightarrow{\mathbb{Z}_2: \, g_A = g_B} \frac{9 g^2 \Lambda^2}{64\pi^2} \left( H_A^\dagger H_A +  H_B^\dagger H_B\right),
\eea
where $g_A$ and $g_B$ are the gauge couplings of the ${\textrm{SM}}_A \times {\textrm{SM}}_B$  gauge group.
Here if the $\mathbb{Z}_2$ symmetry is imposed, the leading corrections to the quadratic part of the scalar potential {\it accidentally} respect 
the original $U(4)$ symmetry.
Thus corrections from the gauge sector cannot contribute to the masses of the Goldstone bosons.
Similarly, consider the Yukawa sector by focusing on the top Yukawa couplings, which takes the form
\bea
	-{\mathcal L} \supset  y_A H_A q_A t_A +  y_B H_B q_B t_B + h.c.,
\eea
where $q_{A,B}$ and $t_{A,B}$ are the left-handed $SU(2)_{A,B}$ doublet quark and right-handed $SU(2)_{A,B}$ singlet top quark in the SM and twin sectors.
The leading corrections take the form
\bea
	\Delta V \supset \frac{-3\Lambda^2}{8\pi^2} \left( y_A^2 H_A^\dagger H_A + y_B^2 H_B^\dagger H_B\right) \xrightarrow{\mathbb{Z}_2: \, y_A = y_B} \frac{-3 y^2 \Lambda^2}{8\pi^2} \left( H_A^\dagger H_A +  H_B^\dagger H_B\right).
\eea
Similarly the $\mathbb{Z}_2$ symmetry ensures that the leading corrections respect the $U(4)$ symmetry. 
Therefore, there is no quadratically divergent contribution to the Higgs boson mass at one loop order.

Although the $\mathbb{Z}_2$ symmetry ensures the quadratically divergent corrections respect the $U(4)$ symmetry, 
the gauge and Yukawa interactions still breaks the $U(4)$ symmetry via the logarithmically divergent corrections. 
The leading logarithmically divergent corrections takes the form
\bea
	\Delta V \supset \frac{3 y^4}{16\pi^2} \left(|H_A|^4 \log\frac{\Lambda^2}{y^2 |H_A|^2} + |H_B|^4 \log\frac{\Lambda^2}{y^2 |H_B|^2} \right).
	\label{eq:quarticYuk}
\eea
The sub-leading corrections proportional to $g^4$ takes the similar form with opposite sign. 
However, since both the squared mass and quartic coupling come from the same loop-suppressed corrections,
the VEV is obtained to be at the scale $f$ as mentioned in Introduction.
This fact can be seen if we write the scalar potential including the radiative corrections, to good approximation, as
\bea
V_{\rm tot} = - \mu^2 (|H_A|^2 + |H_B|^2) + \lambda (|H_A|^2 + |H_B|^2)^2 + \delta (|H_A|^4 + |H_B|^4),
\eea
Here $\delta$ denotes the small $U(4)$-violating but $\mathbb{Z}_2$-preserving loop corrections on the quartic potential, with $\delta \ll \lambda$. 
According to the Eq.~\ref{eq:quarticYuk}, the Yukawa interactions lead to $\delta \simeq  \frac{3 y^4}{16\pi^2}  \log\frac{\Lambda^2}{y^2 f^2} $,
while the gauge interactions give
$\delta \simeq - \frac{9 g^4}{256\pi^2}\log\frac{\Lambda^2}{g^2 f^2} $.
It is interesting to note that the symmetry breaking structure is controlled by the sign of the $\delta$:
\bit
\item if $\delta < 0$ (such as, only including loop corrections from the gauge interactions), the potential induces
\bea
	\langle H_A \rangle = 0, \qquad \langle H_B \rangle = f,
\eea 
which breaks the $\mathbb{Z}_2$ symmetry spontaneously.
\item if $\delta > 0$ (such as, adding loop corrections from the Yukawa interactions), the potential induces
\bea
	\langle H_A \rangle = \frac{f}{\sqrt2}, \qquad \langle H_B \rangle = \frac{f}{\sqrt2},
\eea 
which preserves the $\mathbb{Z}_2$ symmetry.
\eit 
The original twin Higgs belongs to the second case: the vacuum is equally aligned with the two sectors.

In order to realize the symmetry breaking at the electroweak scale, the VEV must be misaligned to be asymmetry
$\langle H_A \rangle = v \ll f$. 
This requires explicit $\mathbb{Z}_2$ symmetry breaking by adding
\bit
\item either a soft $\mathbb{Z}_2$-breaking mass term
\bea
	V_{\rm soft} \supset m^2_A |H_A|^2, \qquad {\textrm{with} } \;\, m^2_A \sim {\mathcal O}\left( \frac{f^2}{16\pi^2} \right) \ll \mu^2,
\eea
\item or a hard $\mathbb{Z}_2$-breaking quartic term
\bea
	V_{\rm hard} \supset \lambda_A |H_A|^4, \qquad {\textrm{with} } \;\, \lambda_A \sim  {\mathcal O}(0.1) \ll \lambda.
\eea
\eit
The approximate $U(4)$ global symmetry is still valid since $\mu^2 \gg m^2_A$ and $\lambda \gg \lambda_A$.
The $\mathbb{Z}_2$-breaking term pushes the VEVs: $\langle H_A \rangle \to v$ and $\langle H_B \rangle \to f$, 
which gives the vacuum misalignment. 
To obtain the correct VEV $v$, one needs to tune the $\mathbb{Z}_2$-breaking parameter.

In the case of the soft mass $m^2_A$, let us rewrite the scalar potential in terms of the Higgs doublet $\bm{h}$.
Taking expansion on the Higgs doublets
\bea
	|H_A|^2 \equiv f^2 \sin^2\left( \frac{\sqrt{\bm{h}^{\dag}\bm{h}}}{f} \right)\simeq \bm{h}^\dagger \bm{h}, \qquad |H_B|^2 \equiv f^2 \cos^2\left( \frac{\sqrt{\bm{h}^{\dag}\bm{h}}}{f} \right) \simeq f^2 - \bm{h}^\dagger \bm{h}.
\eea
we obtain the dominant Higgs potential 
\bea
	V(\bm{h}) \simeq \left(m^2_A - \frac{3 y^4 f^2}{8\pi^2}\log \frac{\Lambda^2}{y^2 f^2} \right) \bm{h}^\dagger \bm{h} 
	+ \frac{3 y^4}{16\pi^2} \left(\log\frac{\Lambda^2}{y^2 f^2} + \log\frac{\Lambda^2}{y^2 \bm{h}^\dagger \bm{h}} \right) ( \bm{h}^\dagger \bm{h})^2 
	+ {\mathcal O}\left(\frac{( \bm{h}^\dagger \bm{h})^3}{f^2}\right).
\eea
If the $m^2_A$ is smaller than $\delta f^2$ term, the mass term could be negative, which induces electroweak symmetry breaking, and the Higgs boson $h$ obtains its mass. 
We minimize the potential and obtain the electroweak VEV from the tadpole condition
\bea
	\frac{v^2}{f^2} =  1 - \frac{m_A^2}{\delta f^2},
\eea
with $\delta \simeq  \frac{3 y^4}{16\pi^2}$.
To realize electroweak VEV, $m^2_A$ should be comparable to the $\delta f^2$ term. 
This implies a moderate tuning between $\delta f^2$ and $m_A^2$. 
We estimate the tuning using the following approximation:
\bea
	\Delta_m = \left|\frac{2\delta m^2}{m_h^2}\right|^{-1}  \simeq \frac{m_h^2}{2\delta f^2} \sim \frac{2v^2}{f^2},
\eea
where $m_h = 125$ GeV.
For a TeV scale $f$, this corresponds to around $15$\% tuning.


\section{Two Twin Higgs Doublet Model}
\label{sec:model}


\subsection{General Twin Higgs Potential}

In this work, the visible sector is extended to the two Higgs doublet model, which is denoted as the 2HDM sector. The twin sector is exactly the mirror copy of the 2HDM sector and it is related to the 2HDM sector by the twin mirror parity $\mathbb{Z}_2$.
It is convenient to label the 2HDM sector and its twin sector as A and B respectively.
In the 2HDM, there are two Higgs doublets $H_{1A}$ and $H_{2A}$. In the twin sector, two twin Higgs doublets $H_{1B}$ and $H_{2B}$ are introduced and they are mapped into the 2HDM Higgses via the twin parity: $H_{1B} \xrightarrow{\mathbb{Z}_2} H_{1A}$, $H_{2B} \xrightarrow{\mathbb{Z}_2} H_{2A}$.
Similar to the original twin Higgs model, it is convenient to define the $U(4)$ invariant fields
\bea
	H_1 \equiv \left(\begin{array}{c} H_{1A} \\ H_{1B} \end{array}\right),\qquad H_2 \equiv \left(\begin{array}{c} H_{2A} \\ H_{2B} \end{array}\right).
\eea
which respect the twin parity $\mathbb{Z}_2$.

The scalar pontential of the fields $H_1$ and $H_2$ is similar to the two Higgs doublet model. 
In the generalized two Higgs doublet framework, we write the general twin Higgs potential 
\bea
	V(H_1, H_2) &=& 
	-\mu_1^2 |H_1|^2 - \mu_2^2 |H_2|^2
	+ \lambda_1 (|H_1|^2)^2 + \lambda_2 (|H_2|^2)^2 
	+ \lambda_3 |H_1|^2 |H_2|^2 \nn \\
	&&
	+  m_{12}^2 \left[ H_1^\dagger H_2 +h.c.\right] 
	+ \lambda_4 |H_1^\dagger H_2 |^2
	+ \frac{\lambda_5}{2} \left[(H_1^\dagger H_2)^2 + h.c.\right]\nn\\
	&&
	+  \left[  (\lambda_6|H_1|^2 + \lambda_7|H_2|^2) H_1^\dagger H_2 + h.c. \right].
	\label{eq:2HDMPot}
\eea
Here all the parameters are taken to be real for simplicity. 
Note that Refs.~\cite{Chacko:2005vw, Goh:2006wj} only contains $\lambda_{1,2,4}$ terms in the potential. 
The symmetries of the potential are recognised as follows:
\bit
\item First, of course, all the terms in the potential preserves the twin parity ${\mathbb{Z}_2}$ symmetry: $A \leftrightarrow B$.
\item The first line of the potential Eq.~\ref{eq:2HDMPot} has the global $U(4)_1 \times U(4)_2$ symmetry. 
\item While the second and the third lines of the Eq.~\ref{eq:2HDMPot} explicitly breaks the global symmetry $U(4)_1 \times U(4)_2 \to U(4)_V$.
If $\lambda_5$ is zero but $\lambda_4$ is non-zero, an additional global $U(1)$ symmetry exists.
\eit
To avoid tree-level Higgs mediated flavor changing
neutral current, similar to 2HDM, 
a softly-broken discrete symmetry $Z'_2:$ $H_1 \to H_1$ $H_2 \to - H_2$
are imposed on the quartic terms, which 
implies that $\lambda_6 = \lambda_7 = 0$, whereas
$m_{12}^2 \neq 0$ is still allowed.

The two Higgs sector is weakly gauged under the mirror SM gauge group. The gauge symmetry is applied~\footnote{
To avoid massless twin photon, sometimes $U(1)_B$ gauge symmetry is not applied.
Here we take the gauged $U(1)_B$. } 
under
\bea
	\left(\begin{array}{cc} SU(2)_A \times U(1)_A & 0 \\ 0 & SU(2)_B \times U(1)_B\end{array}\right)   \subset U(4).
\eea
The covariant kinetic terms of the Higgs fields are written as
\bea	
{\mathcal L}_{\rm kin} = D_\mu H_1^\dagger D^\mu H_1 + D_\mu H_2^\dagger D^\mu H_2,
\eea
where the covariant derivative is $D^\mu H_i = \partial^\mu H_i + i g W^\mu H_i + i g' B^\mu H_i$,
with 
\bea
	W^\mu \equiv \left(\begin{array}{cc} W_A^{a\mu} \tau^a & 0 \\ 0 & W_B^{a\mu} \tau^a \end{array}\right), 
	\quad B^\mu \equiv \left(\begin{array}{cc} \frac12 B_A^\mu & 0 \\ 0 & \frac12 B_B^\mu \tau^a \end{array}\right).
\eea
The global symmetry is weakly broken by the loop effects from the gauge interactions.

If the mass terms $\mu^2_{1,2}$ are positive, the fields $H_1$ and $H_2$ vacua take the form
\bea
	\langle H_1 \rangle \equiv \left(\begin{array}{c} 0 \\ 0 \\ 0 \\ f_1 \end{array}\right), \qquad
	\langle H_2 \rangle \equiv \left(\begin{array}{c} 0 \\ 0 \\ 0 \\ f_2 \end{array}\right).
\eea
Similar to the 2HDM model, let us define the mixing angle $\beta$  and scale $f$
\bea
\tan\beta \equiv \frac{f_2}{f_1}, \quad f = \sqrt{f_1^2 + f_2^2}. 
\eea
Depending on the global symmetry before the symmetry breaking, 
there could be seven or fourteen Goldstone bosons. 
In the following, we discuss the nonlinear parametrization of the fields $H_i$ in $U(4)/U(3)$ and 
$[U(4) \times U(4)]/[U(3) \times U(3)]$ breaking patterns.

\noindent {\underline{(1) $U(4)/U(3)$ Symmetry Breaking}}

The most general scalar potential in Eq.~\ref{eq:2HDMPot} exhibits the global $U(4)$ symmetry. 
The VEVs will break the symmetries of the Lagrangian spontaneously:
\bea
  \textrm{global   symmetry:} && \quad U(4) \to U(3), \nn\\
  \textrm{gauge    symmetry:}  && \quad SU(2)_A \times U(1)_A\times SU(2)_B \times U(1)_B \to SU(2)_A \times U(1)_A.
\eea 
The SUSY twin Higgs model~\cite{Chang:2006ra, Craig:2013fga} belongs to this breaking pattern.

Similar to the original twin Higgs, there are seven Goldstone bosons. 
%
To isolate the Goldstone bosons in the fields, 
similar to 2HDM, it is convenient to work in the Higgs basis by rotating the fields
\bea
	H = H_1 \cos\beta + H_2 \sin\beta, \qquad H' = - H_1 \sin\beta + H_2 \cos\beta,
\eea  
After rotation, only the field $H$ obtain VEV. 
Similar to the original twin Higgs, the field $H$ can be parametrized non-linearly. 
After rotation, the two fields becomes
\bea
	H = \exp\left[\frac{i}{f}\left(\begin{array}{c|cc}
		\bm{0}_{2\times 2}&\bm{0}_{1 \times 2}&\bm{h}\\\hline
		\bm{0}_{2 \times 1}&0&C\\ 
		\bm{h}^{\ast}&C^{\ast}&N
		\end{array} \right)\right]\left(\begin{array}{c} \bm{0}_{1 \times 2} \\\hline 0 \\ f \end{array}\right),
	\quad	
	H' = \left(\begin{array}{c} H^+ \\ H^0 + i A^0 \\ H'^+ \\ H'^0 + i A'^0 \end{array}\right).
	\label{eq:u4u3higgs}
\eea
where the field $\bm{h}$ denotes the SM Higgs doublet $\bm{h} = \left(\begin{array}{c}  h^+  \\  h^0  \end{array}\right)$, 
and $C^\pm$ and $N$ are Goldstone bosons in the B sector, which is absorbed by the twin gauge bosons.
Therefore, similar to the original twin Higgs, taking the expansion, the field $H$ takes the form
\bea
H=
\left(\begin{array}{c}
f\frac{i \bm{h}}{\sqrt{\bm{h}^{\dag}\bm{h}}}\sin\left( \frac{\sqrt{\bm{h}^{\dag}\bm{h}}}{f} \right)\\
0\\
f \cos\left( \frac{\sqrt{\bm{h}^{\dag}\bm{h}}}{f} \right)
\end{array} \right)
\simeq 
\left(\begin{array}{c}
i \bm{h}\\
0\\
f - \frac{1}{2  f}\bm{h}^{\dag}\bm{h}
\end{array} \right),
\eea
Here the field $H$ plays the role of  the twin Higgs as the original twin Higgs model. 
Another field $H'$ does not obtain VEV, and thus it is just another scalar quadruplet  in this model.

\noindent {\underline{(2) $[U(4) \times U(4)]/[U(3) \times U(3)]$ Symmetry Breaking}}

Now let us consider the special scalar potential with larger global symmetry. 
If only the first line exists,
The potential exhibits the exact global $U(4) \times U(4)$ symmetry. 
Here the soft mass term $m_{12}^2$ and the quartic $\lambda_{4}$ and $\lambda_5$ terms are taken to be small, 
and thus the $U(4) \times U(4)$ symmetry becomes approximate.
The VEVs will break the symmetries of the Lagrangian spontaneously:
\bea
  \textrm{global   symmetry:} && \quad U(4) \times U(4) \to U(3) \times U(3), \nn\\
  \textrm{gauge    symmetry:}  && \quad SU(2)_A \times U(1)_A\times SU(2)_B \times U(1)_B \to SU(2)_A \times U(1)_A.
\eea

In this case, the approximate global symmetry breaking is
$U(4)_1 \times U(4)_2 \to U(3)_1 \times U(3)_2$.
Let us parametrize the fields $H_1$ and $H_2$ nonlinearly in terms of the nonlinear sigma field.
Assuming the radial models $\rho_1$ and $\rho_2$ in $H_1$ and $H_2$ are heavy, the fields $H_1$ and $H_2$ are parametrized nonlinearly  as
\bea
	H_1 = \exp\left[\frac{i}{f_1}\left(\begin{array}{c|cc}
		\bm{0}_{2\times 2}&\bm{0}_{1 \times 2}&\bm{h}_1\\\hline
		\bm{0}_{2 \times 1}&0&C_1\\ 
		\bm{h}_1^{\ast}&C_1^{\ast}&N_1
		\end{array} \right)\right]\left(\begin{array}{c} \bm{0}_{1 \times 2} \\\hline 0 \\ f_1 \end{array}\right), \quad 
	H_2 = \exp\left[\frac{i}{f_2}\left(\begin{array}{c|cc}
		\bm{0}_{2\times 2}&\bm{0}_{1 \times 2}&\bm{h}_2\\\hline
		\bm{0}_{2 \times 1}&0&C_2\\ 
		\bm{h}_2^{\ast}&C_2^{\ast}&N_2
		\end{array} \right)\right]\left(\begin{array}{c} \bm{0}_{1 \times 2} \\\hline 0 \\ f_2 \end{array}\right).
		\label{eq:H1H2fields}
\eea
Expanding out the exponential we obtain the explicit form
\bea
H_i =
\left(\begin{array}{c}
f_i\frac{i \bm{h}_i}{\mathbb{H}_i}\sin\left( \frac{\mathbb{H}_i}{f_i} \right)\\
f_i\frac{i C_i}{\mathbb{H}_i}\sin\left( \frac{\mathbb{H}_i}{f_i} \right)\\
f_i \cos\left( \frac{\mathbb{H}_i}{f_i} \right) + f\frac{i N_i}{\mathbb{H}_i}\sin\left( \frac{\mathbb{H}_i}{f_i} \right)
\end{array} \right)
\simeq 
\left(\begin{array}{c}
i \bm{h}_i\\
i C_i\\
f_i - \frac{1}{2  f}\bm{h}_i^{\dag}\bm{h}_i + i N_i
\end{array} \right),
\eea
where $\mathbb{H}_i = \sqrt{\bm{h}_i^\dagger \bm{h}_i + C_i^* C_i + N_i^2}$.
Here the doublets $\bm{h}_i = \left(\begin{array}{c}  h_i^+  \\  h_i^0  \end{array}\right)$ are Goldstone bosons 
in the sector A, and $C_i, N_i$ are Goldstone bosons in the sector B. 
When $U(4) \times U(4)$-breaking terms exist, one combination of the $h_i^\pm$, and one combination of the $C_i^\pm$ becomes pseudo Goldstone bosons.


\subsection{Fermion Assignments}

In the twin Higgs model, the SM fermions are extended to include mirror fermions:
\bea
	q_A (3,2,1/6; 1,1,0)  &\xrightarrow{\mathbb{Z}_2}& q_B (1,1,0; 3,2,1/6), \nn\\
	u_A (3,1,2/3; 1,1,0)  &\xrightarrow{\mathbb{Z}_2}& u_B (1,1,0; 3,1,2/3), \nn\\
	d_A (3,1,-1/3; 1,1,0) &\xrightarrow{\mathbb{Z}_2}& d_B (1,1,0; 3,1,-1/3),
\eea
where the quantum number assignments are $(SU(3)_A, SU(2)_A, U(1)_A; SU(3)_B, SU(2)_B, U(1)_B)$. 
If there are two twin Higgses, the general Yukawa interactions could be written as
\bea
	- {\mathcal L}_{\rm Yuk} = y_1 \left(q_A H_{1A} t_A + q_B H_{1B} t_B\right) + (1 \leftrightarrow 2) + h.c.
\eea
Similar to 2HDM, it is possible to induce Higgs mediated FCNC processes in visible sector. 
To avoid such problem, the discrete $Z'_2$ symmetry $H_1 \to H_1, H_2 \to - H_2$
can also be applied to the fermion contents,
which are identified as the Type-I, II, X, Y 2HDMs~\cite{Branco:2011iw}.
Here for simplicity, we adopt the type-I Yukawa structure: all fermions only couple with  $H_1$. 
Similar to 2HDM, it is straightforward to extend type-I Yukawa structure  to other Yukawa structures. 

\noindent {\underline{(1) Fermion Assignment: Mirror Fermions}}

In this setup, similar to the original twin Higgs model, the 2HDM top Yukawa interactions are
\bea
	- {\mathcal L}_{\rm Yuk} = y \left(q_A H_{1A} t_A + q_B H_{1B} t_B\right) + h.c.
\eea
In the above Lagrangian the $U(4)$ symmetry is explicitly broken by the Yukawa terms.
Similar to the SM fermions, the mirror fermions are treated as the chiral fermions. 
The fermion masses are
\bea
	m_{t_A}^2 = y^2 H_{1A}^2, \qquad m_{t_B}^2 = y^2 H_{1B}^2 \simeq y^2 f^2 - y^2 H_{1A}^2,
\eea
where the relation $m_{t_A}^2 + m_{t_B}^2 \simeq y^2 f^2$ indicates the quadratically divergent cancellation. 
Of course, it is also possible to treat the mirror fermions vector-like~\cite{Craig:2016kue} with
\bea
	- {\mathcal L}_{\rm mass} = M (q_B q_B + t_B t_B) + h.c.
\eea
Here additional fermion degree of freedoms are introduced to make the mirror fermions vector-like.
This will lift the mirror fermion masses but not affect the quadratically divergent cancellation in the Higgs potential.
%
%
Here we only take chiral fermion case.

\noindent {\underline{(2) Fermion Assignment: $SU(6) \times SU(4)$ Fermions}}

To keep the $U(4)$ invariant form, the following fermions~\cite{Chacko:2005pe} are introduced:
\bea
	Q &=& q_A \,(3,2,1/6; 1,1,0) + \tilde{q}_A \,(3,1,2/3; 1,2,-1/2) + q_B \,(1,1,0; 3,2,1/6) + \tilde{q}_{B} (1,2,-1/2;3,1,2/3),\nn\\
	U &=& t_A \,(3,1,2/3; 1,1,0) + t_B \,(1,1,0; 3,1,2/3),\nn\\
	D &=& b_A \,(3,1,-1/3; 1,1,0) + b_B \,(1,1,0; 3,1,-1/3).
\eea
In the $SU(6) \times SU(4)$ invariant form, the fermions are assembled as
\bea
	Q = \left(\begin{array}{cc} q_{A} & \tilde{q}_A \\ \tilde{q}_{B} & q_B \end{array}\right),
	\qquad 
	U = \left(\begin{array}{c} t_{A}   \\ t_B \end{array}\right).
\eea
Similar for the leptons. 
The $U(4) \times U(4)$ invariant top Yukawa interactions are written as
\bea
	- {\mathcal L}_{\rm Yuk} = y H_1^\dagger Q U + h.c. 
	= \left(\begin{array}{cc} H_{1A}  & H_{1B} \end{array}\right) \left(\begin{array}{cc} q_{A} & \tilde{q}_A \\ \tilde{q}_{B} & q_B \end{array}\right)
	\left(\begin{array}{c} t_{A}   \\ t^B \end{array}\right) + h.c.
\eea
To lift the non-SM fermions masses, additional vector-like fermion mass terms are introduced as
\bea
	- {\mathcal L}_{\rm mass}  = M (q_B q_B + t_B t_B) + \tilde{M} (\tilde{q}_A \tilde{q}_A + \tilde{q}_B \tilde{q}_B) + h.c.
\eea
The vector-like mass terms exhibit $U(4) \times U(4)$ breaking effects in the Yukawa sector.

Expanding the Yukawa interactions, we obtain
\bea
	- {\mathcal L}_{\rm Yuk} = y \left(q_A H_{1A} t_A + q_B H_{1B} t_B  + H_{1A} \tilde{q}_B t_{B}  + H_{1B}\tilde{q}_A t_{A}\right) +h.c.
\eea
Thus the mass matrices are
\bea
	- {\mathcal L}_{\rm mass} = \left(\begin{array}{cc} q_A  & \tilde{q}_A  \end{array}\right)  \left(\begin{array}{cc} H_{1A}  & 0 \\ H_{1B} & \tilde{M} \end{array}\right)  \left(\begin{array}{c} t_A  \\ \tilde{q}_A \end{array}\right)
	 +
	  \left(\begin{array}{cc} q_B  & \tilde{q}_B \end{array}\right)  \left(\begin{array}{cc} H_{1B}  & 0 \\ H_{1A} & \tilde{M} \end{array}\right)  \left(\begin{array}{c} t_B  \\ \tilde{q}_B \end{array}\right) + h.c.
\eea


\subsection{Radiative Corrections}

The gauge and Yukawa interactions break the global symmetry explicitly, which generates the scalar potential for the pseudo-Goldstonbe bosons.
The one-loop Coleman-Weinberg potential in Landau gauge is
\bea
	V_{\rm CW}(H_1, H_2) = \frac{1}{64\pi^2} {\rm STr}\left[ m^4(H_{1,2})\log\frac{m^2(H_{1,2})}{\Lambda^2} - \frac32\right],
\eea
where the super-trace STr is taken among all the dynamical fields that have the Higgs dependent masses.
The Higgs dependent gauge boson masses are
\bea
	m_{W_A}^2 = \frac{g^2}{2} \left( |H_{1A}|^2 + |H_{2A}|^2 \right), \qquad m_{W_B}^2 = \frac{g^2}{2} \left( |H_{1B}|^2 + |H_{2B}|^2 \right),
\eea
for the $SU(2) \times SU(2)$ gauge bosons, and 
similarly for the  $U(1) \times U(1)$ gauge boson masses. 
The Higgs dependent top sector masses in the fermion assignment I are
\bea
	m_{t_A}^2 =  y^2 |H_{1A}|^2 , \qquad m_{t_B}^2 =  y^2 |H_{1B}|^2 .
\eea
The field dependent top sector masses  in the fermion assignment II are 
\bea
	m^2_{t_A, t'_A} &=& \frac{y^2|H_{1A}|^2 + y^2 |H_{1B}|^2 + M^2}{2} \mp \frac12\sqrt{(y^2|H_{1A}|^2 + y^2 |H_{1B}|^2 + M^2)^2 - 4 y^2|H_{1A}|^2 M^2},\nn\\
	m^2_{t_B, t'_B} &=& \frac{y^2|H_{1A}|^2 + y^2 |H_{1B}|^2 + M^2}{2} \mp \frac12\sqrt{(y^2|H_{1A}|^2 + y^2 |H_{1B}|^2 + M^2)^2 - 4 y^2|H_{1B}|^2 M^2}.
\eea

Let us examine that how the quadratic divergence cancels at the one-loop again due to the $\mathbb{Z}_2$ symmetry. 
The leading corrections to the quadratic part of the scalar potential are
\bea
	\Delta V \supset \frac{9 g^2 \Lambda^2}{64\pi^2} \left( H_{1A}^\dagger H_{1A} +  H_{1B}^\dagger H_{1B} + H_{2A}^\dagger H_{2A} +  H_{2B}^\dagger H_{2B}\right), 
\eea
from the gauge sector,
and 
\bea
	\Delta V \supset \frac{-3 y^2 \Lambda^2}{8\pi^2} \left( H_{1A}^\dagger H_{1A} +  H_{1B}^\dagger H_{1B}\right),
\eea
due to the Yukawa interactions in the top sector.
Note that both quadratic contributions respect the original $U(4)$ symmetry, 
and thus there is no quadratically divergent contribution to the Higgs boson masses. 
Therefore the leading corrections are the quartic terms in the effective potential.
The radiative corrections to the gauge sector is
\bea
	\Delta V \supset \frac{3 g^4}{16\pi^2} \left[\left(|H_{1A}|^2 + |H_{2A}|^2\right)^2 \log\frac{\Lambda^2}{g^2(|H_{1A}|^2 + |H_{2A}|^2)} + \left(|H_{1B}|^2 + |H_{2B}|^2\right)^2 \log\frac{\Lambda^2}{g^2(|H_{1B}|^2 + |H_{2B}|^2)} \right].
	\label{eq:gaugecorr}
\eea
Similar for the $U(1)$ sector.
The radiative corrections to the top sector in the mirror fermion model
\bea
	\Delta V \supset \frac{3 y^4}{16\pi^2} \left[\left(|H_{1A}|^2 \right)^2 \log\frac{\Lambda^2}{g^2 |H_{1A}|^2} + \left(|H_{1B}|^2 \right)^2 \log\frac{\Lambda^2}{g^2|H_{1B}|^2  } \right].
	\label{eq:fermioncorr}
\eea

In most general case, the dominant contributions of the radiative corrections could be parametrized as 
\bea
	V_{\rm rad.cor.} &=& \delta_1 \left(|H_{1A}|^4 +  |H_{1B}|^4\right)
	+ \delta_2 \left(|H_{2A}|^4 +  |H_{2B}|^4\right)
	+ \delta_3 \left(|H_{1A}|^2 |H_{2A}|^2 +  |H_{1B}|^2|H_{2B}|^2\right) \nn\\
	&+& \delta_4 \left(|H_{1A}^\dagger H_{2A}|^2   +  |H_{1B}^\dagger H_{2B}|^2 \right)
	+ \frac{\delta_5}{2} \left[(H_{1A}^\dagger H_{2A})^2   +  (H_{1B}^\dagger H_{2B})^2 + h.c. \right].
	\label{eq:loopcorr}
\eea
Note that there could have $\delta_{6,7}$ terms in the scalar potential (just like the $\lambda_{6,7}$ terms in 2HDM). 
However, since we have taken the $\lambda_{6,7}$ terms to be zero, and we adopt the Type-I Yukawa structure, the radiative corrections could not generate 
$\delta_{6,7}$ terms. 
According to the effective potential, we list the  the coefficients in Eq.~\ref{eq:loopcorr}:
\bea
	\delta_1 &\simeq &  -\frac{1}{16 \pi^2} \left( \frac94 g^4 + \frac32 g^2 g'^2 + \frac34 g'^4 \right) \log\frac{\Lambda^2}{f^2},\nn\\
	\delta_2 &\simeq &  -\frac{1}{16 \pi^2} \left( \frac94 g^4 + \frac32 g^2 g'^2 + \frac34 g'^4 \right) \log\frac{\Lambda^2}{f^2},\nn\\
	\delta_3 &\simeq &  -\frac{1}{16 \pi^2} \left( \frac92 g^4 -       3 g^2 g'^2 + \frac32 g'^4 \right) \log\frac{\Lambda^2}{f^2},\nn\\
	\delta_4 &\simeq &  -\frac{1}{16 \pi^2} \left(       6 g^2 g'^2  \right) \log\frac{\Lambda^2}{f^2},\nn\\
	\delta_5 &= & 0.
\eea
from gauge interactions~\cite{Chacko:2005vw}. 
In the Type-I Yukawa structure,  
the Yukawa interactions induce
\bea
	\delta_1 &\simeq & + \frac{1}{16 \pi^2} (3 y^4)\log\frac{\Lambda^2}{f^2}, \quad \delta_{2,3,4,5} = 0,
\eea
for the fermion assignment I and
\bea
	\delta_1 &\simeq & -  \frac{3}{16\pi^2} \frac{y^2 M^2/f^2}{M^2 - y^2 f^2} \left( M^2 \log \frac{M^2 + y^2 f^2 }{ M^2} 
	- y^2 f^2 \log \frac{M^2 + y^2 f^2}{y^2 f^2}\right),  \quad \delta_{2,3,4,5} = 0,
\eea
for the fermion assignment II~\cite{Chacko:2005pe}.
In other Yukawa structures, the Yukawa radiative corrections could be different.
Here we neglect other non-logarithm contributions and small radiative contributions from scalar self-interactions.

The overall radiative corrections are the sum over gauge boson and fermion contributions. 
Note that the above radiative corrections are independent of the breaking patterns. 
It is valid for both both $U(4)/U(3)$ and $\left[U(4) \times U(4)\right]/\left[U(3) \times U(3)\right]$ pattern.
Given the gauge and fermion assignments, the radiative corrections is completely determined 
by gauge and Yukawa couplings. 
In the following, we take general form of $\delta_{1-5}$. 
In the numerical calculation, we take the values from the fermion assignments I:
\bea
	&& \delta_1 = 0.09, \quad \delta_2 = - 0.004, \quad \delta_{3} = - 0.005 \quad \delta_{4} = - 0.002 \quad \delta_{5} = 0, \quad \textrm{(benchmark point)}.
	\label{eq:benchmark}
\eea
This serves as  our benchmark point in the following discussions.


\section{Symmetry Breaking and Vacuum Misalignment}
\label{sec:misalignment}

The radiative corrections further trigger spontaneous symmetry breaking, 
and induce VEVs for the $h^0_1$ and $h^0_2$ components in $H_{1A,2A}$ defined in Eq.~\ref{eq:H1H2fields}. 
We could determine the VEVs of $h_{1,2}^0$ in terms of general tadpole conditions from most general scalar potential.
We find that the symmetry breaking and vacuum misalignment is quite sensitive to the global symmetry breaking patterns.
Thus we will discuss the symmetry breaking and vacuum misalignment in both $U(4)/U(3)$  and $\left[U(4) \times U(4)\right]/\left[U(3) \times U(3)\right]$ breaking 
patterns.


\subsection{$U(4)/U(3)$ Breaking Pattern}

In this breaking pattern, due to the existence of the $m_{12}^2$ term and $\lambda_{4-5}$ terms in the potential,
the global symmetry breaking pattern is $U(4) \to U(3)$, 
with seven Goldstone bosons generated. 
The  $\delta_{1-5}$ terms further trigger spontaneous symmetry breaking,  and cause the Goldstone bosons become PGBs. 

The radiative corrections from the gauge and Yukawa interactions trigger electroweak symmetry breaking. 
According to Eq.~\ref{eq:u4u3higgs},  only one combination of the twin Higgses $H_{1,2}$ obtain the VEV. 
Denoting the VEV $\theta \equiv \frac{\langle h^0 \rangle}{f}$
we have the field VEVs in the Higgs basis, or the $H_{1,2}$ basis:
\bea
	\langle H \rangle = \left(\begin{array}{c} 0 \\ f \sin\theta \\ 0 \\ f \cos\theta \end{array}\right), \quad  
	\langle H' \rangle \equiv 0,  \qquad \textrm{or} \qquad 
	\langle H_1 \rangle \equiv \left(\begin{array}{c} 0 \\ f_1 \sin\theta \\ 0 \\ f_1 \cos\theta \end{array}\right), \quad
	\langle H_2 \rangle \equiv \left(\begin{array}{c} 0 \\ f_2 \sin\theta \\ 0 \\ f_2 \cos\theta \end{array}\right).
\eea

Let us calculate the VEV $\theta = \langle h^0\rangle/f$ using the tadpole conditions.
The full tadpole conditions are listed in the Appendix A. 
The tadpole conditions not only determine the mass-squared parameters $\mu_{1,2}^2$, but also tell us the value of the VEV $\theta$. 
Here we only list the tadpole conditions which determines the VEV:
\bea
       \left(f_1^2 \delta_1 + f_2^2 \delta_{345}\right) \cos2\theta & = &  0, \nn\\
	 \left(f_2^2 \delta_2 + f_1^2 \delta_{345}\right) \cos2\theta & = &  0.
\eea
If $f_1 \neq f_2$ and $\delta_1 \neq \delta_2$, the two conditions lead to 
\bea
	\cos(2\theta) = 0 \qquad \Rightarrow \qquad \theta = \frac{\pi}{4} \qquad \Rightarrow \qquad \langle H_A \rangle = \langle H_B \rangle = f/\sqrt2. 
\eea
The VEVs are equally aligned because of the $\mathbb{Z}_2$ symmetry. 
Similar to the original twin Higgs model, adding the soft or hard breaking terms could realize vacuum misalignment.  
Here we add the soft mass breaking terms in the scalar potential
\bea
	V_{\rm soft} = - m_{1A}^2 |H_{1A}|^2 - m_{2A}^2 |H_{2A}|^2 - m_{12A}^2 \left[H_{1A}^\dagger H_{2A} + h.c.\right].
\eea

\begin{figure}[!t]
\begin{center}
\includegraphics[width=4.8cm]{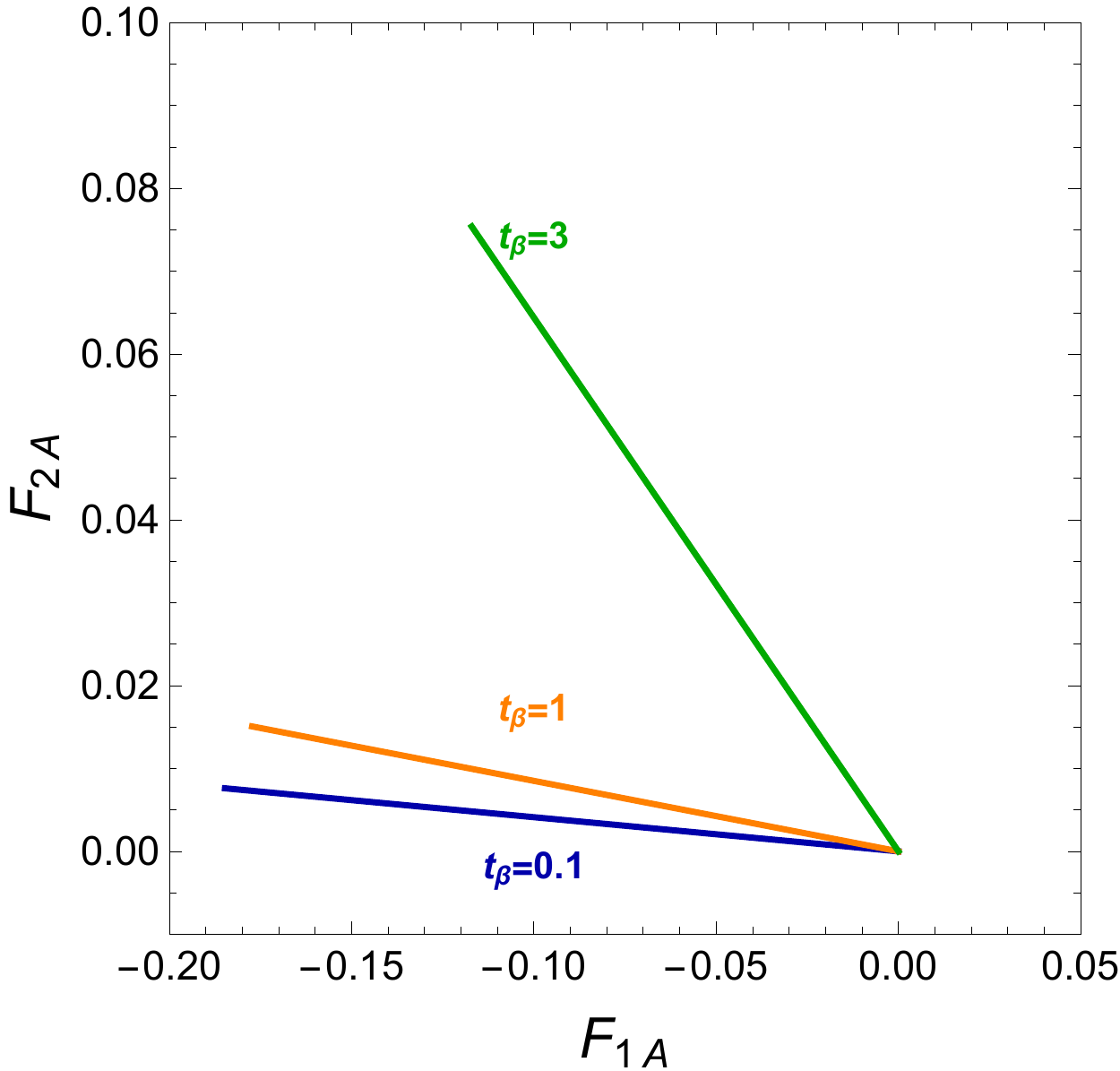}
\includegraphics[width=4.8cm]{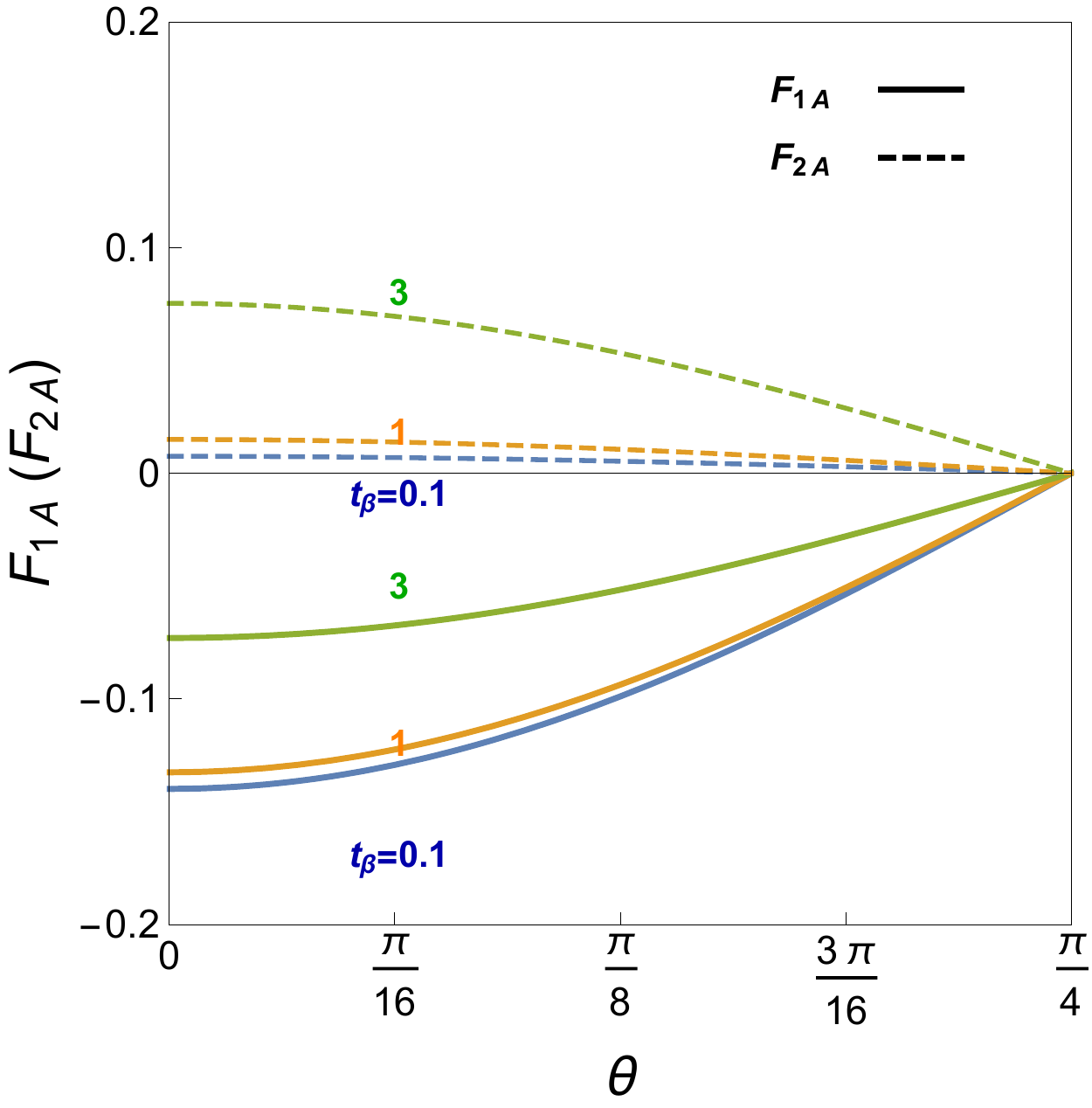}
\includegraphics[width=4.8cm]{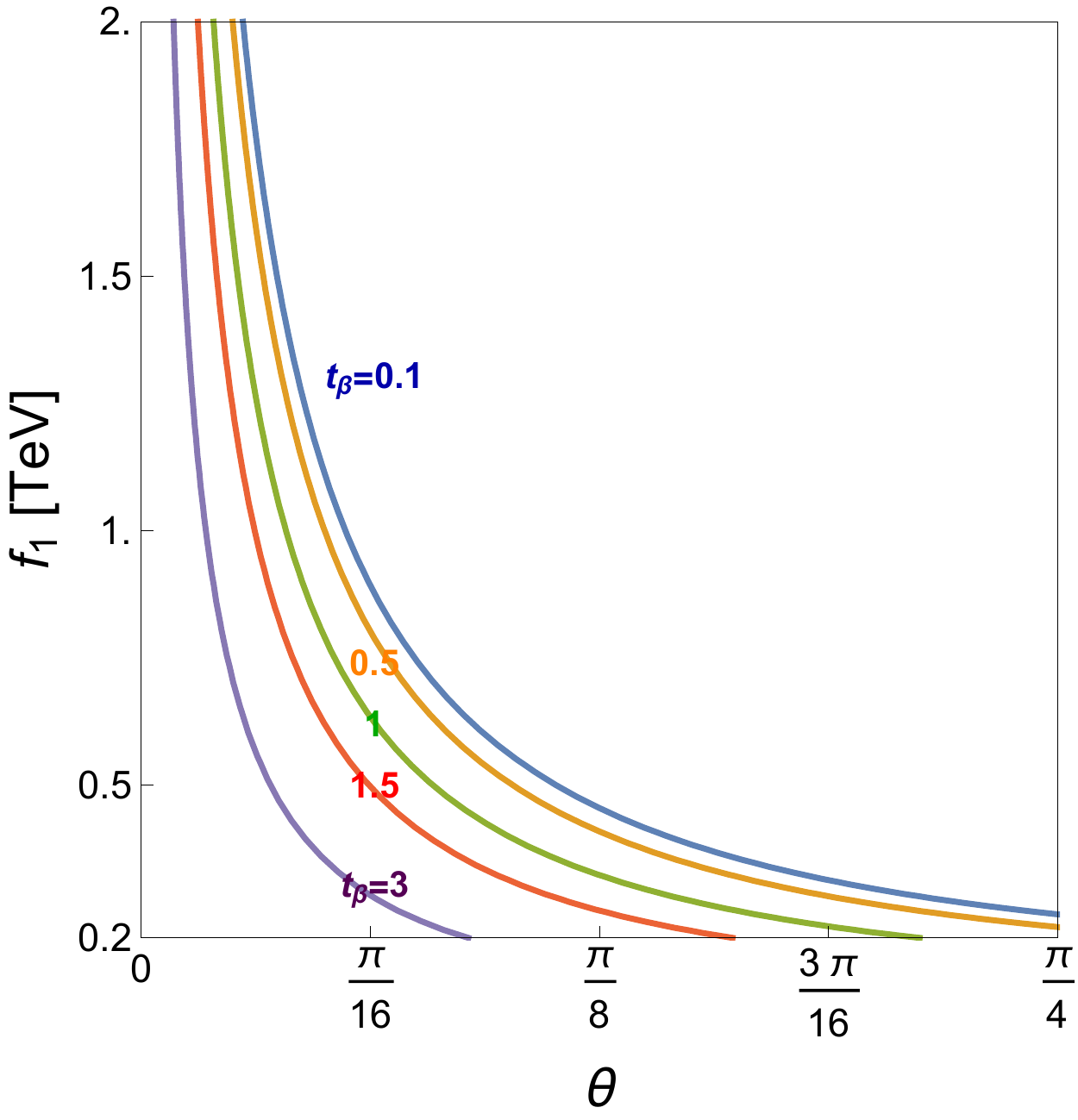}
\caption{
The left sub-figure shows the correlation between $F_{1A}$ and $F_{2A}$ for different $t_\beta$
On the middle, 
it shows the $F_{1A}$ (solid lines) and $F_{2A}$ (dashed lines) as function of $\theta$ 
for different $t_\beta$.
On the right, it shows the scale $f_1$ as function of $\theta$ for different $t_\beta$.  
Here the benchmark parameters $\delta_{1-5}$ in Eq.~\ref{eq:benchmark} are used. 
}
\label{fig:u4FFx}
\end{center}
\end{figure}

Taking the soft breaking terms into account, the relevant tadpole conditions become
\bea
        F_{1A}+ \left(2 \delta_1 + t_\beta^2 \delta_{345}\right) \cos2\theta & = &  0, \nn\\
	  F_{2A} + \left(2 t_\beta^2 \delta_2 +  \delta_{345}\right) \cos2\theta & = &  0.
	  \label{eq:tadpoleu4}
\eea
where $F_{1A} = \frac{m_{1A}^2 +  m_{12A}^2 t_\beta^2}{f_1^2} $ and $F_{2A} = \frac{m_{2A}^2 + t^{-1}_\beta m_{12A}^2}{f_1^2}$. 
From the above relations we see that $\theta$ could be less than $\pi/4$ 
only if there are relations  
\bea
	F_{1A}  \left(2 t_\beta^2 \delta_2 +  \delta_{345}\right) = F_{2A}\left(2 \delta_1 + t_\beta^2 \delta_{345}\right),
	\qquad  -F_{1A} < \left(2 \delta_1 + t_\beta^2 \delta_{345}\right).
\eea 
These could be easily satisfied, 
because both $F_{1A}$ and $F_{2A}$ are free parameters. 
Fig.~\ref{fig:u4FFx} (left) shows the relations between between $F_{1A}$ and $F_{2A}$ given the $t_\beta$ and benchmark parameters in Eq.~\ref{eq:benchmark}.
Given the soft mass term  $F_{1A}$ or $F_{2A}$, we could determine the VEV $\theta$ using tadpole conditions in Eq.~\ref{eq:tadpoleu4}.
Fig.~\ref{fig:u4FFx} (middle) shows that given the $F_{1A}$ or $F_{2A}$, values of the $\theta$ for different $\tan\beta$. 
The Figure shows the solution  $\theta < \pi/4$ exist, and thus the vacuum misalignment is realized.

Although the tadpole conditions in Eq.~\ref{eq:tadpoleu4} determines $\theta$, 
but to obtain VEV $v$, we need to know the scale $f$.  
To obtain the VEV at electroweak scale, the following condition should be imposed
\bea
	f \sin\theta = f_1 \sqrt{1 + t^2_\beta}\sin\theta = v = 174\, {\rm GeV}.
\eea
Given fixed $\tan\beta$, there are relations between $\theta$ and $f_1$. The relation is shown in Fig.~\ref{fig:u4FFx} (right), 
which plots the $(\theta, f_1)$ contours for various $t_\beta$. 
In summary, given appropriate values of $ m_{1A}^2$ (or $m_{2A}^2$) and $t_\beta$, $\theta$ and $f_1$ are totally determined, and the 
vacuum misalignment with electroweak vacuum is realized.
%


\subsection{$\left[U(4) \times U(4)\right]/\left[U(3) \times U(3)\right]$ Breaking Pattern}

If the tree-level breaking terms are small, the  potential exhibits approximate $U(4) \times U(4)$ global symmetry
 and exact $\mathbb{Z}_2$ symmetry.
In the global symmetry breaking pattern $\left[U(4) \times U(4)\right]/\left[U(3) \times U(3)\right]$, 
14 Goldstone bosons are generated after symmetry breaking. 
The  $\delta_{1-5}$ terms further trigger spontaneous symmetry breaking,  and cause the Goldstone bosons become PGBs.

The gauge and Yukawa interactions radiatively generate the symmetry breaking for the Goldstone bosons $\langle h_{1,2} \rangle$. 
Denoting 
\bea
	\theta_1 \equiv \frac{\langle h_{1} \rangle}{f_1}, \quad \theta_2 \equiv \frac{\langle h_{2} \rangle}{f_2},
\eea 
we parametrize  the  VEVs of the fields $H_{1,2}$ as
\bea
	\langle H_1 \rangle \equiv \left(\begin{array}{c} 0 \\ f_1 \sin \theta_1 \\ 0 \\ f_1 \cos \theta_1 \end{array}\right), \qquad
	\langle H_2 \rangle \equiv \left(\begin{array}{c} 0 \\ f_2 \sin \theta_2 \\ 0 \\ f_2 \cos \theta_2 \end{array}\right).
\eea

The tadpole conditions not only determines the mass-squared parameters $\mu_{1,2}^2$, 
but also VEVs $\theta_{1,2}$. 
The full tadpole conditions are presented in Appendix B. 
Here we only list the two tadpole conditions which determine the VEVs:
\bea
	 && \sin 4\theta_1 + \Omega_1 \sin 4 \theta_2  +  \Omega_2  \sin 2(\theta_1 + \theta_2) = 0,\nn \\
	 && \sin 4\theta_1 - \Omega_1 \sin 4 \theta_2  +   \Omega_2 \sin 2(\theta_1 - \theta_2)  
	 + 2 F_\lambda \sin 2(\theta_1 - \theta_2) -  4 F_m \sin(\theta_1 - \theta_2) = 0, 
	\label{eq:TadpoleU4U4}
\eea
where we denote
\bea
\Omega_1  &\equiv& t_\beta^4 \frac{\delta_2}{\delta_1} , \quad \Omega_2 \equiv t_\beta^2 \frac{\delta_{345}}{\delta_1}, \quad
F_\lambda \equiv  t_\beta^2 \frac{\lambda_{45}}{\delta_1}, \quad F_m \equiv  t_\beta \frac{m_{12}^2}{\delta_1 f_1^2}. 
\eea
In Ref.~\cite{Beauchesne:2015lva}, only $\Omega_1$ and $m_{12}^2$ terms are included in the tadpole conditions.
Thus the tadpole conditions in Ref.~\cite{Beauchesne:2015lva} could be treated as a special case of this general discussion.
Note that parameters $(\Omega_1, \Omega_2)$ only depend on $t_\beta$ and radiative corrections, denoted as ``radiative breaking parameters''.  
$(\Omega_1, \Omega_2)$ are uniquely determined once we know the gauge and fermion assignments.   
In Type-I fermion assignment, we have $\delta_1 >0 $ and $\delta_{2-5} < 0$.
This indicates $\Omega_1 < 0$ and $\Omega_2 < 0$. 
In the following, we will focus on the region $\Omega_1 < 0$ and $\Omega_2 < 0$~\footnote{%
If $\Omega_{1} < 0, \Omega_{2} > 0$ or $\Omega_{1} > 0, \Omega_{2} < 0$, 
the $\mathbb{Z}_2$ symmetry breaking could also be realized. 
For example, if $\Omega_{1} < 0, \Omega_{2} > 0$, the $\mathbb{Z}_2$ symmetry breaking happens when $\theta_1 < \theta_2 < \pi/4$.
This could happen in a different fermion assignments. 
}. 
On the other hand, $(F_\lambda, F_m)$ depend on both radiative parameters and tree-level $U(4)\times U(4)$ breaking terms $\lambda_{45}$ and $m_{12}^2$, 
denoted as ``tree breaking parameters''. 
Given radiative and tree-level breaking parameters, $(\theta_1, \theta_2)$ are uniquely determined by the tadpole conditions. 

\begin{figure}[!t]
\begin{center}
\includegraphics[width=4.5cm]{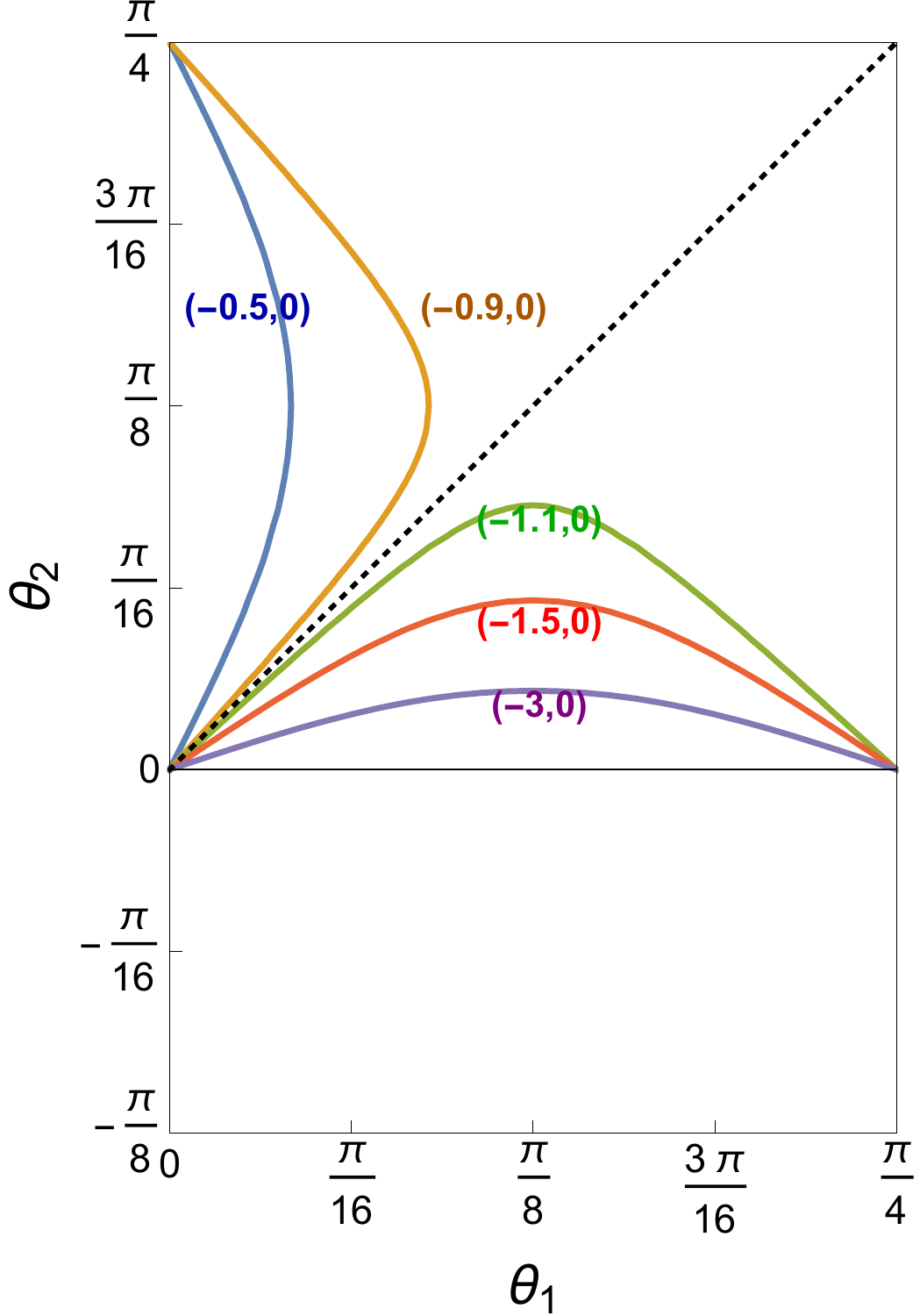}
\includegraphics[width=4.5cm]{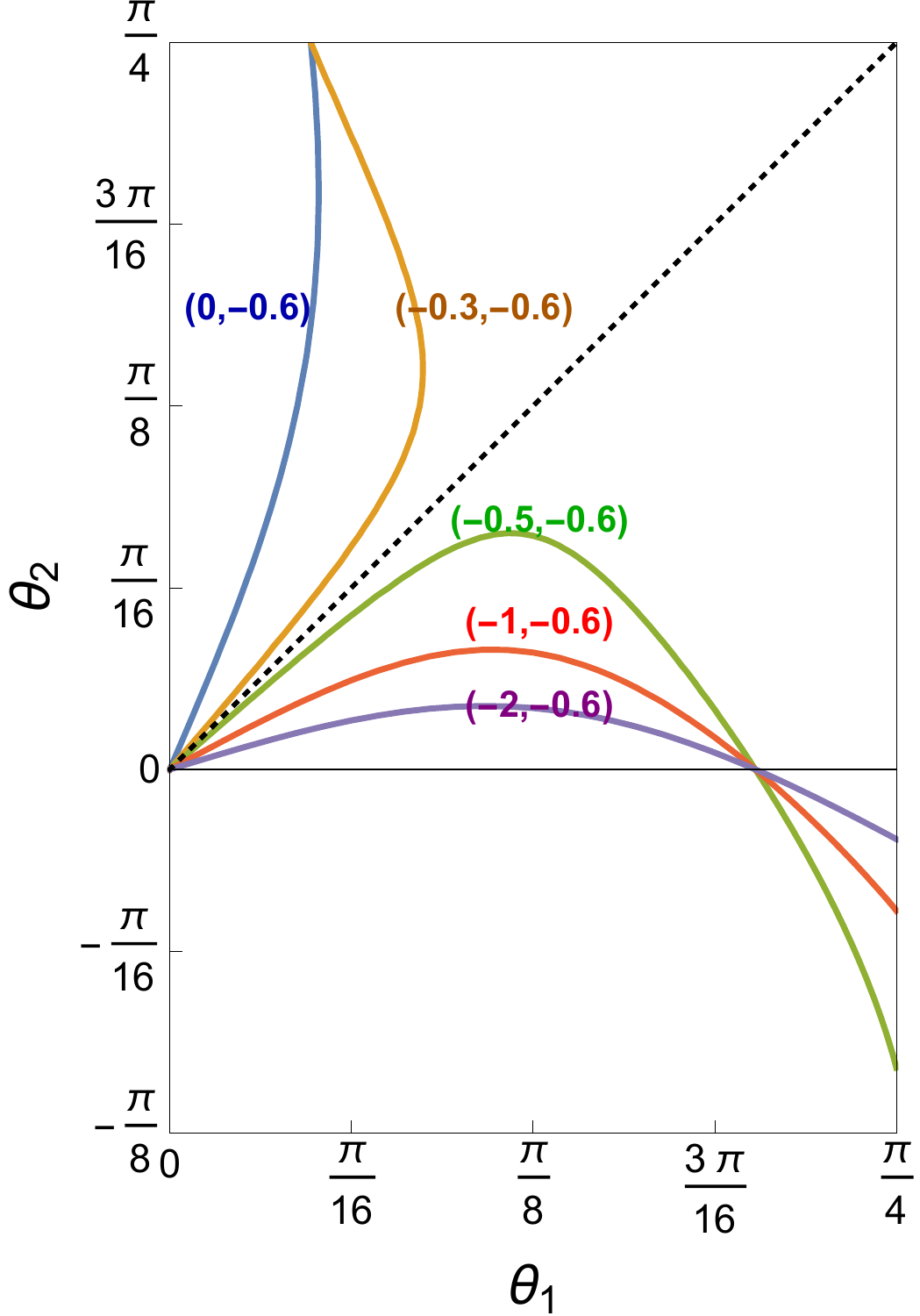}
\includegraphics[width=4.5cm]{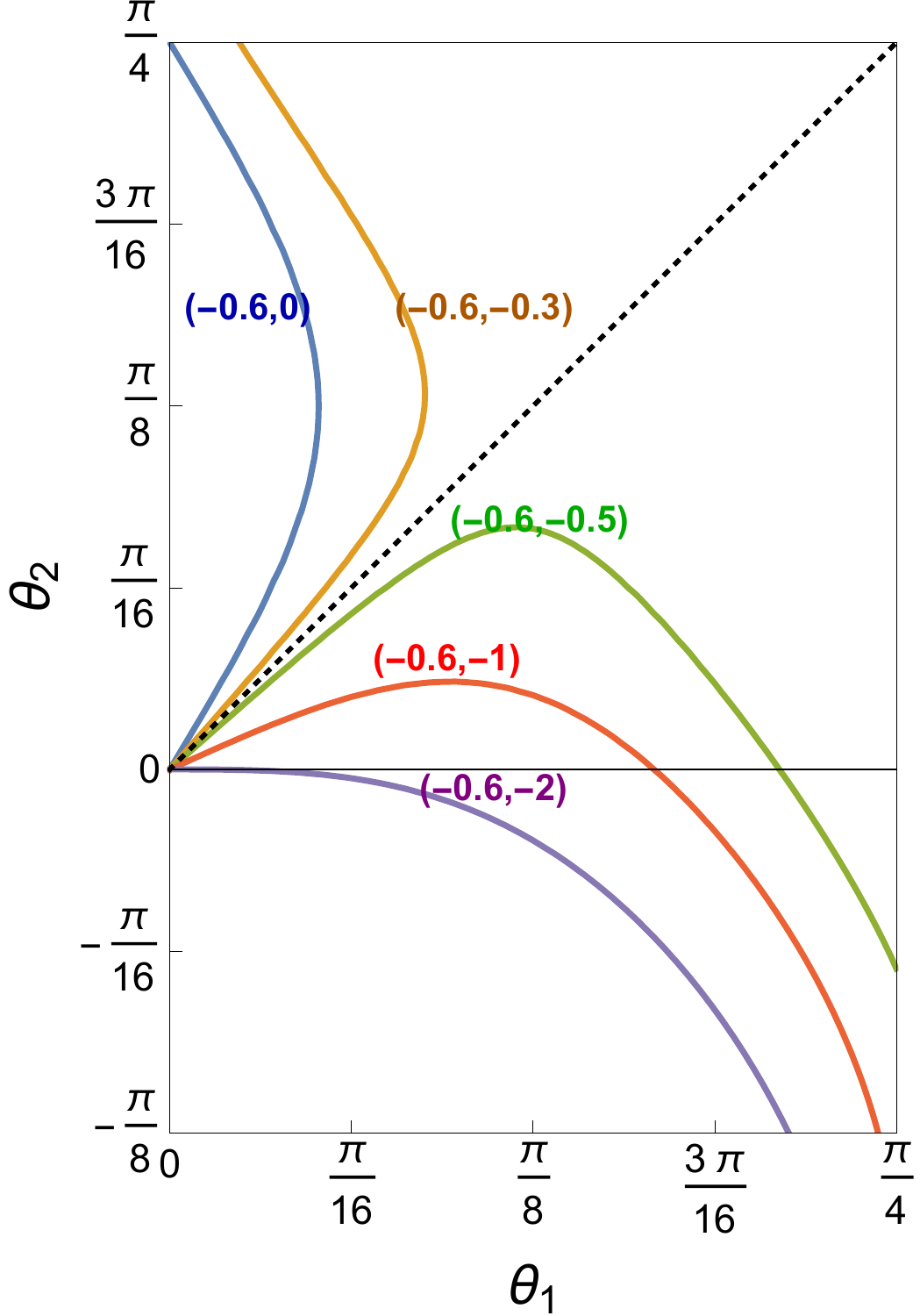}
\caption{
The contour lines show the relation between $(\theta_1, \theta_2)$ imposed by the first tadpole condition in Eq.~\ref{eq:TadpoleU4U4}. 
Each contour is labeled by the radiative parameters $(\Omega_1, \Omega_2)$.
The left (middle) panel shows contours for different $\Omega_1$ with $\Omega_2 = 0 (-0.6)$.
The right panel shows contours for different  $\Omega_2$ with $\Omega_1 = -0.6$. 
}
\label{fig:theta12}
\end{center}
\end{figure}

The first tadpole condition in Eq.~\ref{eq:TadpoleU4U4} tells us the relation between $\theta_1$ and $\theta_2$. 
In Fig.~\ref{fig:theta12} we plot the correlation between $(\theta_1, \theta_2)$ for different  $(\Omega_1, \Omega_2)$ .  
Several features are in order. 
First, depending on the size of $|\Omega_1 +\Omega_2|$, the contours live in regions: 
$\theta_2 < \theta_1$ for $|\Omega_1 +\Omega_2| >1$, and  $\theta_2 > \theta_1$ for $|\Omega_1 +\Omega_2| <1$. 
Second, $\Omega_2$ determines intersection point $\theta_1^*$ between the contour curve and the $x$-axis, or $\theta_2^*$ between the curve and y-axis. 
If  $\Omega_2$ is zero, the intersection point is either $\theta_1^* = \pi/4$ or  $\theta_2^* = \pi/4$.
From Fig.~\ref{fig:theta12} (right), the smaller $\Omega_2$, the smaller $\theta_1^*$ if $|\Omega_1 +\Omega_2| >1$, 
while the larger $\theta_2^*$ if $|\Omega_1 +\Omega_2| <1$. 
Third, $\Omega_1$ only controls the convex behaviour of the contours. 
From Fig.~\ref{fig:theta12} (left and middel), the smaller $\Omega_1$, the larger convex contour if $|\Omega_1 +\Omega_2| >1$, 
while vice versa for  $|\Omega_1 +\Omega_2| <1$.

\begin{figure}[!t]
\begin{center}
\includegraphics[width=4.8cm]{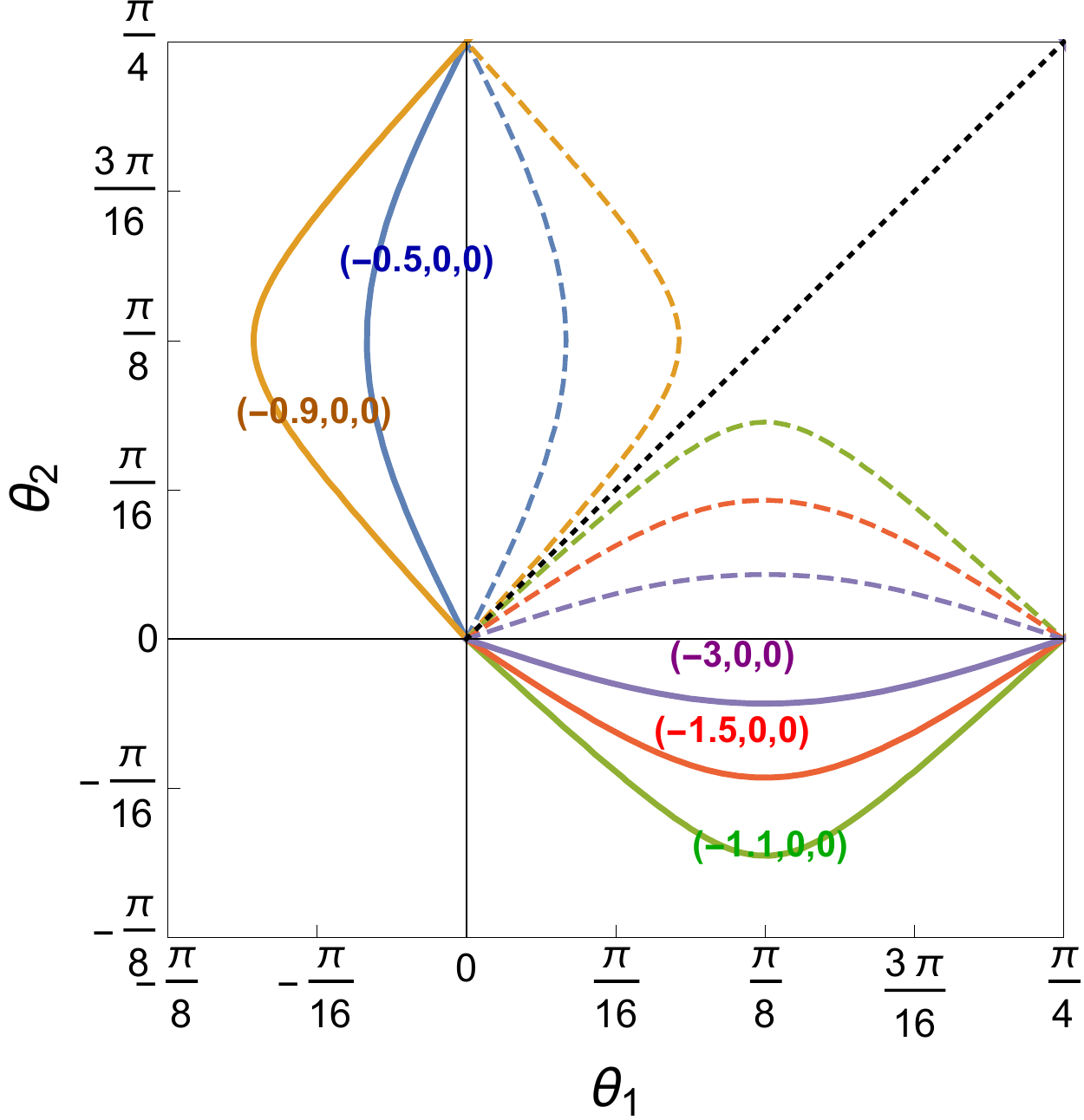}
\includegraphics[width=4.8cm]{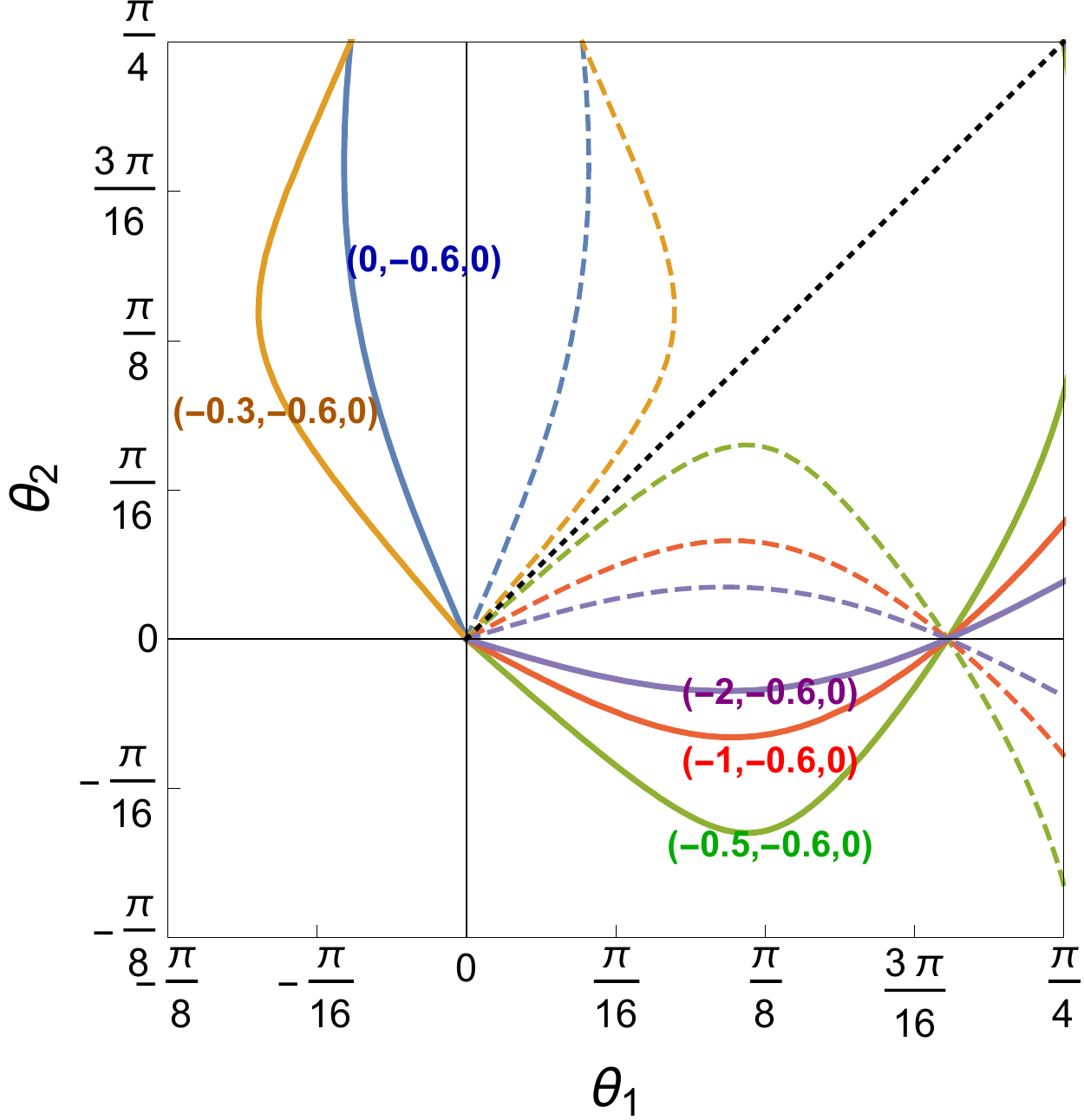}
\includegraphics[width=4.8cm]{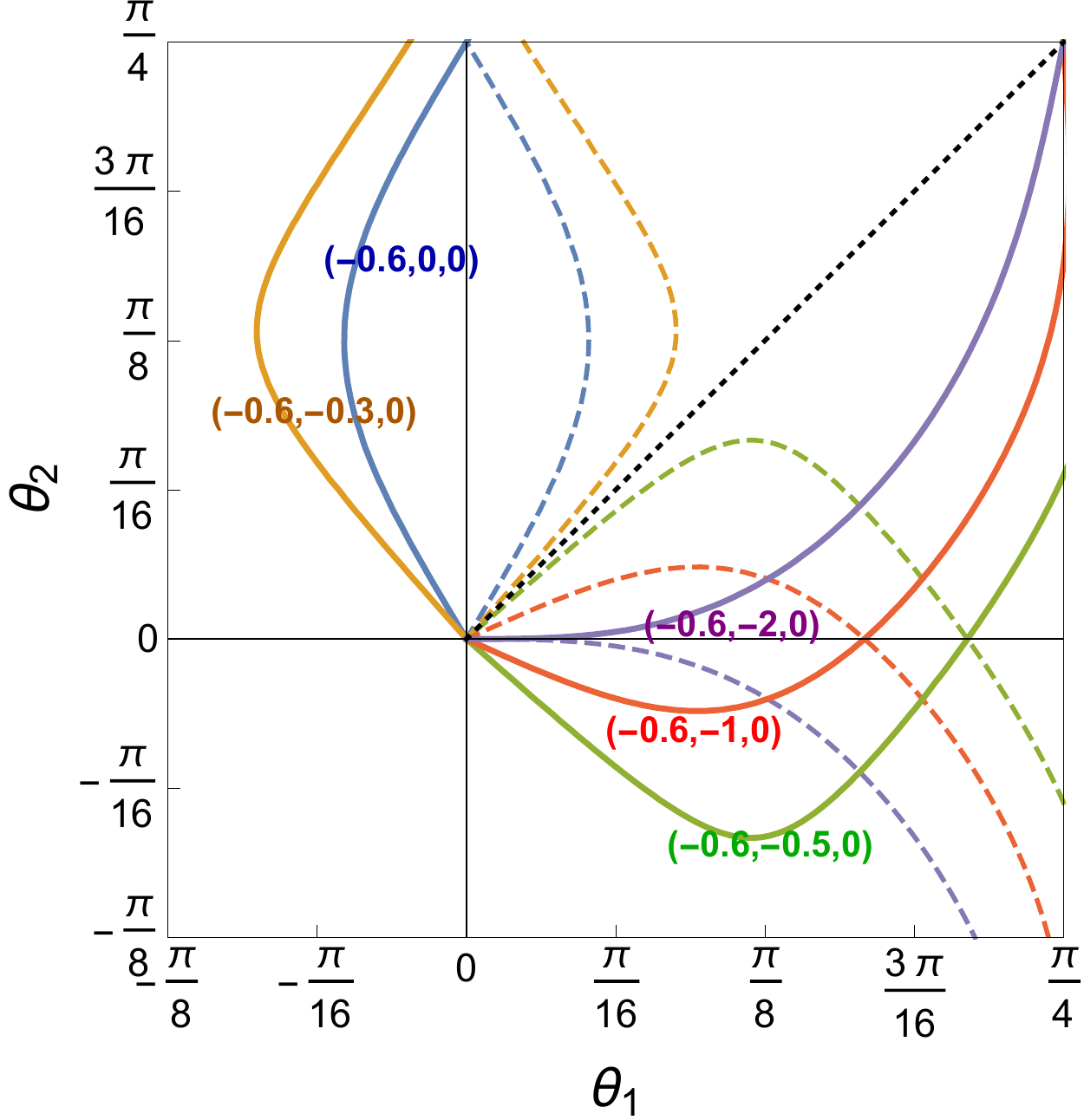}\\
\includegraphics[width=4.5cm]{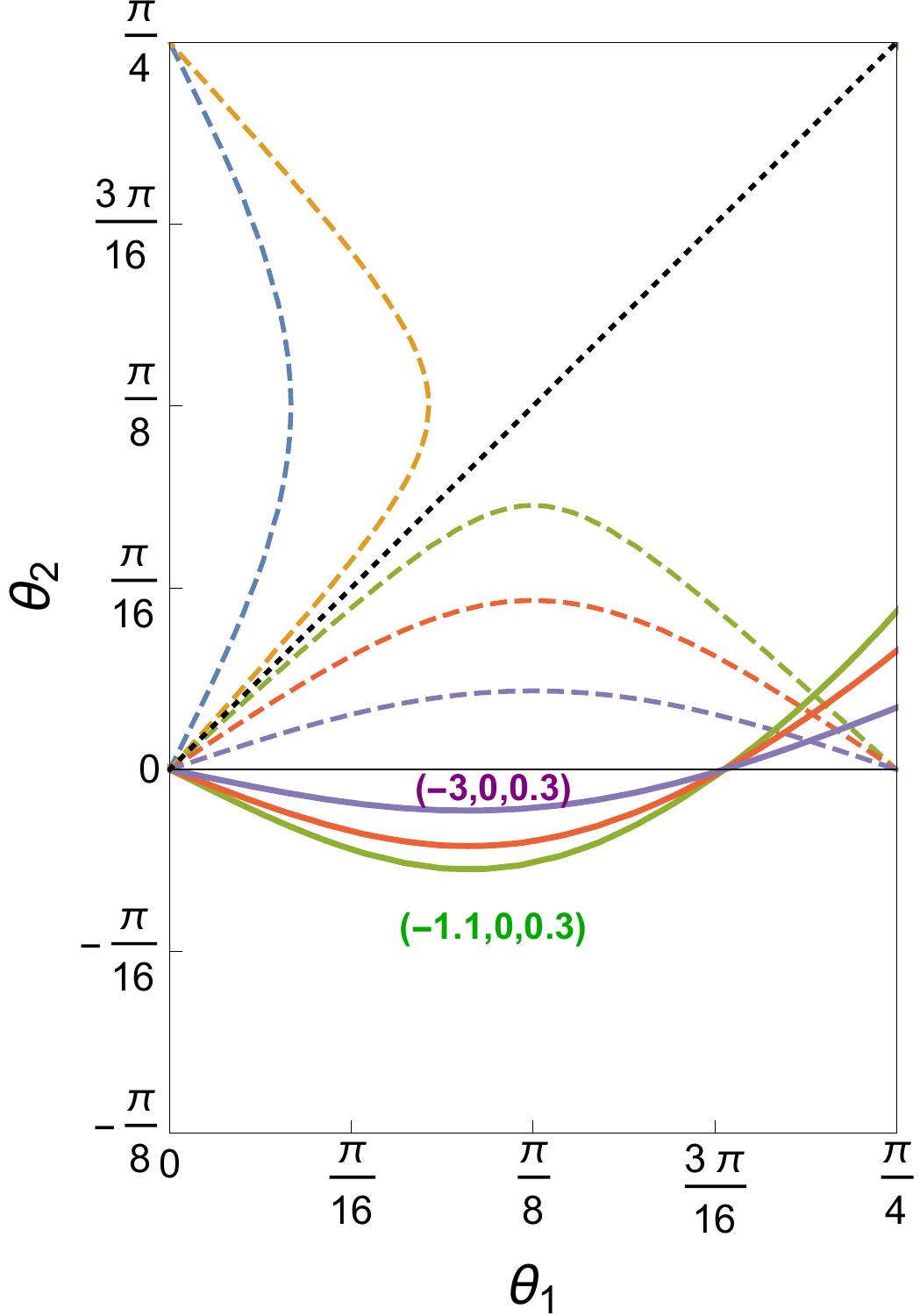}
\includegraphics[width=4.5cm]{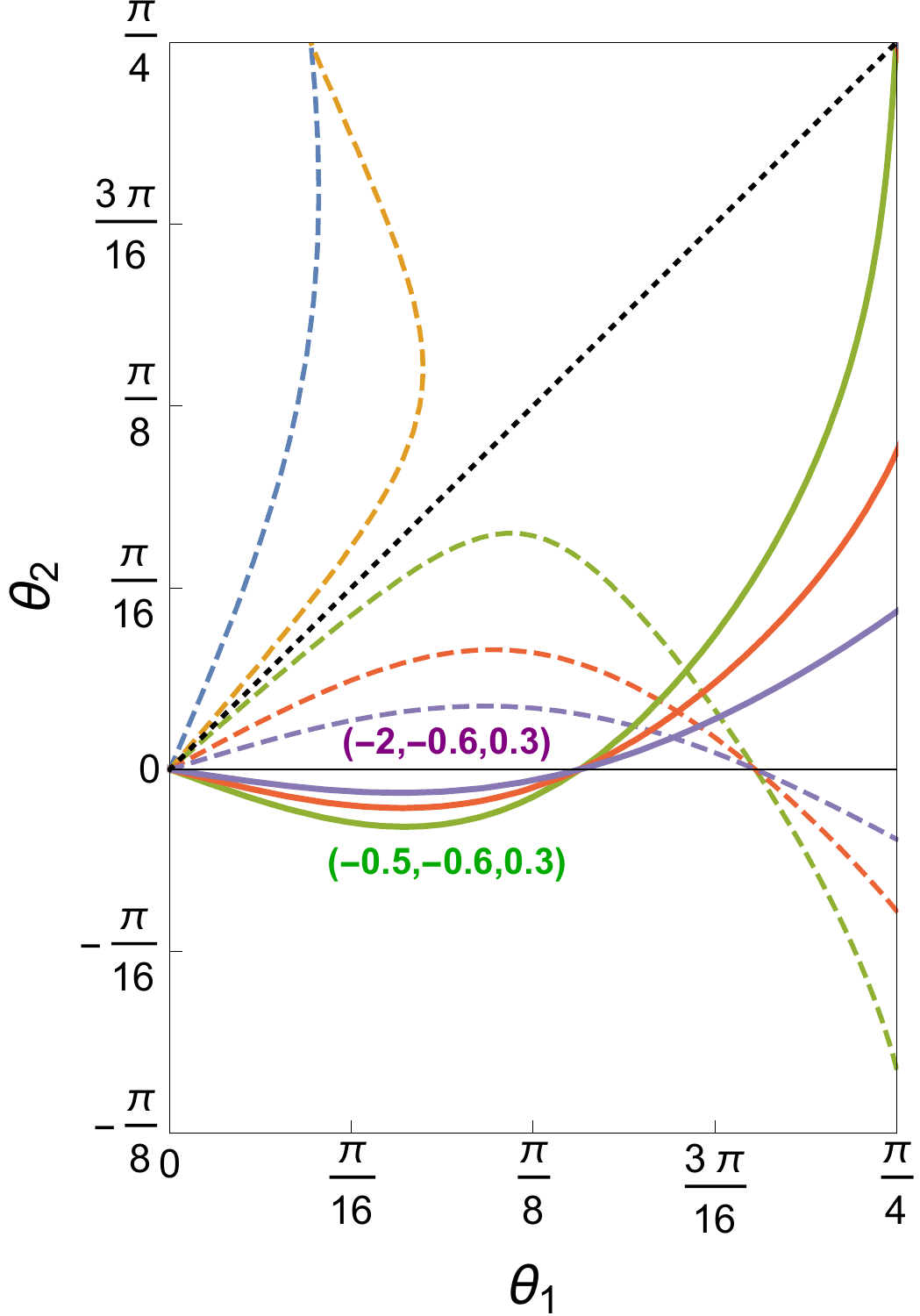}
\includegraphics[width=4.5cm]{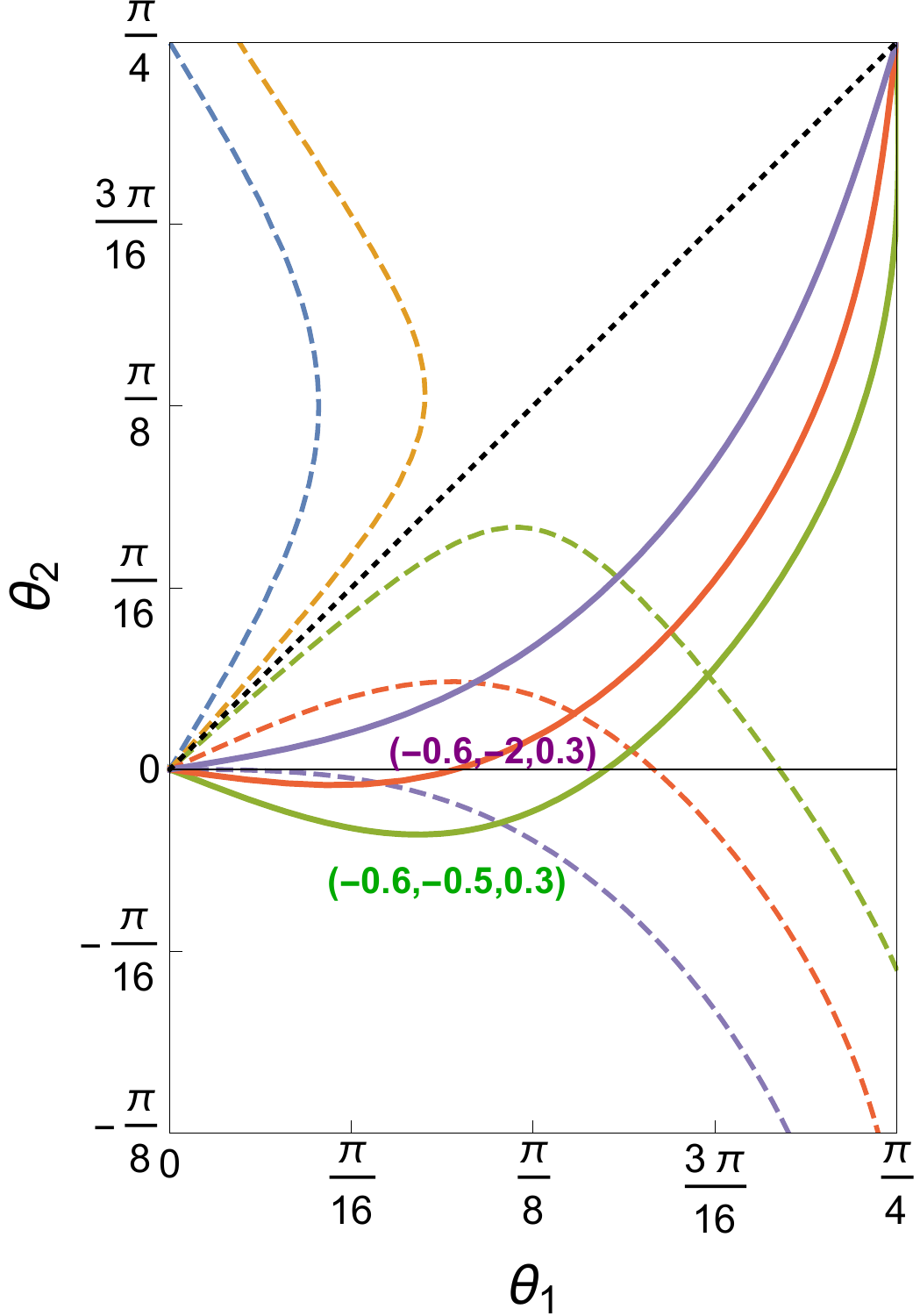}
\caption{
The contour lines show the relation between $(\theta_1, \theta_2)$ imposed by the first (dashed lines) and second (solid lines) tadpole conditions. 
Each contour is labeled by the radiative parameters $(\Omega_1, \Omega_2, F_m)$.
The left (middle) panel shows contours for different $\Omega_1$ with $\Omega_2 = 0 (-0.6)$.
The right panel shows contours for different  $\Omega_2$ with $\Omega_1 = -0.6$. 
The upper and lower panels correspond to $F_m = 0$ and $0.3$ respectively.
}
\label{fig:FF12}
\end{center}
\end{figure}

The second tadpole condition provides us another relation on $(\theta_1, \theta_2)$, which is shown as another contour in the $(\theta_1, \theta_2)$ plane. 
Together with the contour from first tadpole condition, the two contours uniquely determine value of $(\theta_1, \theta_2)$ which is the intersection point
between two contours. 
Similar to  Fig.~\ref{fig:theta12}, let us plot the $(\theta_1, \theta_2)$ contours imposed by the second tadpole condition. 
To clearly present the effects of each parameters, we first turn off tree-level breaking parameters $(F_\lambda, F_m)$. 
In this case, the two conditions reduce to
\bea
	 && \sin 4\theta_1 + \Omega_1 \sin 4 \theta_2  +  \Omega_2  \sin 2(\theta_1 + \theta_2) = 0,\nn \\
	 && \sin 4\theta_1 - \Omega_1 \sin 4 \theta_2  +   \Omega_2 \sin 2(\theta_1 - \theta_2) = 0.  
\eea 
Fig.~\ref{fig:FF12} shows the $(\theta_1, \theta_2)$ contours imposed by two conditions for different  $(\Omega_1, \Omega_2)$. 
We note that the two conditions are symmetric under $\theta_1 \leftrightarrow -\theta_1$ if $|\Omega_1 +\Omega_2| >1$,
while they are symmetric under $\theta_2 \leftrightarrow -\theta_2$ if $|\Omega_1 +\Omega_2| <1$. 
This symmetric behaviour can be seen from Fig.~\ref{fig:FF12}. 
Therefore we can determine the solution for $(\theta_1, \theta_2)$:
\bea
	\begin{cases}		
	\theta_2  = 0,   \theta_1  \le \pi/4,  &  \text{for } \, |\Omega_1 +\Omega_2| >1 \\
	\theta_1  = 0,   \theta_2  \ge \pi/4,  &  \text{for } \, |\Omega_1 +\Omega_2| <1.
	\end{cases}
\eea
This indicates only one Higgses $H_i$ obtain VEV. 
From the left panel of the Fig.~\ref{fig:FF12}, 
if $\Omega_2 = 0$, we have either $\theta_1 = \pi/4$ (if $|\Omega_1 +\Omega_2| >1$) or 
$\theta_2 = \pi/4$ (if $|\Omega_1 +\Omega_2| <1$). 
According to the middle panel, 
when $\Omega_2 < 0$, we have either $\theta_1 < \pi/4$ (if $|\Omega_1 +\Omega_2| >1$) or 
$\theta_2 > \pi/4$ (if $|\Omega_1 +\Omega_2| <1$). 
On the right panel, it shows as $\Omega_2$ decreases, the value of $\theta_1$ decreases.
Thus we could obtain appropriate asymmetric vacua $\theta_1$ when we vary $\Omega_2$. 
When we take $|\Omega_1 +\Omega_2| >1$, $\theta_1$ could be smaller than $\pi/4$ as we vary $\Omega_2$. 
Thus even without tree-level breaking parameters, the vacuum misalignment could still happen. 
This is the scenario of radiative $\mathbb{Z}_2$ symmetry breaking~\cite{Yu:2016bku}.

Turning on tree-level breaking terms $(F_\lambda, F_m)$ will change the contour curve between $\theta_1$ and $\theta_2$ imposed by the second tadpole condition. 
For simplicity, let us turn on one tree-level breaking term: $F_\lambda$ or $F_m$. 
Fig.~\ref{fig:FF12} (lower panel) shows the $(\theta_1, \theta_2)$ contours imposed by two conditions for different  $(\Omega_1, \Omega_2, F_m)$. 
For comparison, we use the same $(\Omega_1, \Omega_2)$ in both the upper and lower panels of Fig.~\ref{fig:FF12}. 
We find that turning on $F_m$  shifts the intersection point between the contour and the x-axis to lower $\theta_1$, 
and also change the convex behavior of the contour. 
$F_m$ plays a similar role as $\Omega_2$. 
Fig.~\ref{fig:FF12} (left) show that even $\Omega_2$ is zero, turning on $F_m$ will obtain the following solution:
\bea		
	\theta_2  <   \theta_1  \le \pi/4,  &  \text{for } \, |\Omega_1 +\Omega_2| >1 .
\eea
Thus vacuum misalignment could be realized via the bilinear term $m_{12}^2$. 
This is the scenario of tadpole induced $\mathbb{Z}_2$ symmetry breaking~\cite{Beauchesne:2015lva, Harnik:2016koz}.
Furthermore, Fig.~\ref{fig:FF12} (middle and right) show that turning on $\Omega_2$ will 
also obtain the viable solution, with   
the feature: the larger $\Omega_2$, the smaller $\theta_1$. 
Finally, our discussion on the tree-level breaking term $F_m$ could also apply to the case with only $F_\lambda$. 
The results are quite similar to the one in Fig.~\ref{fig:FF12}. 
In this case, the $\lambda_{45}$ plays the role to obtain vacuum misalignment. 
This is the scenario of quartic induced $\mathbb{Z}_2$ symmetry breaking.

\begin{figure}[!t]
\begin{center}
\includegraphics[width=5cm]{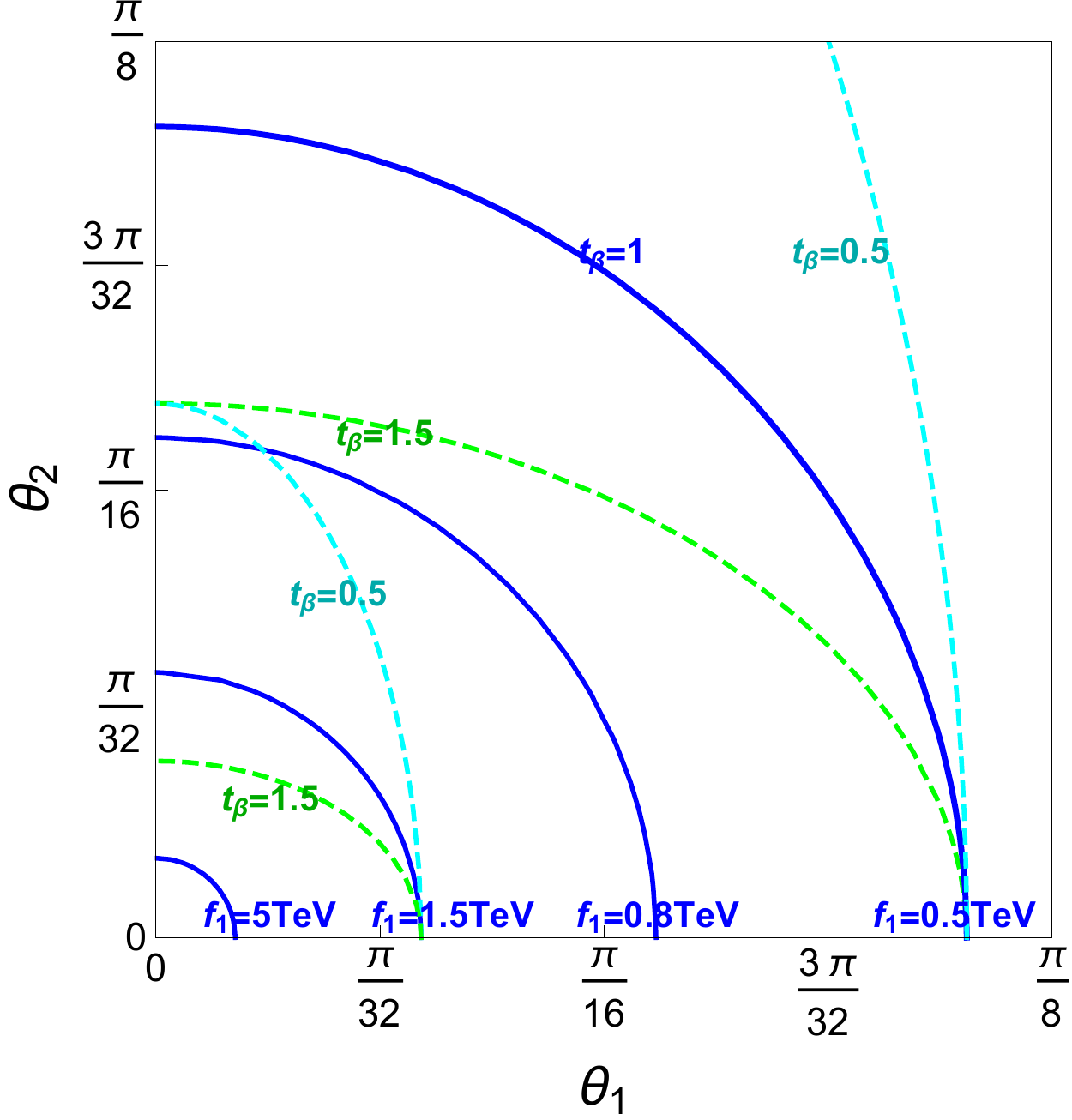}
\includegraphics[width=4.4cm]{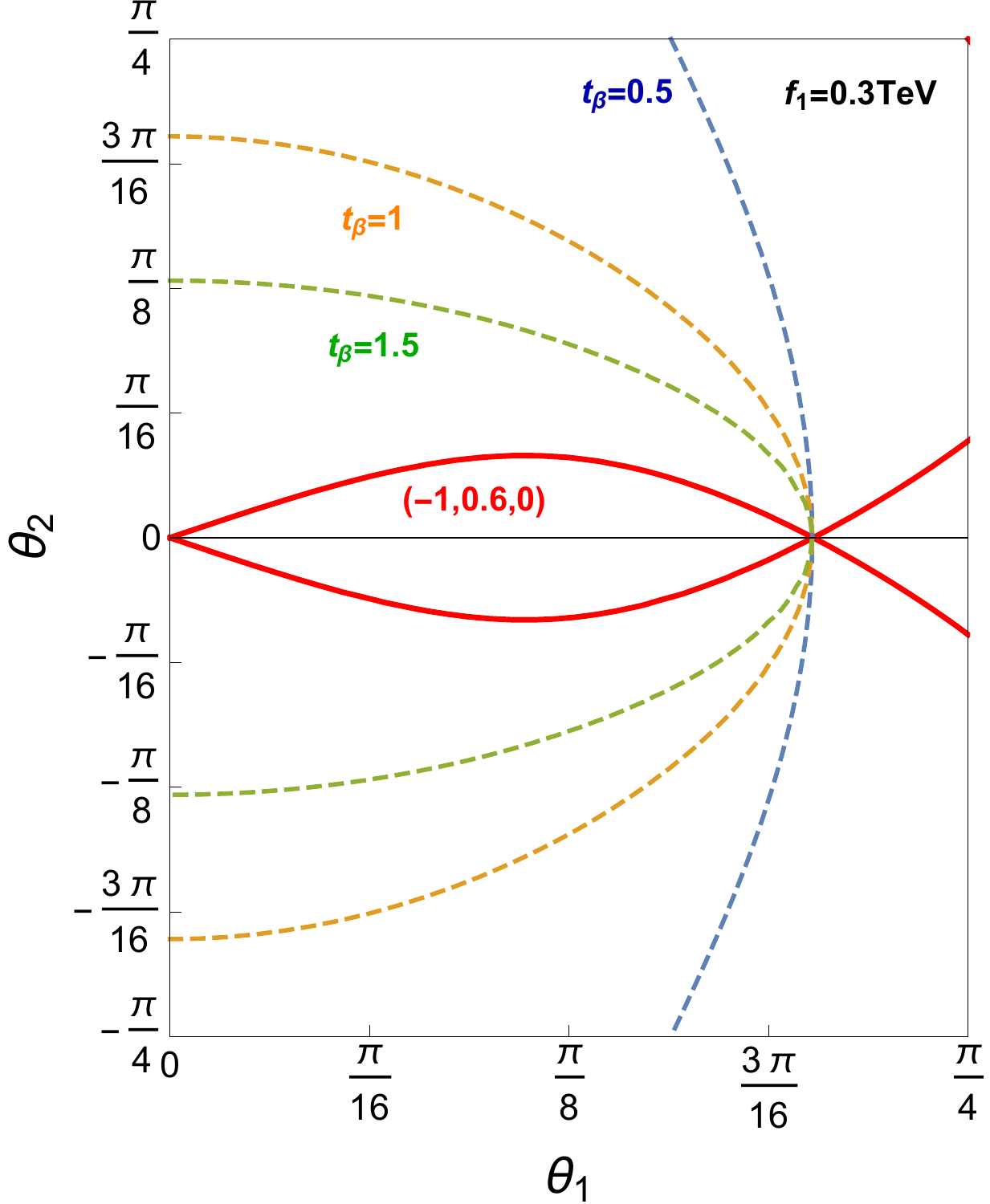}
\includegraphics[width=4.4cm]{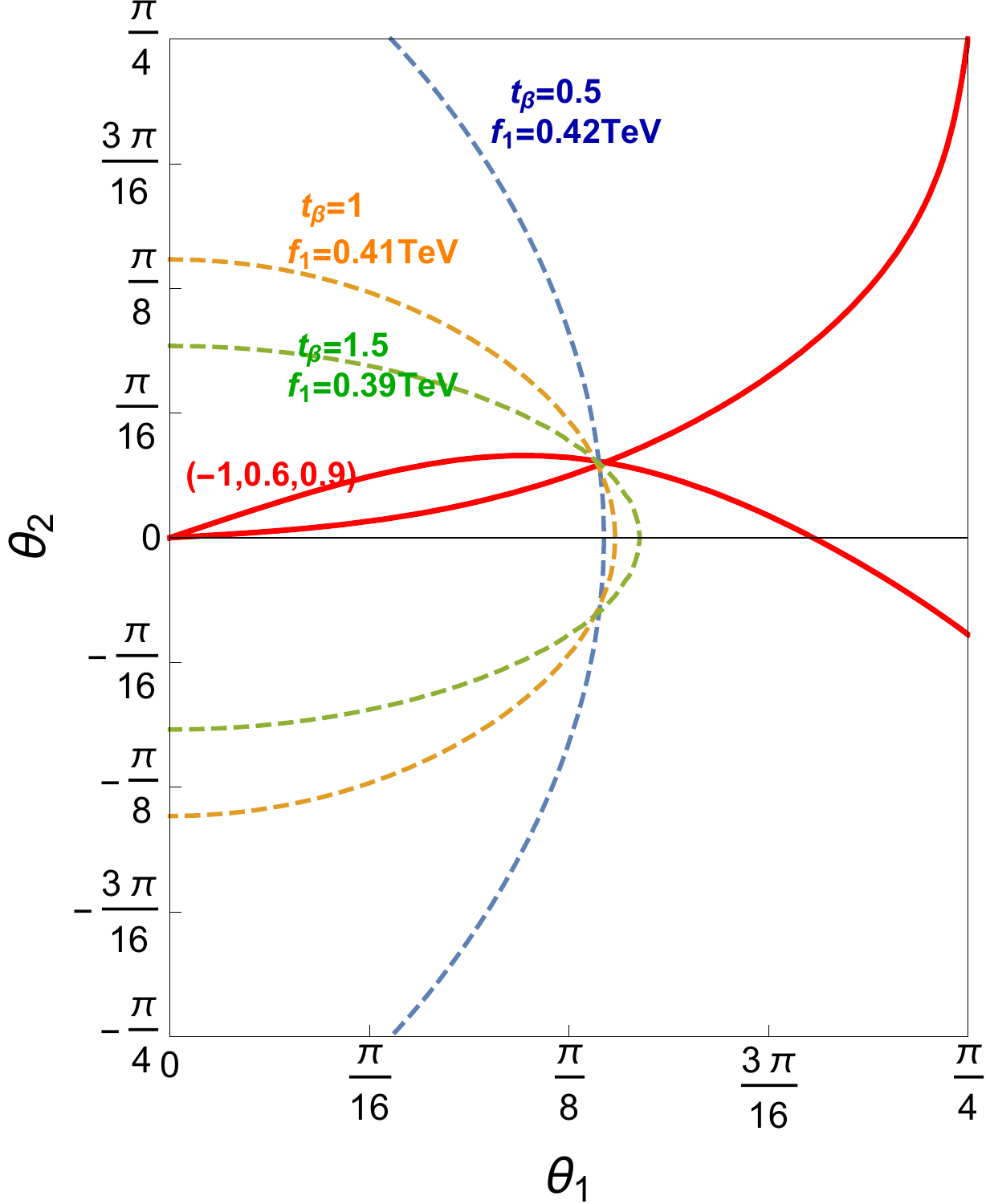}
\caption{
The solid contours on the $(\theta_1, \theta_2)$ planes shows the VEV relations for fixed $t_\beta = 1.5$
and different $f_1 = 0.5, 0.75, 1, 2.5 $ TeV. 
The dashed lines shows the VEV relations for different $t_\beta = 0.5, 1, 1.5, 2$ with fixed $f_1 = 0.5 $ TeV (the left panel) and $f_1 = 1 $ TeV (the  right panel).
}
\label{fig:vev12}
\end{center}
\end{figure}

The set of parameters $(\Omega_1, \Omega_2, F_\lambda, F_m)$ only uniquely determine $(\theta_{1}, \theta_2)$, but not 
the VEVs $(v_1, v_2)$. 
To obtain the electroweak VEVs, additional condition (the VEV condition) needs to be imposed:
\bea
	f_1^2 \sin^2\theta_1 + f_2^2 \sin^2 \theta_2 = v^2,
	\label{eq:VEVcondition}
\eea
where $v = 174$ GeV. 
Given $t_\beta$ and $\theta_{1,2}$, we could determine $f_1$ and $f_2$. 
Fig.~\ref{fig:vev12} (left) shows the VEV contours on the $\theta_1$ and $\theta_2$ plane for different $f_1$ and $t_\beta$. 
$f_1$ determines the intersection point between the curve and  the $x$-axis, 
while $t_\beta$ determines the bending behaviour of the curve. 
Fig.~\ref{fig:vev12} (middle and right) shows that once $t_\beta$ is fixed, $f_1$ could be determined and thus the VEV contour is fixed.
There is one special case. In Fig.~\ref{fig:vev12} (middle), when the tree-level breaking term is off, 
$f_1$ is the same for different $t_\beta$. 
Thus in radiative $\mathbb{Z}_2$ case, $f_1$ is  determined by  $(\Omega_1, \Omega_2)$.

\begin{figure}[!t]
\begin{center}
\includegraphics[width=5.8cm]{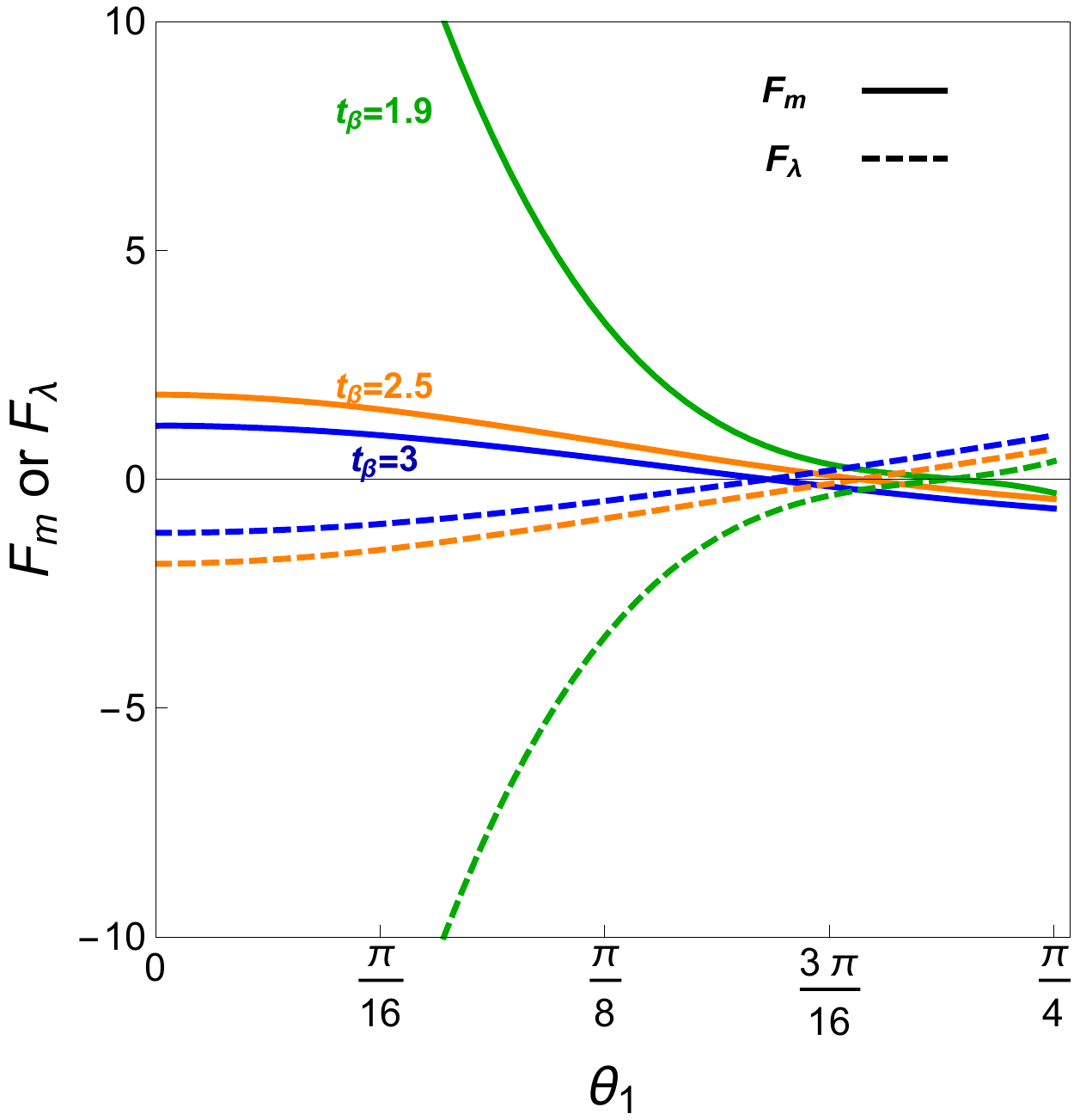}
\includegraphics[width=5.8cm]{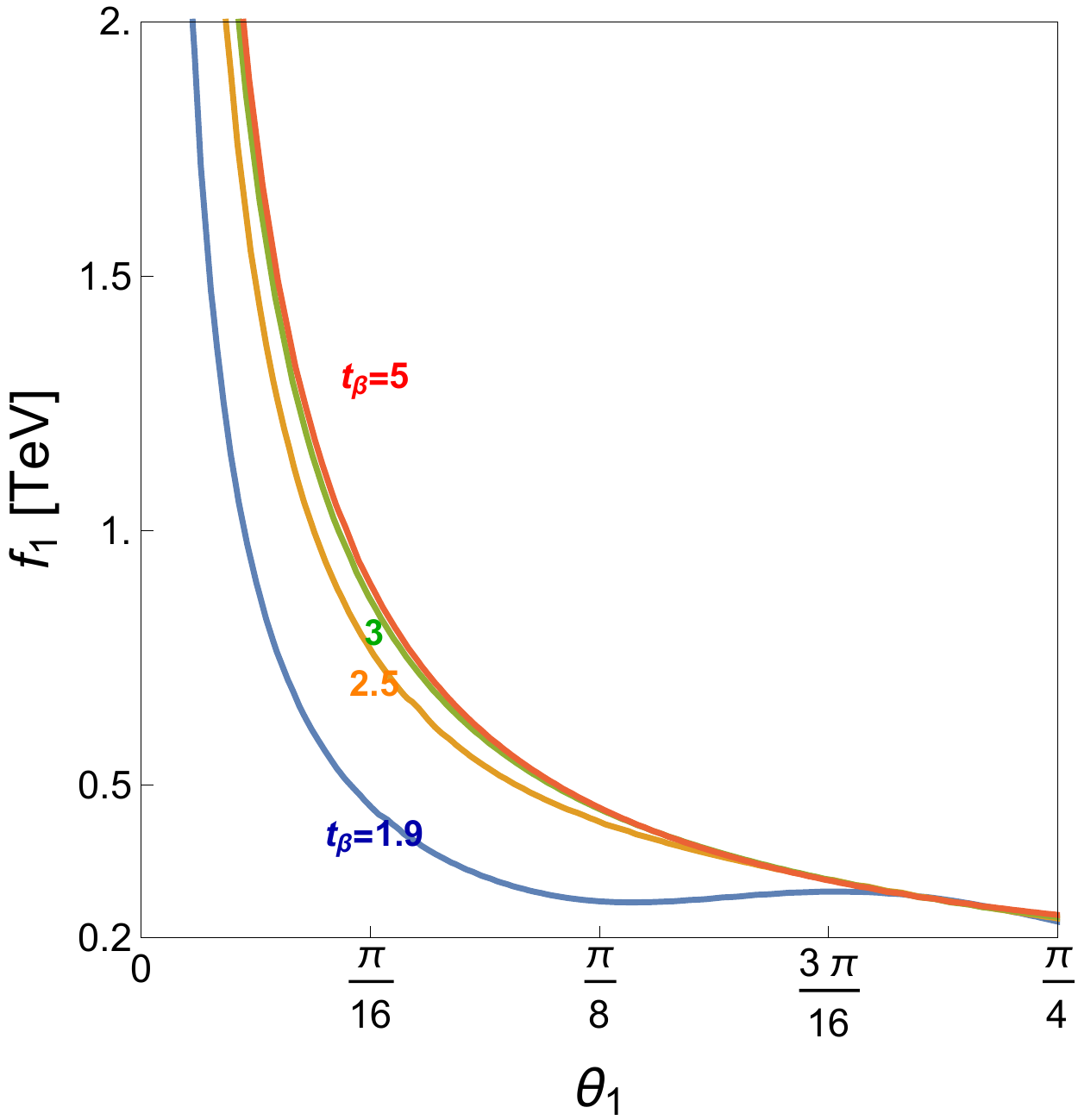}
\caption{
On the left panel, it shows the relations between $\theta_1$, and $F_m$ (solid lines) or $F_\lambda$ (dashed lines). 
On the right panel, the contours shows the relations between $\theta_1$ and $f_1$
imposed by the VEV condition for different $t_\beta$.   
}
\label{fig:Flamx1}
\end{center}
\end{figure}

Let us estimate 
the range of the tree-level breaking parameters $F_m, F_\lambda$ and global symmetry breaking scales $f_{1,2}$
if we would like to obtain $\theta_1 < \pi/4$. 
Fig.~\ref{fig:Flamx1} (left) shows that value of $F_m$ or $F_\lambda$ which determines $\theta_1$ for different $t_\beta$. 
Interestingly, even when $F_m$ or $F_\lambda$ is absent, 
we could still obtain $\theta_1 < \pi/4$, which corresponds to the radiative breaking scenario.
Fig.~\ref{fig:Flamx1} (right) shows that once we know $\theta_1$ (and $t_\beta$), $f_1$ is determined. 
As $t_\beta$ gets larger, the needed $f_1$ becomes larger.
Note that this relation is quite general and does not depend on scenarios.
Thus in tadpole or quartic induced symmetry breaking, only two independent parameters are needed, which are typically taken to be $F_m (F_\lambda)$ and $t_\beta$. 
If there is no tree-level breaking term, only one parameter $t_\beta$ could determine the VEVs.

Finally let us summarize what we have obtained so far from the tadpole conditions. 
The tadpole conditions determine $(\theta_1, \theta_2)$, which depend on $(\Omega_1, \Omega_2)$ and/or $(F_m, F_\lambda)$.
Given  $\Omega_{1,2} \le 0$, the vacuum misalignment requires  $\theta_2 < \theta_1 < \pi/4$ with $|\Omega_1 +\Omega_2| > 1$. 
Three scenarios are discussed to obtain this misalignment.
We classify these scenarios according to parameters $(\Omega_1, \Omega_2)$ and $(F_m, F_\lambda)$:
\bit
\item Radiative $\mathbb{Z}_2$ breaking~\cite{Yu:2016bku}, when $\Omega_1 \neq 0, \Omega_2 \neq 0$.   
Since there is no tree-level breaking term, the tree-level potential is $U(4)\times U(4)$ invariant:
\bea
	V_{U(4)\times U(4)} &=& 
	-\mu_1^2 |H_1|^2 - \mu_2^2 |H_2|^2
	+ \lambda_1 (|H_1|^2)^2 + \lambda_2 (|H_2|^2)^2 
	+ \lambda_3 |H_1|^2 |H_2|^2.
\eea
The radiative corrections to the scalar potential are shown in Eq.~\ref{eq:loopcorr}. 
The $\Omega_1$ determines whether the electroweak symmetry breaking could happen, while 
the $\Omega_2$ determines whether vacuum misalignment could happen.
Since $\Omega_1 < 0, \Omega_2 < 0$, the solution of the asymmetric vacua have $ \theta_1 < \pi/4$ and $\theta_2 \equiv 0$. 
The two tadpole conditions reduce to one
\bea
	 && \sin 4\theta_1  +  \Omega_2  \sin 2(\theta_1 ) = 0. 
\eea 
Thus the $\theta_1$ only depends on $\Omega_{2}$: the larger $\Omega_{2}$ the smaller $\theta_1$. 
Although the gauge corrections are much smaller than the Yukawa corrections, $\Omega_2$ could be large if $t_\beta \gg 1$. 
When $\Omega_{2}$ approaches one, $\theta_1$ approaches to zero. 

\item $m_{12}^2$-induced $\mathbb{Z}_2$ breaking~\cite{Beauchesne:2015lva, Harnik:2016koz}, when $\Omega_1 \neq 0, m_{12}^2 \neq 0$. 
The tree-level potential is 
\bea
	V_{\rm Tadpole} &=& 
	V_{U(4)\times U(4)} +  m_{12}^2 \left[ H_1^\dagger H_2 +h.c.\right]. 
\eea
The dominant radiative corrections to the scalar potential are the same as the radiative $\mathbb{Z}_2$ breaking case. 
Similarly $\Omega_1$ determines whether the electroweak symmetry breaking could happen. 
However, if $t_\beta$ is not so large, only $\Omega_2$ could not obtain small enough $\theta_1$.
In certain case $\Omega_2$ could even disappear.
The $m_{12}^2$ is needed to obtain appropriate $\theta_1$. 
In this case it is the $m_{12}^2$ which determines whether vacuum misalignment could happen. 
As explained in Ref. and next section, the $m_{12}^2$ plays the role of the tadpole terms, which
transit the value of $\theta_1 $ to $\theta_2$, and thus the $\theta_1 < \pi/4$ is produced.

\item $\lambda_{45}$-induced $\mathbb{Z}_2$ breaking, when $\Omega_1 \neq 0, \lambda_{45} \neq 0$. 
The tree-level potential is 
\bea
	V_{\rm Quartic} &=& 
	V_{U(4)\times U(4)}  
	+ \lambda_4 |H_1^\dagger H_2 |^2
	+ \lambda_5 \left[(H_1^\dagger H_2)^2 + h.c.\right].
\eea
Similarly $\Omega_1$ determines whether the electroweak symmetry breaking could happen. 
Although $\Omega_{2}$ could exist, it is the $\lambda_{4,5}$ determines whether vacuum misalignment could happen.
Negative $\lambda_{4,5}$ is favored to obtain appropriate $\theta_1$.

\eit
It is also possible that both both $m_{12}^2$ and $\lambda_{4,5}$ terms  exist in the potential. 
In this case, it is the $m_{12}^2$ and $\lambda_{4,5}$ which determine whether vacuum misalignment could happen. 
This is the mixture of the tadpole and quartic induced $\mathbb{Z}_2$ breaking scenarios. 
%


\section{Spontaneous $\mathbb{Z}_2$ Breakings in 2HDM Framework}
\label{sec:pheno}

In the above section, we discussed how the tadpole conditions determine the electroweak vacua $(\theta_1, \theta_2)$. 
Three different mechanisms could lead to the vacuum misalignment $\theta_2 < \theta_1 < \pi/4$, and thus the spontaneous $\mathbb{Z}_2$ breaking.
Let us understand the physics behind these $\mathbb{Z}_2$ breaking scenarios.

Since the electroweak symmetry breaking only involves in the PGBs in visible $A$ sector,
we will simplify the original scalar potential in Eq.~\ref{eq:2HDMPot} and Eq.~\ref{eq:loopcorr} by setting
\bea
	H_{1B} \simeq \left(\begin{array}{c}
0\\
f_1 \cos\left( \frac{\mathbb{H}_1}{f_1} \right) 
\end{array} \right), \qquad 
H_{2B} \simeq \left(\begin{array}{c}
0\\
f_2 \cos\left( \frac{\mathbb{H}_2}{f_2} \right) 
\end{array} \right). 
\eea 
Expanding the potential to the quartic order, we obtain the approximated visible sector potential 
in the 2HDM framework
\bea
	V(H_{1A}, H_{2A}) &=& 
	-\mu_{1A}^2 |H_{1A}|^2 - \mu_{2A}^2 |H_{2A}|^2
	+ \lambda_{1A} |H_{1A}|^4 + \lambda_{2A} |H_{2A}|^4
	+ \lambda_{3A}  |H_{1A}|^2 |H_{2A}|^2 \nn \\
	&&
	+  m_{A12}^2 \left[ H_{1A}^\dagger H_{2A} +h.c.\right] 
	+ \lambda_{4A} |H_{1A}^\dagger H_{2A} |^2
	+ \frac{\lambda_{5A}}{2}  \left[(H_{1A}^\dagger H_{2A})^2 + h.c.\right]\nn\\
	&&
	+  \left[  (\lambda_{6A}|H_{1A}|^2 + \lambda_{7A}|H_{2A}|^2)H_{1A}^\dagger H_{2A} + h.c. \right].
	\label{eq:2HDMPotA}
\eea
All the coefficients in the potential are proportional to the tree-level and loop-induced breaking terms: 
\bea
	\mu_{1A}^2 & = & 2 \delta_1 f_1^2 + (\delta_{345}+\lambda_{45})f_2^2 + m_{12}^2 t_\beta,\nn\\
	\mu_{2A}^2 & = & 2 \delta_2 f_2^2 + (\delta_{345}+\lambda_{45})f_1^2 + m_{12}^2 t_\beta^{-1},\nn\\
	\lambda_{1A} &=& \frac{8 \delta_1}{3} + \frac{\delta_{345} + \lambda_{45}}{3} t_\beta^{2} + \frac{m_{12}^2}{12 f_1^2} t_\beta,\nn\\
	\lambda_{2A} &=& \frac{8 \delta_2}{3} + \frac{\delta_{345} + \lambda_{45}}{3} t_\beta^{-2} + \frac{m_{12}^2}{12 f_2^2} t_\beta^{-1},\nn\\
	\lambda_{3A} &=& \delta_{345} + \lambda_{45} + \frac{m_{12}^2}{2 f_1 f_2},\nn\\
	m_{A12}^2 &=& m_{12}^2 + \lambda_{45}f_1 f_2,\quad
	\lambda_{4A} = \lambda_4 + \delta_4,\quad \lambda_{5A} = \lambda_5 + \delta_5, \nn\\
	\lambda_{6A} &=& \frac{2 \lambda_{45} }{3} t_\beta + \frac{m_{12}^2}{6 f_1^2},\quad \lambda_{7A} =  \frac{2 \lambda_{45} }{3} t_\beta^{-1} + \frac{m_{12}^2}{6 f_2^2}. 
\eea
Note that there is no dependence on the tree-level parameters $\lambda_{1-3}$.

\subsection{Radiative $\mathbb{Z}_2$ Breaking}

In this scenario, the tree-level breaking terms $m_{12}^2$ and $\lambda_{4,5}$ do not exist. 
From the above potential, the Higgs mass squared terms reduce to 
\bea
	\mu_{H_{1A}}^2 &=&  f_1^2 \left(2 \delta_1 + \delta_{345} t_\beta^2\right) , \quad
	\mu_{H_{2A}}^2 =  f_2^2 \left(2 \delta_2 + \delta_{345} t_\beta^{-2}\right),
\eea
and the quartic terms reduce to 
\bea
\lambda_{1A} = \frac{8 \delta_1}{3} + \frac{\delta_{345}}{3} t_\beta^{2}, 
\quad
\lambda_{2A} =  \frac{8 \delta_2}{3} + \frac{\delta_{345}}{3} t_\beta^{-2}.
\eea
Since $H_{1A}$ has negative mass-squared $-\mu_{H_{1A}}^2<0$, $H_{1A}$ gets VEV. 
While $H_{2A}$ has positive mass-squared  $-\mu_{H_{2A}}^2>0$, there is no VEV for $H_{2A}$. 
The potential reduces to the inert Higgs doublet potential~\cite{Barbieri:2006dq}:
\bea
	V_{\rm inert} &\supset& 
		\left[-|\mu_{1A}^2| |H_{1A}|^2 + \lambda_{1A} |H_{1A}|^4  \right]
		+\left[|\mu_{2A}^2| |H_{2A}|^2 + \lambda_{2A} |H_{2A}|^4 + \delta_{345} |H_{1A}|^2 |H_{2A}|^2\right],
\eea
where we neglect the $\delta_{4,5}$ contributions.
The first two terms in the potential, determine the electroweak vacuum:
\bea
	v^2 = \mu_{1A}^2/\lambda_{1A}  \simeq  \frac38 \left(2  + \delta_{345} t_\beta^2/\delta_1 \right) f_1^2.
\eea
To obtain the electroweak VEV $v = 174$ GeV, the two terms should cancel with each other. 
We know although contributions $\delta_1$ (from Yukawa corrections) and $\delta_{345}$ (from gauge
corrections) have opposite sign, the adequate cancellation will not happen because typically 
$\delta_1 \gg \delta_{345}$. 
To have required cancellation, we could utilize large $t_\beta$ in the second term to enhance the 
second term in the  mass-squared  $\mu_{H_{2A}}^2$. 
At the same time, if $\lambda_{1A}$ is not reduced compared to the mass-squared  $\mu_{H_{2A}}^2$,
the electroweak VEV is obtained.

Finally, we could read out the masses of the PGBs.
The Higgs mass is
\bea
	m_{h}^2 = 2\mu_{1A}^2 = \left(4 \delta_1 + 2\delta_{345} t_\beta^2\right) f_1^2.
\eea
We could obtain the masses of the charged and neutral scalars in the inert doublet:
\bea
	m_{H^\pm}^2 &=& \mu_{2A}^2 + \delta_{345} v^2, \quad
	m_{A^0}^2  =  \mu_{2A}^2 + \delta_{345} v^2. 
\eea
In typical 2HDM model, the masses of the charged and CP-odd neutral scalar 
are only proportional to $\delta_{4,5}$, which is very small. 
In this scenario, the inert scalar masses also have $\delta_{2,3}$ dependence, 
which induces a large mass for the inert scalars. 
Therefore, the radiative  $\mathbb{Z}_2$  breaking scenario can be viewed as a natural UV completion of the inert Higgs doublet model.

\subsection{Tadpole Induced $\mathbb{Z}_2$ Breaking}

The radiative $\mathbb{Z}_2$ breaking scenario can only realize the electroweak symmetry breaking 
when $\delta_{345}$ is non-zero, and if the enhancement from $t_\beta \gg 1$ exists. 
Otherwise, the vacuum misalignment cannot be obtained by purely radiative $\mathbb{Z}_2$ breaking.
The tadpole induced $\mathbb{Z}_2$ breaking scenario is suitable to the case with $\delta_{345}$ is zero, or $t_\beta \sim 1$. 
However, the price to pay is introducing additional $m_{12}^2$ term.

Let us turn on $m_{12}^2$ gradually to see how the VEVs $\theta_{1,2}$ vary. 
When $m_{12}^2$ term is off, from the radiative breaking scenario, the VEVs have
\bea
	\langle H_{1A} \rangle \simeq f_1, \quad \langle H_{2A} \rangle \simeq 0.
\eea
If $t_\beta \sim 1$, we obtain $\mu_{1A}^2 \gg \mu_{2A}^2$ due to $\delta_1 \gg -\delta_{345}$. 
Thus $m_{h_1}$ is much heavier than $m_{h_2}$.   
When gradually turning on $m_{12}^2$ term, the $h_2$ starts to obtain small VEV. 
This can be seen from the potential assuming $h_1$ is too heavy and decoupled. 
After integrating out $h_1$, the potential generates an effective tadpole term. 
The $h_2$ potential is dominated by the tadpole and quadratic terms
\bea
	V(H_{2A}) &\supset&  \mu_{2A}^2 h_2^2 + m_{12}^2 f_1  h_2 
\eea
Thus $h_2$ obtain VEV 
\bea
\langle H_{2A} \rangle  \sim m_{12}^2 f_1/\mu_{2A}^2,
\eea
which gradually becomes large as $m_{12}^2$ increases. 
At the same time, the VEV of $h_1$ decreases. 
This can be seen from the $H_{1A}$ potential.  
Assuming the VEV  $\langle h_2 \rangle$ is small, the relevant $H_{1A}$ potential is
\bea
	V(H_{1A}) &\simeq& 
	-\left(2 \delta_1 f_1^2 + \delta_{345} f_2^2 + m_{12}^2 t_\beta\right) |H_{1A}|^2  
	+ \left( \frac{8 \delta_1}{3} + \frac{\delta_{345}}{3} t_\beta^{2} + \frac{m_{12}^2}{12 f_1^2} t_\beta\right) |H_{1A}|^4. 
\eea
Here the tadpole contribution is negligible due to $\langle h_1 \rangle > \langle h_2 \rangle$. 
From the potential, we see that as the $m_{12}^2$ becomes larger, there 
are large cancellation in the quadratic term, which cause the VEV $\langle h_1 \rangle$ becomes smaller.
Therefore, the bilinear term $m_{12}^2$ plays the role of an effective tadpole. 
As the effective tadpole term increases, the VEV $\theta_1$ decreases from $\pi/4$, while
the VEV $\theta_2$ increases from $0$. 
The vacuum misalignment $\theta_2 < \theta_1 < \pi/4$ could be realized when an appropriate $m_{12}^2$ term is taken.

\subsection{Quartic Induced $\mathbb{Z}_2$ Breaking}

In this scenario, only quartic breaking terms $\lambda_{4,5}$ are in the tree level potential. 
Unlike the $m_{12}^2$ case, the quartic breaking scenario could be all the range of the $t_\beta$. 
The $\lambda_{4,5}$ terms appear in both quadratic term $\mu_{1A}^2$ and the bilinear term. 
The quadratic term has 
\bea
	\mu_{1A}^2 &=&  f_1^2 \left(2 \delta_1 + \left(\delta_{345} + \lambda_{45}\right) t_\beta^2\right).
\eea
We see that both $\delta_{345}$ and $\lambda_{4,5}$ contribute to cancel the 
opposite $\delta_1$ corrections.
At the same time, it generates the effective tadpole term, which transits the value of the VEV $\theta_1$ to the one of the $\theta_2$. 
Since it is similar to the above cases, it could generate the appropriate $\mathbb{Z}_2$ breaking.


\section{Higgs Phenomenology}
\label{sec:mass}

\subsection{Higgs Mass Spectra}

In this natural 2HDM framework, it contains two Higgs doublets $H_{1A}, H_{2A}$ in $A$ sector, 
and another two Higgs doublets $H_{1B}, H_{2B}$ (with two neutral radial mode decoupled) in $B$ sector.
There are six exact GBs: three ($z^{\pm,0}$) from $H_{iA}$ and three ($C^\pm, N^0$) from $H_{iB}$. 
All of them are eaten by gauge bosons in $A$ and $B$ sectors.
Depending on the breaking pattern, the other particles in the scalar multiplets could be 
PGBs or just scalar particles.
We present the details of the mass spectra in two breaking pattern in Appendix A and B.

\begin{figure}[!t]
\begin{center}
	\includegraphics[width=7cm]{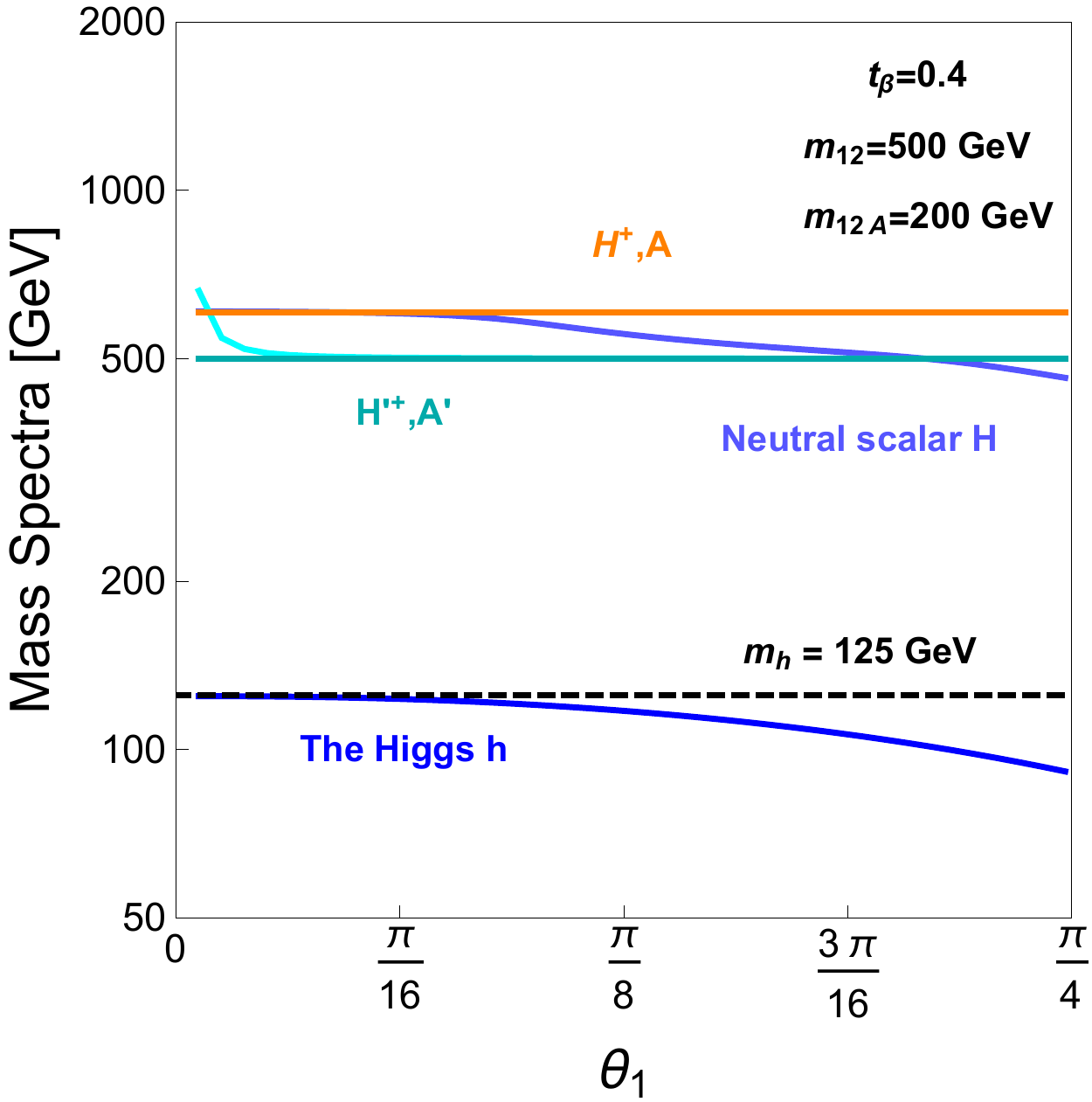} 
	\includegraphics[width=7cm]{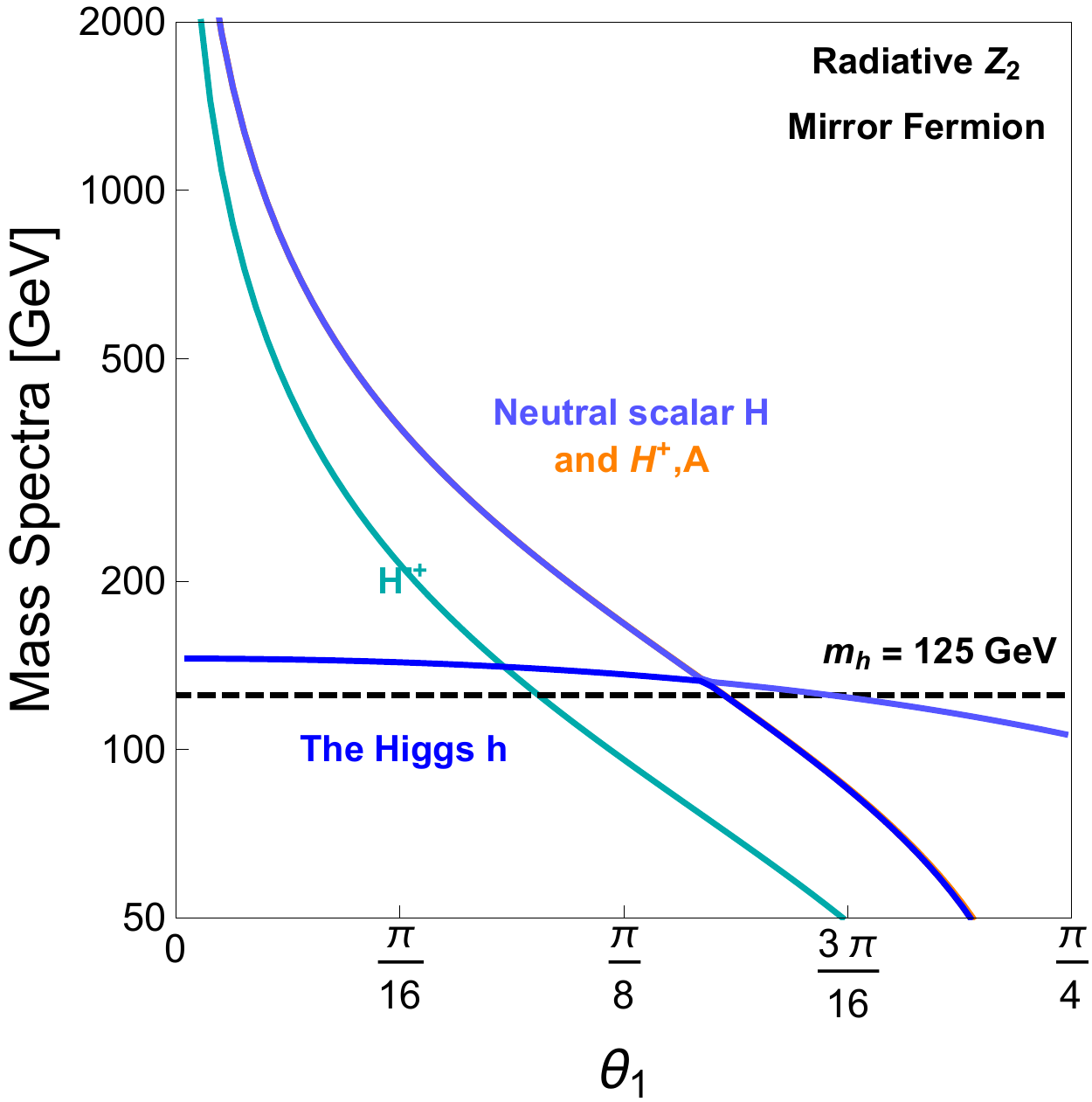} \\
	\includegraphics[width=7cm]{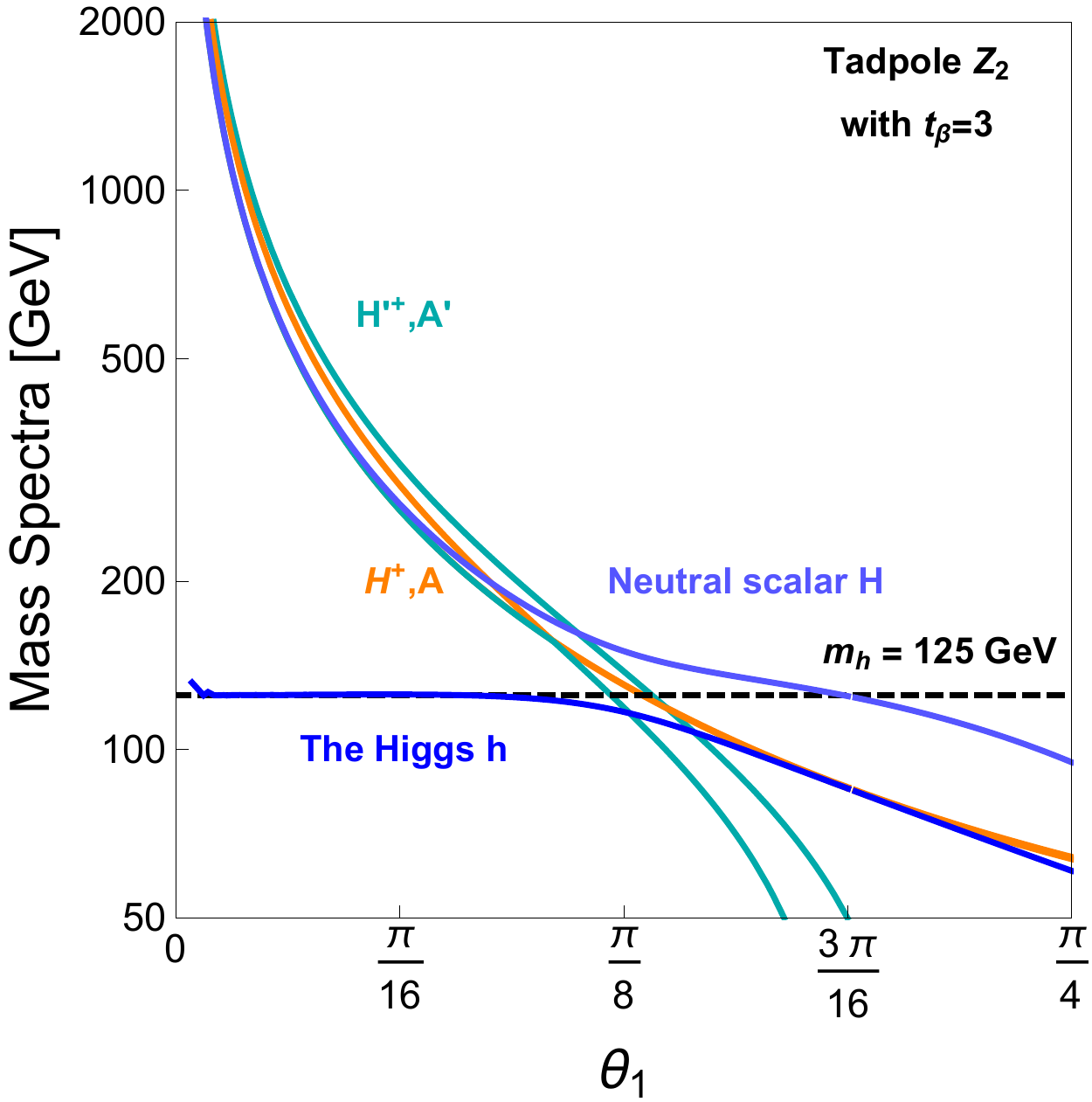} 
	\includegraphics[width=7cm]{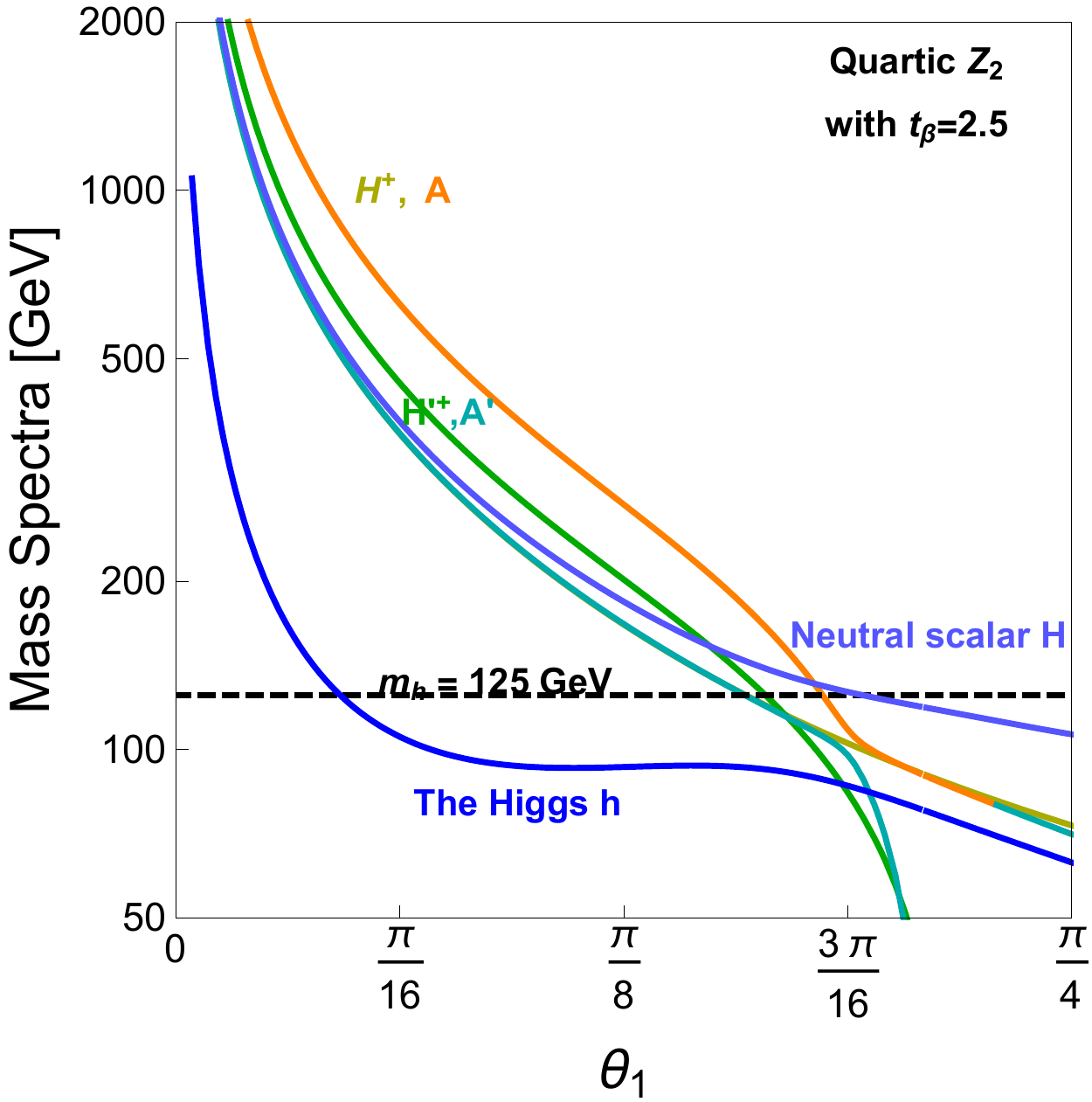} 
	\caption{
  The masses spectra as the function of $\theta_1$ in four  $\mathbb{Z}_2$ breaking scenarios. 
  The particles in the mass spectra are charge and neutral CP odd Higgses $(H^\pm, A^0)$ in visible sector, 
  charged and neutral CP odd Higgses $(H'^\pm, A'^0)$ in twin sector, and 
  two CP even Higgses $(h, H)$ in visible sector. 
}
\label{fig:massspectra}
\end{center}
\end{figure}

Here we summarize main features of the mass spectra based on results in Appendix A and B. 
\bit

\item {\underline{Explicit Soft $\mathbb{Z}_2$ Breaking}} 

In the Higgs basis, the field $H$ plays the role of twin Higgs, while another field $H'$ is just additional scalar $U(4)$ multiplet. 
Thus among seven GBs, six are eaten by gauge bosons, and one PGB is the Higgs boson. 
For the additional scalars in $H'$, the masses are 
\bea
	m_{H^\pm}^2  &=&    m_{A}^2  = \frac{m_{12}^2 + m_{12A}^2}{s_\beta c_\beta} - \frac12 (\lambda_4 + \lambda_5) f^2, \nn\\
      m_{H'^\pm}^2 &=&    m_{A'}^2 = \frac{m_{12}^2}{s_\beta c_\beta} - \lambda_5 f^2.
\eea
which only depends on $U(4)$ breaking parameters $m_{12}^2, m_{12A}^2$ and $\lambda_{4,5}$ in the potential. 
If the tree-level terms $\lambda_{4,5}$ do not exist, then all the new scalars have degenerate masses. 
In SUSY extension of the twin Higgs model, the mass spectra are much simplified.

\item {\underline{Radiative $\mathbb{Z}_2$ Breaking}}

The global symmetry breaking is $[U(4)\times U(4)] \to [U(3) \times U(3)]$. 
All the scalar components in visible sector are PGBs. 
Furthermore, since only $H_{1A}$ obtains VEV, $H_{2A}$ is an inert Higgs doublet. 
In the twin sector, since both  $H_{1B}$ and $H_{2B}$ have VEVs, 
the PGBs in twin sector have mixing. 
The PGBs mass eigenstates have
\bea
	m_{H^\pm}^2 & = &  
	 - 2 \delta_2 f_2^2 - \delta_{345}f_1^2 \cos 2\theta_1
	 	- \delta_{45}f_1^2\sin^2 \theta_1, \qquad
	 m_{H'^\pm}^2 =  -\delta_{45} (f_1^2 \cos^2\theta_1 + f_2^2),\nn\\
	m_{A^0}^2 & = &  
        - 2 \delta_2 f_2^2 - \delta_{345}f_1^2 \cos 2\theta_1
      	- 2\delta_{5}f_1^2\sin^2 \theta_1, \qquad 
	 m_{A'^0}^2 = -2\delta_{5} (f_1^2 \cos^2\theta_1 + f_2^2).
\eea

\item {\underline{Tadpole-Induced $\mathbb{Z}_2$ Breaking}}

Similar to the radiative breaking scenario, 
all the scalars except the radial modes in two two Higgs doublets are PGBs. 
The difference is that there are mixing between two Higgs doublet in $A$ sector, with mixing angle $\beta_A$, defined in Appendix B. 
All the masses of the charged Higgses and CP odd Higgses depend on $m_{12}^2$ and 
are nearly degenerate when $\theta_{1,2}$ are much smaller than $\pi/4$.
The mass spectra read
\bea
	m_{H^\pm}^2 & = &  m_{A^0_1}^2  = 
	\frac{m_{12}^2}{f_1\sin\theta_1 f_2\sin\theta_2} \left(f_1^2 \sin^2\theta_1 + f^2_2 \sin^2 \theta_2\right) 
	,\nn\\
	m_{H'^\pm}^2 & = & m_{A_2^0}^2 = 
	\frac{m_{12}^2}{f_1\cos\theta_1 f_2\cos\theta_2}\left(f_1^2 \cos^2\theta_1 + f^2_2 \cos^2 \theta_2\right).
\eea

\item  {\underline{Quartic-Induced $\mathbb{Z}_2$ Breaking}}

Similar to the tadpole induced breaking scenario, 
all the masses of the charged Higgses and CP odd Higgses depend on $\lambda_{4,5}$. 
The difference is that there are 
mass splitting between charged and neutral CP odd Higgses unless $\lambda_4 = \lambda_5$. 
The charged scalar masses are
\bea
	m_{H^\pm}^2 & = &  
	- \lambda_{45}(1 + \cot\theta_1\cot\theta_2) \left(f_1^2 \sin^2\theta_1 + f^2_2 \sin^2 \theta_2\right),\nn\\
	m_{H'^\pm}^2 & = & 
	- \lambda_{45}(1 + \tan\theta_1\tan\theta_2)\left(f_1^2 \cos^2\theta_1 + f^2_2 \cos^2 \theta_2\right).
\eea
and the CP-odd scalar masses are presented in Appendix B. 

\eit
In all scenarios, the SM Higgs boson origins from the mixing between $h_1$ and $h_2$ in visible sector. 
The exception is that in radiative breaking scenario, there is no mixing between $h_1$ and $h_2$. 
The Higgs mass is proportional to all the breaking parameters $\delta_{1-5}$ and/or $m_{12}^2 (\lambda_{4,5})$. 
We present the mass matrices of the Higgs boson in Appendix A and B.
Fig.~\ref{fig:massspectra} shows the mass spectra in the above four scenarios. 
The independent parameters in four scenarios are taken to be $(\theta_1, t_\beta, m_{12}=500, m_{12A}=200)$ (explicit $Z_2$ breaking),
$\theta_1$ (radiative breaking), $\theta_1, t_\beta = 3$ (tadpole breaking), and $\theta_1, t_\beta = 2.8$ (quartic breaking).
Fig.~\ref{fig:massspectra} shows that typically charge and neutral CP odd Higgses $(H^\pm, A^0)$ in visible sector have degenerate 
masses, and similarly for charged and neutral CP odd Higgses $(H'^\pm, A'^0)$ in twin sector. 
If we want to obtain 125 GeV Higgs boson mass, 
this will give us additional constraint on the model parameters. 
Fig.~\ref{fig:massspectra} shows that once we fix other parameters in the model, 
$\theta_1$ is determined by the requirement of the 125 GeV mass of the Higgs boson.


%

\subsection{Collider Constraints}

Let us first consider the visible sector. 
The visible sector contains the same particle contents as the 2HDM. 
The phenomenology in visible sector should be very similar to the one in 2HDM, 
except that  there could be additional decay channels to the twin particles.
For simplicity, we take the Type-I Yukawa structure in this work, although other Yukawa structure, such as Type-II, X, Y, are possible.  
Let us setup the notation similar to 2HDM. 
According to the Appendix A and B, the 2HDM mixing angles and electroweak VEV are different in two breaking patterns, defined as
\bea
	 {U(4)/U(3)} &\equiv& \begin{cases}		
	\beta = \frac{f_2}{f_1},  &  \text{mixing angle between charged/CP-odd scalars}, \\
	v  = f \sin\theta, &  \text{electroweak vacuum}, \\
	\alpha,  &  \text{mixing angle between CP even scalars},
	\end{cases}
	\\
	 {U(4)^2/U(3)^2} &\equiv&
	\begin{cases}		
	\beta_A = \frac{f_2 \sin\theta_2}{f_1 \sin\theta_1},  &  \text{mixing angle between charged/CP-odd scalars}, \\
	v = \sqrt{f_1 \sin\theta_1 + f_2 \sin\theta_2}, &  \text{electroweak vacuum}, \\
	\alpha,  &  \text{mixing angle between CP even scalars}.
	\end{cases}
\eea
Note that the definition of the mixing angle $\alpha$ is opposite from the typical notation, such as Ref.~\cite{Branco:2011iw}. 
The normalized Higgs couplings to the SM gauge bosons and fermions are 
\bea
	\kappa_{hVV} \equiv \frac{g_{hVV}}{g_{hVV}^{\rm SM}}&=& \begin{cases}		
	\cos\theta c_{\alpha - \beta}, &  \text{Explicit $\mathbb{Z}_2$ Breaking}, \\
	\cos\theta_1,  &  \text{Radiative $\mathbb{Z}_2$ Breaking}, \\
	\cos\theta_1 c_\alpha c_{\beta_A}  + \cos\theta_2  s_\alpha s_{\beta_A} , & \text{Tadpole and Quartic $\mathbb{Z}_2$ Breaking},
	\end{cases}\\
	\kappa_{hff} \equiv \frac{y_{hff}}{y_{hff}^{\rm SM}} &=& \begin{cases}		
	 \cos\theta \frac{c_\alpha}{c_\beta}, &  \text{Explicit $\mathbb{Z}_2$ Breaking}, \\
	\cos\theta_1,  &  \text{Radiative $\mathbb{Z}_2$ Breaking}, \\
	\cos\theta_1\frac{c_\alpha}{c_{\beta_A}},  & \text{Tadpole and Quartic $\mathbb{Z}_2$ Breaking}.
	\end{cases}
\eea
Here the SM couplings are taken to be $g_{hVV}^{\rm SM} = \frac{2m_V^2}{v}$ and $g_{hff}^{\rm SM} = \frac{m_f}{v}$. 
These Higgs couplings are constrained by the Higgs coupling measurements at the LHC.
The charged and CP-odd neutral scalars in visible sector have the same constraints as the one in 2HDM. 
On the other hand, the CP-even neutral scalars need to take the decays to twin particles into account.

The twin sector contains another two Higgs doublet $H_{1B,2B}$, the mirror gauge bosons, and mirror fermions. 
The mirror gauge bosons are mirror photon, and mirror $W_B, Z_B$, which absorb 
three GBs in two Higgs doublets.
For simplicity, two radial modes in $H_{1B,2B}$ are assumed to be decoupled. 
The physical scalars in twin sector are charged and neutral CP odd scalars $H'^\pm, A'^0$. 
The mirror fermions could induce very rich twin hadron phenomenology~\cite{Craig:2015xla} because they are 
charged under the mirror QCD. 
Since the twin fermions are mirror copy of the SM particles, the mirror fermion phenomenology should be similar to the original twin Higgs. 
For simplicity, we take the fermion setup in the fraternal twin Higgs model~\cite{Craig:2015xla}, 
and leave more general discussion for future.
The fermionic ingredients of the fraternal twin Higgs setup are summarized as follows:
\bit
\item To avoid the twin $SU(3)$ and twin $SU(2)$ anomalies, 
the whole third generation twin fermions are introduced: twin top, bottom, tau, and twin tau neutrino, but not first two generations;
\item The fermion Yukawa interactions are taken to be the fermion assignment I in our discussion;
\item The twin $SU(3)$ has confinement, which indicates the existence of the twin glue-balls, and twin bottomonium and hadrons
below confinement scalar $\Lambda'_3 \sim {\mathcal O}(10) \Lambda_{\rm QCD}$. 
\eit 
To be specific, we take the twin bottom Yukawa coupling the same as the bottom Yukawa coupling,
which indicates $m_{b^B} \simeq m_b \frac{f}{v}$.
Thus the Higgs boson could decay into $b_B$: $h \to b_B \bar{b}_B$.
Because twin fermions are SM charge neutral, some of them could be dark matter candidate. 
This has been discussed in Refs.~\cite{Craig:2015xla, Garcia:2015loa, Farina:2015uea}. 
%

%
The Higgs boson and the heavier CP even neutral scalar provide connection between visible and twin sector. 
The Higgs boson also couples to the twin particles because it is a PGB. 
Here we denote the VEV in twin sector $ v' \equiv f\cos\theta$ ($ v' \equiv \sqrt{f_1^2\cos\theta_1^2 + f_2^2\cos\theta_2^2}$), and mixing angle $\beta$ ($\beta_B$) in explicit (spontaneous) breaking pattern. 
The normalized Higgs couplings to the twin gauge bosons and fermions are 
\bea
	\kappa'_{hVV} \equiv \frac{g_{hV_BV_B}}{g_{hV_BV_B}^{\rm SM}}&=& \begin{cases}		
	\sin\theta c_{\alpha - \beta}, &  \text{Explicit $\mathbb{Z}_2$ Breaking}, \\
	\sin\theta_1,  &  \text{Radiative $\mathbb{Z}_2$ Breaking}, \\
	\sin\theta_1 c_\alpha c_{\beta_B}  + \sin\theta_2  s_\alpha s_{\beta_B} , & \text{Tadpole and Quartic $\mathbb{Z}_2$ Breaking},
	\end{cases}\\
	\kappa'_{hff} \equiv \frac{y_{hf_Bf_B}}{y_{hf_Bf_B}^{\rm SM}} &=& \begin{cases}		
	 \sin\theta \frac{c_\alpha}{c_\beta}, &  \text{Explicit $\mathbb{Z}_2$ Breaking}, \\
	\sin\theta_1,  &  \text{Radiative $\mathbb{Z}_2$ Breaking}, \\
	\sin\theta_1\frac{c_\alpha}{c_{\beta_B}},  & \text{Tadpole and Quartic $\mathbb{Z}_2$ Breaking}.
	\end{cases}
\eea
Here the SM-like couplings are taken to be $g_{hV_BV_B}^{\rm SM} = \frac{2m_{V_B}^2}{v'}$ and $g_{hf_Bf_B}^{\rm SM} = \frac{m_{f_B}}{v'}$. 
The Higgs invisible decay channels are $h\to f_B \bar f_B, V_B V_B, A_B A_B$. 
Since in general the normalized couplings of the Higgs boson to the twin gauge bosons and fermions are different, 
we can not do a simple scaling on the signal strength. We calculate  the Higgs invisible decay widths based on the above couplings.

\begin{figure}[!t]
\begin{center}
	\includegraphics[width=5cm]{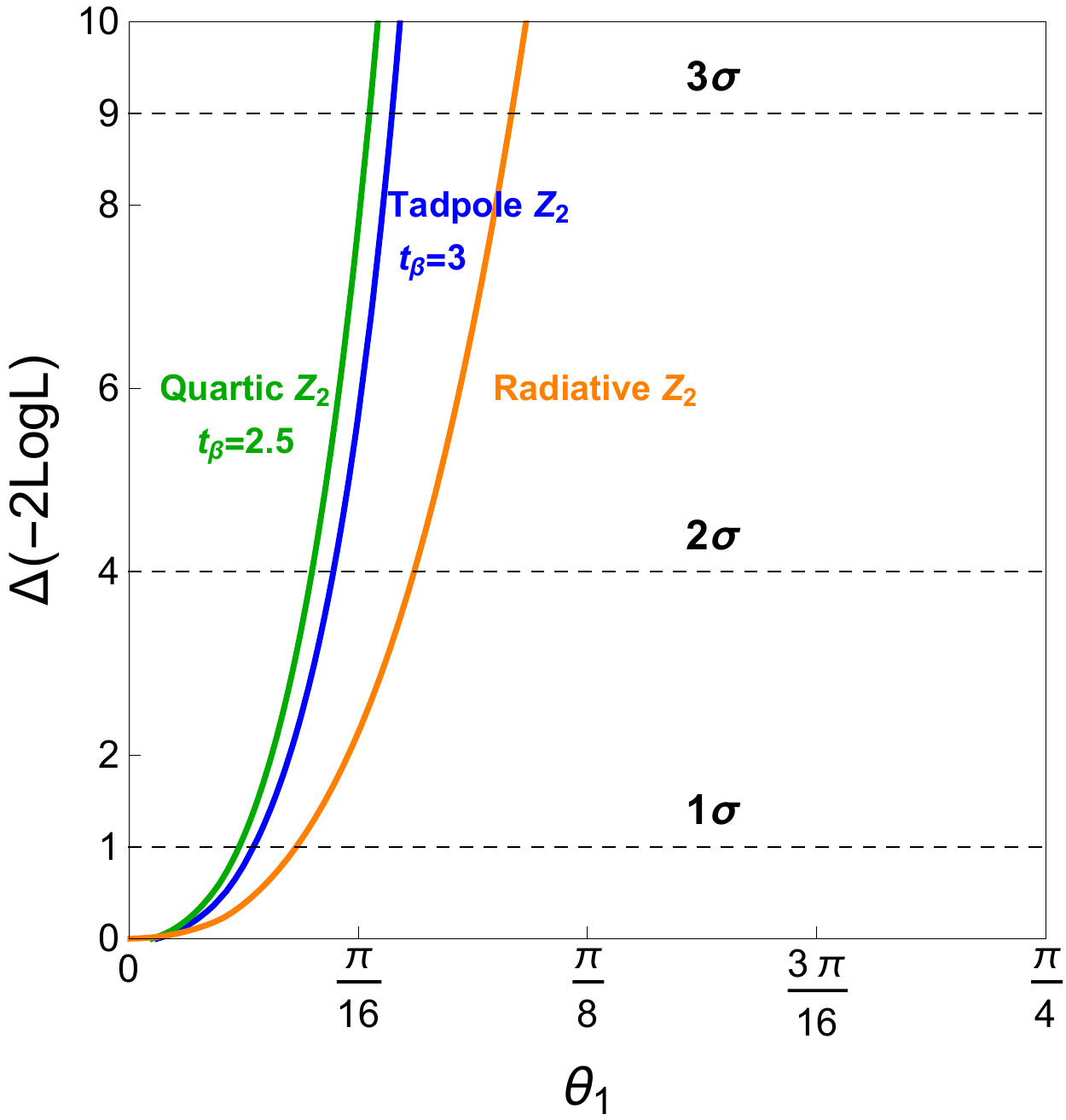} 
	\includegraphics[width=5cm]{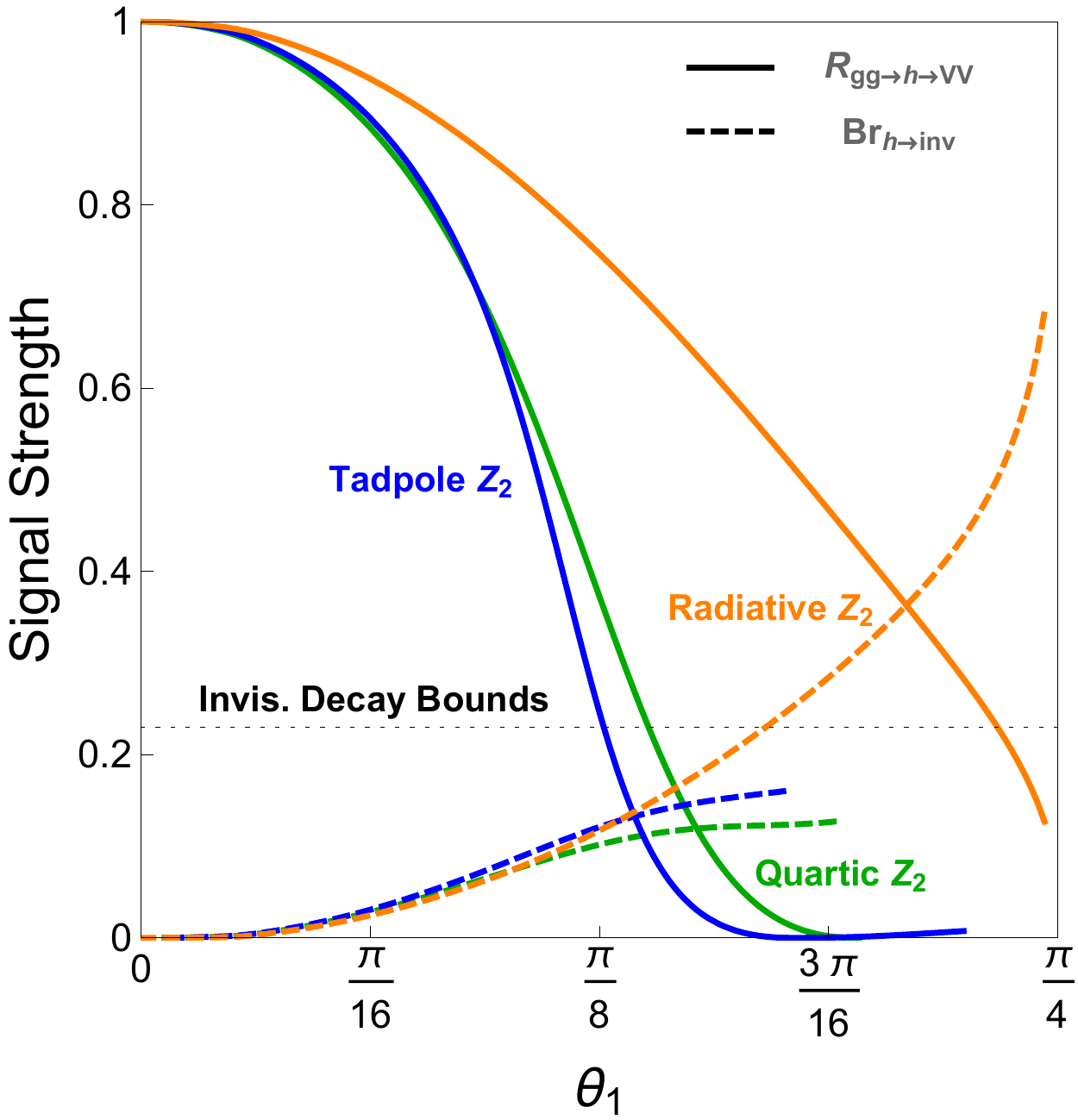} 
	\includegraphics[width=5.3cm]{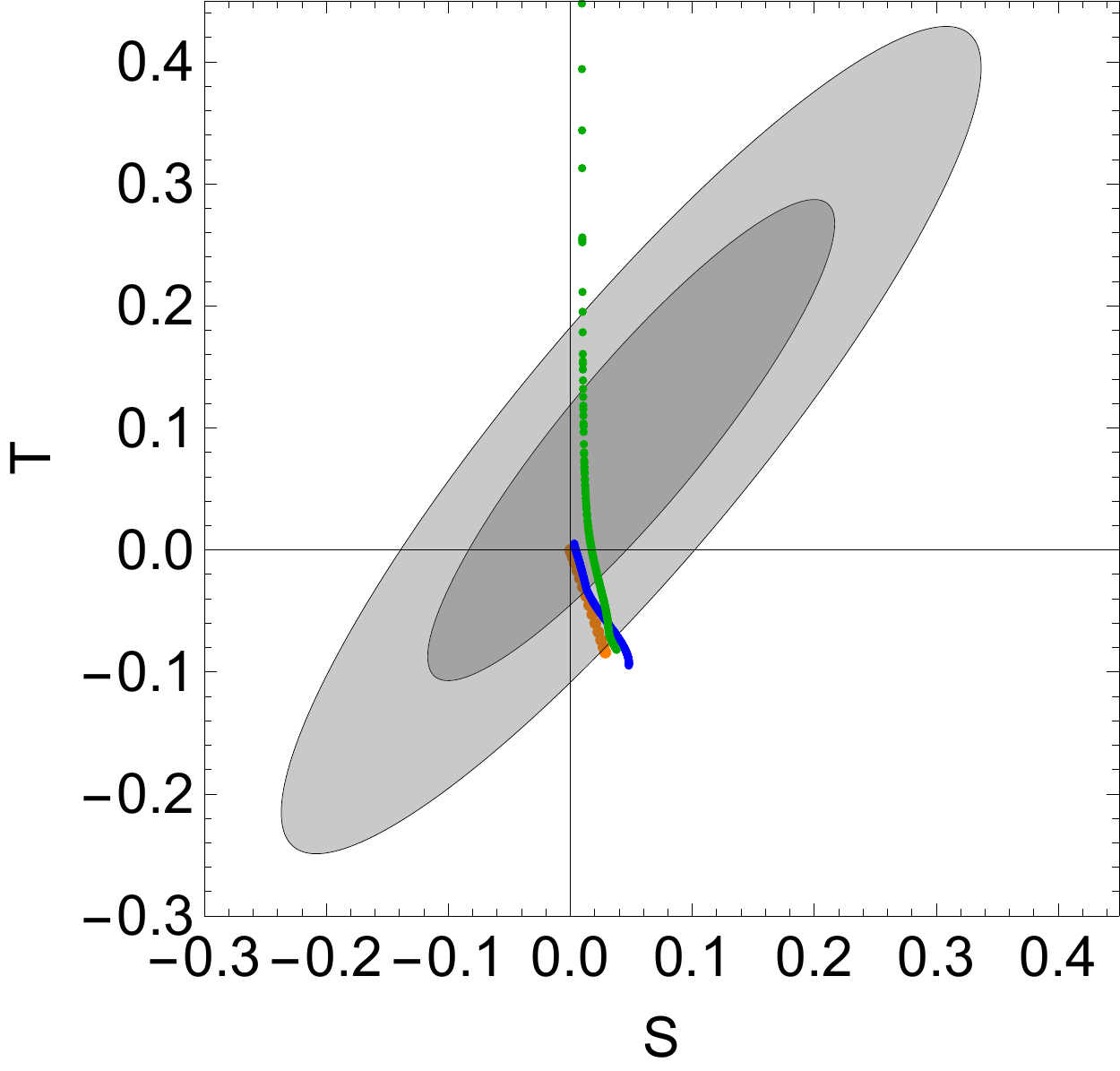} 
	\caption{
  On the left, it shows the log-likelihood profile $\Delta (-2\log\mathcal{L})$ as the function of $\theta_1$ in three
  scenarios. Here $1\sigma, 2\sigma, 3\sigma$ errors are also shown. 
  On the middle, it shows the signal strength in gluon fusion production and subsequent $VV$ decays, 
and Higgs invisible branching ratio as function of $\theta_1$ in three scenarios. 
The bound on the invisible decay branching ratio is ${\textrm Br}_{\rm inv} < 0.23$.  
 On the right, the $S-T$ oblique parameter contours at the  $1\sigma, 2\sigma$ levels are shown. 
 The dotted points are the parameter points in three scenarios: radiative (orange color), tadpole (blue), 
 and quartic (green) $\mathbb{Z}_2$ breaking scenarios. 
}
\label{fig:Higgsfits}
\end{center}
\end{figure}

We take the latest LHC results on the Higgs coupling measurements~\cite{Aad:2015gba, Khachatryan:2014jba} and Higgs invisible decays~\cite{Aad:2015pla}, 
and perform a global fit on the model parameters. 
In the tadpole and quartic $\mathbb{Z}_2$ breaking scenarios, 
if we fix the parameter $t_\beta$, 
only one free parameter exists. Thus in all the spontaneous $\mathbb{Z}_2$ breaking scenarios,
we will vary  $\theta_1$ and fix other parameters. 
Furthermore, we will not consider the explicit breaking scenario, since it should be less constrained than the other three scenarios. 
In the following, we  utilize the Lilith package~\cite{Bernon:2015hsa} to perform a global fitting 
on the Higgs signal strength. 
In the case where the Higgs coupling measurements are well within the Gaussian statistical regime, the 
likelihood function is defined 
\bea
	- 2 \log {\mathcal{L}}({\mathbf{\mu}}) = \chi^2({\mathbf{\mu}}) = \sum_{i=1}^{n} \frac{[\mu_i - \hat{\mu}_i(\theta_1)]^2}{\Delta \mu_i}.
\eea
Based on Higgs signal strengths at the 8 TeV LHC with 20.7 fb$^{-1}$ data~\cite{Aad:2015gba}, a statistical analysis is
performed by the Lilith package. 
Fig.~\ref{fig:Higgsfits} (left panel) shows the log-likelihood profile $\Delta (-2\log\mathcal{L})$ as the function of $\theta_1$, in three scenarios. 
Here  in tadpole and quartic scenarios we fix the parameter $t_\beta = 3$ and $t_\beta  =2.5$. 
Up to the $2\sigma$ level, the exclusion limits in three scenarios are that $\theta_1$ should typically be less than $0.2$.  
This put very strong constraints on the model parameter. 
Looking back to Fig.~\ref{fig:massspectra}, we note that both this Higgs coupling constraints and the requirement on 125 GeV Higgs mass should be 
satisfied. 
The tadpole and quartic breaking scenarios are viable, but the radiative breaking scenario has the tension between the 
Higgs coupling constraints and the 125 GeV Higgs boson mass requirement.  
If the $U(4)$ fermion assignment is taken in the radiative breaking scenario, there will not have such tension, and 
there are $\theta_1$ which could satisfy both conditions. 
This viable fermion assignment has been discussed in Ref.~\cite{Yu:2016bku}.
Although the invisible decay width has been taken into account indirectly in the above global fitting, 
we would like to consider constraints from the direct searches on the Higgs invisible decays. 
The updated upper limits on the invisible decay branching ratio is ${\textrm Br}_{\rm inv} < 0.23$~\cite{Aad:2015pla}. 
Fig.~\ref{fig:Higgsfits} (middle panel) shows the invisible decay branching ratio as the function of $\theta_1$. 
As a comparison, we also plot the signal strength in the gluon fusion process $gg\to h_1 \to VV$. 
From the Figure, we see that the direct searches on the invisible decays put much weaker constraints than the Higgs coupling measurements. 
The high luminosity LHC will improve sensitivity of signal strengths to around 5\% assuming current uncertainty with  3 ab$^{-1}$ luminosity~\cite{atlas:HLLHC}. 
Thus we should be able to  explore more parameter regions at the high luminosity LHC. 

According to the updated results on oblique parameters via Gfitter package~\cite{Baak:2014ora}, the S, T parameters have
$S = 0.05 \pm 0.09 , T = 0.11 \pm 0.13$, with correlation coefficients of $+0.90$ between $S$ and $T$.
In this model, the $S$ and $T$ parameters contains two contributions: corrections from possible radial modes, and 
corrections from 2HDM scalars. 
The corrections from radial modes takes the form
\bea
	\Delta S \simeq \frac1{6\pi} \sin^2 \theta \log\frac{m_\rho}{m_h}, \quad
	\Delta T \simeq -\frac3{8\pi c_W^2} \sin^2 \theta \log\frac{m_\rho}{m_h},
\eea
with radial modes $\rho$,
while the 2HDM corrections~\cite{Haber:2010bw} are roughly
\bea
	\Delta S \simeq \frac{m_{H}^2 + m_{A}^2 - 2 m_{H^\pm}^2}{24 \pi m_{A}^2}, 
	\quad
	\Delta T \simeq \frac{(m_{H^\pm}^2 - m_{A}^2) ( m_{H^\pm}^2 - m_{H}^2}{48 \pi s_W^2 m_W^2 m_{A}^2}.
\eea 
In our numerical study, the complete form of the $S, T$ parameters~\cite{Haber:2010bw, Branco:2011iw} are used.
From the above, we see that if the radial modes decouple, or if the heavy scalars are degenerate, the first or the second correction will be negligible. 
Fig.~\ref{fig:Higgsfits} (right panel) plots the predicted $S, T$ values in three scenarios, which we vary the parameter $\theta_1$ while fixing
$t_\beta = 3$ in tadpole scenario, and $t_\beta=2.5$ in tadpole scenario. 
According to the $S-T$ oblique parameter contours at the  $1\sigma, 2\sigma$ levels, 
we note that most of the $S, T$ values are within the $2\sigma$ level contour. 
Thus the precision electroweak test can not provide tighter constraints on the model parameters than the one from the Higgs coupling measurements.

Let us briefly discuss the distinct signatures of this model. 
First, like the original twin Higgs model, 
the twin hadron phenomenology~\cite{Craig:2015xla} provides us very distinct signatures from other models.
Furthermore, the additional charged and neutral scalars provide us 
a way to distinguish this model from the original twin Higgs. 
This has been explored in the 2HDM contents for general case~\cite{Branco:2011iw} and inert case~\cite{Blinov:2015qva}. 
Finally, to distinguish it from the typical 2HDM, 
the signatures from the twin $H'^\pm$ and $A'^0$ need to be explored. 
Furthermore, if the radial modes are not so heavy (thus not decoupled), 
exploring the radial mode decay channels could provide us different signatures from the typical 2HDM model.
The detailed collider phenomenology would require studies of their own.
We leave the detailed study in future.


\section{Conclusions}
\label{sec:conclusion}

In this work, we investigated a class of two twin Higgs models, in which 
the Higgs sector is extended to incorporate two twin Higgses and the global symmetry breaking pattern could
be either $U(4)\to U(3)$ or $[U(4)\times U(4)] \to [U(3)\times U(3)]$. 
The SM Higgs boson is identified as one of the pseudo Goldstone Bosons after symmetry breaking.
The discrete $\mathbb{Z}_2$ symmetry protects the Higgs mass term against the quadratically divergent radiative corrections.
However, the $\mathbb{Z}_2$ symmetry needs to be broken to generate electroweak scale, which should be separated  from the new physics scale. 
Typically the soft or hard explicit $\mathbb{Z}_2$ breaking terms are introduced to do so. 
We found that in the two twin Higgs setup, it is possible to realize spontaneous $\mathbb{Z}_2$ breaking, without the need of explicit $\mathbb{Z}_2$ breaking terms.

We performed a systematical study on the general $\mathbb{Z}_2$ breaking conditions in a natural two Higgs doublet framework,
and discussed various possible scenarios which could realize the vacuum misalignment.
In the radiative $\mathbb{Z}_2$ breaking scenario, given the appropriate fermion assignments, the $\mathbb{Z}_2$ symmetry could be spontaneously broken purely due to the radiative corrections to the Higgs potential.
In this scenario, only one Higgs obtains the electroweak vacuum, and the other is just an inert Higgs. 
The tadpole-induced $\mathbb{Z}_2$ breaking scenario can also be classified in this two Higgs doublet framework.  
In this scenario, the bilinear term in the scalar potential triggers the spontaneous $\mathbb{Z}_2$ breaking. 
We also proposed a novel scenario: the quartic-induced $\mathbb{Z}_2$ breaking scenario. 
In this scenario, the $\lambda_{4,5}$ terms  instead of the bilinear term in the scalar potential trigger
the spontaneous $\mathbb{Z}_2$ breaking. 

In the two twin Higgs models, we discussed phenomenology of the Higgs sector in the two Higgs doublet framework. 
Although particle contents in the scalar sector are the same, the Higgs mass spectra are quite distinct for each $\mathbb{Z}_2$ breaking scenarios. 
The radiative $\mathbb{Z}_2$ breaking scenario includes an inert Higgs doublet with degenerated masses.
Both the  tadpole-induced and  quartic-induced $\mathbb{Z}_2$ breaking scenarios 
contain additional scalars in two Higgs doublet model with not so degenerated masses. 
We calculated various Higgs couplings and utilized the the Higgs coupling measurements at the current LHC to constrain the model parameters. 
The additional scalars from the Higgs sector should be able to be probed at the Run-2 LHC and future colliders.

\section*{Acknowledgements}

The author would like to thank Can Kilic and  Nathaniel Craig for valuable discussions. 
This work was supported by DOE Grant DE-SC0011095.


\appendix



\section{Details in the $U(4) / U(3)$ Breaking Pattern}
\label{sec:appen1}

In this breaking pattern, the scalar potential takes the form
\bea
	V_{U(4)} &=& 
	-\mu_1^2 |H_1|^2 - \mu_2^2 |H_2|^2
	+ \lambda_1 (|H_1|^2)^2 + \lambda_2 (|H_2|^2)^2 
	+ \lambda_3 |H_1|^2 |H_2|^2 \nn \\
	&&
	+  m_{12}^2 \left[ H_1^\dagger H_2 +h.c.\right] 
	+ \lambda_4 |H_1^\dagger H_2 |^2
	+ \lambda_5 \left[(H_1^\dagger H_2)^2 + h.c.\right],\nn\\
	V_{\rm rad.cor.} &=& 
	+ \delta_1 \left(|H_{1A}|^4 +  |H_{1B}|^4\right)
	+ \delta_2 \left(|H_{2A}|^4 +  |H_{2B}|^4\right)
	+ \delta_3 \left(|H_{1A}|^2 |H_{2A}|^2 +  |H_{1B}|^2|H_{2B}|^2\right) \nn\\
	&& + \delta_4 \left(|H_{1A}^\dagger H_{2A}|^2   +  |H_{1B}^\dagger H_{2B}|^2 \right)
	+ \frac{\delta_5}{2} \left[(H_{1A}^\dagger H_{2A})^2   +  (H_{1B}^\dagger H_{2B})^2 + h.c. \right],\nn\\
	V_{\rm soft} &=& - m_{1A}^2 |H_{1A}|^2 - m_{2A}^2 |H_{2A}|^2 - m_{12A}^2 \left[H_{1A}^\dagger H_{2A} + h.c.\right].
\eea
As a special case, the supersymmetric extension of the twin Higgs model is one specific realization in this breaking pattern. 
%
%
We identify the specific terms in the scalar potential~\cite{Craig:2013fga} in SUSY twin Higgs model as
\bea
	&&\mu_1^2 = -(m_{H_u}^2 + \mu^2),\quad
	\mu_2^2 = -(m_{H_d}^2 + \mu^2),  \quad m_{12}^2 = - b, \nn\\
	&& \lambda_{1,2,3, 5} = 0, \quad \lambda_4= \lambda^2, \quad  \delta_1 = \delta_2 = \frac{\delta_3}{2} = \frac{g^2 + g'^2}{8}, \quad 
	\delta_{4,5} = 0.
\eea
Thus all our discussion about  $U(4) / U(3)$ breaking pattern can be applied to the SUSY twin Higgs model.

The tadpole conditions are 
\bea
	\mu_{1}^2 &=& 
	\frac{f_2}{f_1} m_{12}^2 + (\delta_1 +  2\lambda_1) f_1^2
	+ \left( \lambda_{345}  + \delta_{345}/2 \right) f_2^2 
	+  \left(f_1^2 \delta_1 + f_2^2 \delta_{345}/2\right)\cos2\theta,\nn\\
	\mu_{2}^2 &=& 
	\frac{f_1}{f_2} m_{12}^2 + (\delta_2 +  2\lambda_2) f_2^2
	+ \left( \lambda_{345} + \delta_{345}/2 \right) f_1^2 
	+ \left( f_2^2 \delta_2 + f_1^2 \delta_{345}/2\right) \cos2\theta ,\nn\\
      m_{1A}^2  & = & -\frac{f_2}{f_1} m_{12A}^2 + \left(f_1^2 \delta_1 + f_2^2 \delta_{345}\right) \cos2\theta, \nn\\
	m_{2A}^2  & = & -\frac{f_1}{f_2} m_{12A}^2 + \left(f_2^2 \delta_2 + f_1^2 \delta_{345}\right) \cos2\theta.
\eea

Similar to the 2HDM, rotating to the Higgs basis
\bea
	H = H_1 \cos\beta + H_2 \sin\beta, \qquad H' = - H_1 \sin\beta + H_2 \cos\beta.
\eea
In the Higgs basis, the masses of the charged gauge bosons are 
\bea
	m_{G^\pm}^2 &=& m_{C^\pm}^2 = 0, \quad \textrm{(Goldstone Bosons)},\nn\\
	m_{H^\pm}^2 &=& \left( \frac{m_{12A}^2 +  m_{12}^2}{f_1 f_2} - \delta_{45} \sin^2\theta -  \lambda_{45}  \right)(f_1^2 + f_2^2),\nn\\
	m_{H'^\pm}^2 &=& \left( \frac{ m_{12}^2}{f_1 f_2} - \delta_{45}\cos^2\theta - \lambda_{45} \right)(f_1^2 + f_2^2).
\eea
The mass matrices of the neutral CP-odd gauge bosons are 
\bea
	m_{G^0}^2 &=& m_{N^0}^2 =0 , \quad \textrm{(Goldstone Bosons)},\nn\\
	m_{A^0A^0}^2 &=& \left( \frac{m_{12A}^2 +  m_{12}^2}{f_1 f_2} - 2 \delta_{5} \sin^2\theta - (\lambda_4 - \lambda_5)\cos^2\theta - 2 \lambda_5    \right)(f_1^2 + f_2^2),\nn\\
	m_{A'^0A'^0}^2 &=& \left( \frac{ m_{12}^2}{f_1 f_2} - 2 \delta_{5} \cos^2\theta -  (\lambda_4 - \lambda_5) \sin^2\theta - 2 \lambda_5    \right)(f_1^2 + f_2^2),\nn\\
	m_{A^0A'^0}^2 &=& \frac{\lambda_4 - \lambda_5}{2} \sin2\theta (f_1^2 + f_2^2),	
\eea
Performing a further rotation on the fields $(A^0, A'^0)$, we obtain the mass eigenvalues 
\bea
	m_{A, A'}^2 = m_{A^0A^0}^2 + m_{A'^0A'^0}^2 \pm \sqrt{(m_{A^0A^0}^2 - m_{A'^0A'^0}^2)^2 - 4 m_{A^0A'^0}^4}.
\eea 
Assuming radial mode $H'^{0}$ is heavy, the mass matrices of the CP-even gauge bosons are
\bea
	m_{h^0 h^0}^2 
	&=& 
	4  \delta_1 f_1^2 \sin^2\theta + \frac{f_2(m_{12}^2 + m_{12A}^2)}{f_1} - \lambda_{45} f_2^2 \cos^2\theta ,\nn\\
	m_{H^0 H^0}^2 
	&=& 
	4  \delta_2 f_2^2 \sin^2\theta + \frac{f_1(m_{12}^2 + m_{12A}^2)}{f_2} - \lambda_{45} f_1^2 \cos^2\theta,\nn\\
	m_{h^0 H^0}^2 
	&=&
	- m_{12}^2 - m_{12A}^2 + \lambda_{45} f_1f_2  + (2 \delta_{345}+ \lambda_{45}) f_1f_2 \sin^2\theta. 
\eea
Similar to 2HDM, we could further rotate the field with a rotation angle $\alpha - \beta$:
\bea
	\left(\begin{array}{c} h \\ H  \end{array}\right) 
	= \left(\begin{array}{cc} c_{\alpha - \beta} &  s_{\alpha - \beta} \\ - s_{\alpha - \beta} & c_{\alpha - \beta} \end{array}\right)
	 \left(\begin{array}{c} h^0 \\ H^0 \end{array}\right),
\eea 
we obtain the mass eigenstates
\bea
	m_{h,H}^2 = m_{h^0h^0}^2 + m_{H^0H^0}^2 \mp \sqrt{(m_{h^0h^0}^2 - m_{H^0H^0}^2)^2 - 4 m_{h^0H^0}^4}.
\eea 
Here we identify $h$ as the SM Higgs boson.

\section{Details in the  $\left[U(4) \times U(4)\right]/\left[U(3) \times U(3)\right]$ Breaking Pattern}
\label{sec:appen2}

The general scalar potential reads
\bea
	V_{U(4) \times U(4)} &=& -\mu_1^2 |H_1|^2 - \mu_2^2 |H_2|^2 
	+ \lambda_1 (|H_1|^2)^2 + \lambda_2 (|H_2|^2)^2 
	+ \lambda_3 |H_1|^2 |H_2|^2,\nn\\
	V_{\rm breaking} &=& m_{12}^2 \left[ H_1^\dagger H_2 +h.c.\right]  + \lambda_4 |H_1^\dagger H_2 |^2
  	+  \frac{\lambda_5}{2} \left[(H_1^\dagger H_2)^2 + h.c.\right] 
	\nn\\
	V_{\rm rad.cor.} &=& \delta_1 \left(|H_{1A}|^4 +  |H_{1B}|^4\right)
	+ \delta_2 \left(|H_{2A}|^4 +  |H_{2B}|^4\right)
	+ \delta_3 \left(|H_{1A}|^2 |H_{2A}|^2 +  |H_{1B}|^2|H_{2B}|^2\right)\nn\\
	&+& \delta_4 \left(|H_{1A}^\dagger H_{2A}|^2   +  |H_{1B}^\dagger H_{2B}|^2 \right)
	+ \frac{\delta_5}{2} \left[(H_{1A}^\dagger H_{2A})^2   +  (H_{1B}^\dagger H_{2B})^2 + h.c. \right].
\eea
Here due to existence of the small tree-level breaking terms, the $U(4) \times U(4)$ symmetry is approximate.

The tadpole conditions are 
\bea
	\mu_1^2 &=&   (\delta_1 + 2 \lambda_1)f_1^2  + \frac{1}{2} (\delta_{345} + 2\lambda_3 +\lambda_{45}) f_2^2
	+ \frac{ \sin 2 \theta_2}{2\sin 2 \theta_1} \lambda_{45}  f_2^2
	- \frac{f_2 \sin(\theta_1 + \theta_2)}{f_1 \sin 2 \theta_1} m_{12}^2 ,\\
	\mu_2^2 &=&  (\delta_2 + 2 \lambda_2) f_2^2 + \frac{1}{2}  (\delta_{345} + 2\lambda_3 +\lambda_{45})f_1^2
	+ \frac{\sin 2 \theta_1}{2\sin 2 \theta_2}\lambda_{45} f_1^2 
	- \frac{f_1 \sin(\theta_1 + \theta_2)}{f_2 \sin 2 \theta_2} m_{12}^2,
\eea
and
\bea
	f_1^4 \delta_1 \sin 4\theta_1 + f_2^4 \delta_2 \sin 4 \theta_2   &=& -  f_1^2 f_2^2 \delta_{345} \sin 2(\theta_1 + \theta_2),\\
	f_1^4 \delta_1 \sin 4\theta_1 - f_2^4 \delta_2 \sin 4 \theta_2  &=&  4 f_1 f_2 m_{12}^2 \sin(\theta_1 - \theta_2) 
	-  f_1^2 f_2^2 (\delta_{345} + 2\lambda_{45}) \sin 2(\theta_1 - \theta_2).
\eea

Similar to the 2HDM, let us define the mixing angles of the VEVs in the $A$ and $B$ sectors
\bea
	\tan \beta_A = \frac{f_2 \sin \theta_2}{f_1 \sin\theta_1}, \qquad
	\tan \beta_B = \frac{f_2 \cos \theta_2}{f_1 \cos\theta_1}.
\eea
Then we will perform a rotation from the $H_1, H_2$ basis
\bea
	\left(\begin{array}{c} h^\pm \\ H^\pm  \end{array}\right) 
	= \left(\begin{array}{cc} \cos \beta_A &  \sin \beta_A \\ - \sin \beta_A & \cos \beta_A \end{array}\right)
	 \left(\begin{array}{c} h^\pm_1 \\ h^\pm_2  \end{array}\right), \qquad
 	\left(\begin{array}{c} C^\pm \\ H'^\pm  \end{array}\right) 
 	= \left(\begin{array}{cc} \cos \beta_B &  \sin \beta_B \\ - \sin \beta_B & \cos \beta_B \end{array}\right)
 	 \left(\begin{array}{c} C^\pm_1 \\ C^\pm_2  \end{array}\right), \nn\\
	\left(\begin{array}{c} z^0 \\ A_1^0  \end{array}\right) 
	= \left(\begin{array}{cc} \cos \beta_A &  \sin \beta_A \\ - \sin \beta_A & \cos \beta_A \end{array}\right)
	 \left(\begin{array}{c} {\rm Im} h^0_1 \\ {\rm Im} h^0_2  \end{array}\right), \qquad
 	\left(\begin{array}{c} N^0 \\ A_2^0  \end{array}\right) 
 	= \left(\begin{array}{cc} \cos \beta_B &  \sin \beta_B \\ - \sin \beta_B & \cos \beta_B \end{array}\right)
 	 \left(\begin{array}{c} N^0_1 \\ N^0_2  \end{array}\right).
\eea
The charged mass spectra have
\bea
	m_{C^\pm}^2 &=& m_{h^\pm}^2 = 0, \quad \textrm{(exact Goldstone bosons)}, \nn\\
	m_{H^\pm}^2 & = &  
	- \left(f_1^2 \sin^2\theta_1 + f^2_2 \sin^2 \theta_2\right) 
	\left[ \delta_{45} + \lambda_{45}(1 + \cot\theta_1\cot\theta_2) - \frac{m_{12}^2}{f_1\sin\theta_1 f_2\sin\theta_2}\right],\nn\\
	m_{H'^\pm}^2 & = & 
	- \left(f_1^2 \cos^2\theta_1 + f^2_2 \cos^2 \theta_2\right)
	\left[ \delta_{45} + \lambda_{45}(1 + \tan\theta_1\tan\theta_2) - \frac{m_{12}^2}{f_1\cos\theta_1 f_2\cos\theta_2}\right].
\eea
Similarly, the CP-odd neutral masses have
\bea
	m_{N^0}^2 &=& m_{z^0}^2 = 0, \quad \textrm{(exact Goldstone bosons)}, \nn\\
	m_{A^0_1}^2 & = & - \left(f_1^2 \sin^2\theta_1 + f^2_2 \sin^2 \theta_2\right) 
	\left[ 2\delta_{5} + 2\lambda_5 + \lambda_{45}\cot\theta_1\cot\theta_2 - \frac{m_{12}^2}{f_1\sin\theta_1 f_2\sin\theta_2}\right],\nn\\
	m_{A_2^0}^2 & = & - \left(f_1^2 \cos^2\theta_1 + f^2_2 \cos^2 \theta_2\right)
	\left[ 2\delta_{5} + 2\lambda_5 + \lambda_{45} \tan\theta_1\tan\theta_2 - \frac{m_{12}^2}{f_1\cos\theta_1 f_2\cos\theta_2}\right],\nn\\
	m_{A_1^0 A_2^0}^2 &=& (\lambda_4 -  \lambda_5)  \sqrt{\left(f_1^2 \sin^2\theta_1 + f^2_2 \sin^2 \theta_2\right)\left(f_1^2 \cos^2\theta_1 + f^2_2 \cos^2 \theta_2\right)}.
\eea
Note that there is still mixing between $(A_1^0, A_2^0)$,
a further rotation from $(A_1^0, A_2^0)$ to the mass eigenstates $(A^0, A'^0)$ are needed with the mass eigenstates
\bea
	m_{A^0, A'^0}^2 = m_{A^0_1}^2 + m_{A_2^0}^2 \pm \sqrt{(m_{A^0_1}^2 - m_{A_2^0}^2)^2 - 4 m_{A_1^0 A_2^0}^4}.
\eea

Finally, we obtain the masses for the SM-like Higgs boson and heavier Higgs boson. 
Assuming the radial modes are heavy, 
we obtain the 
the $2\times 2$ mass matrices in the $\textrm{Re}\, h_1^0, \textrm{Re}\,h_2^0$  basis
\bea
	{\mathcal{M}}^2_{\rm Higgs} = \left(\begin{array}{cc} 
	m_{h_1h_1}^2 & m_{h_1h_2}^2 \\ 
	m_{h_1h_2}^2 & m_{h_2h_2}^2
	\end{array}\right).
\eea
The mass matrices read
\bea
	m_{h_1h_1}^2 &=& \frac{\delta_1 f_1^2}{2} (1 - 3 \cos 4\theta_1) 
	- \frac{\delta_{345} f_2^2}{2} \cos 2\theta_1 \cos 2\theta_2
	- \frac{\lambda_{45} f_2^2}{2} \frac{\sin (4\theta_1 - 2 \theta_2) + 3 \sin 2 \theta_2}{\sin 2\theta_1}
	+ \frac{m_{12}^2 f_2}{f_1}\frac{\sin (\theta_1 + \theta_2)}{\sin 2\theta_1},\nn\\
	m_{h_2h_2}^2 &=& \frac{\delta_2 f_2^2}{2} (1 - 3 \cos 4\theta_2) 
	- \frac{\delta_{345} f_1^2}{2} \cos 2\theta_1 \cos 2\theta_2
	- \frac{\lambda_{45} f_1^2}{2} \frac{\sin (4\theta_2 - 2 \theta_1) + 3 \sin 2 \theta_1}{\sin 2\theta_2}
	+ \frac{m_{12}^2 f_1}{f_2}\frac{\sin (\theta_1 + \theta_2)}{\sin 2\theta_2},\nn\\
	m_{h_1h_2}^2 &=& \frac{\delta_{345} f_1 f_2}{2} \cos 2(\theta_1 + \theta_2)
	+ \frac{(\delta_{345}+\lambda_{45})f_1 f_2}{2} \cos 2(\theta_1 - \theta_2)
	- m_{12}^2 \cos(\theta_1 - \theta_2).
\eea
Similar to 2HDM, let us rotate the $(h_1, h_2)$ to the mass eigenstates $(h, H)$ with rotation angle $\alpha$:
\bea
	\left(\begin{array}{c} h \\ H  \end{array}\right) 
	= \left(\begin{array}{cc} \cos \alpha &  \sin \alpha \\ - \sin \alpha & \cos \alpha \end{array}\right)
	 \left(\begin{array}{c} h_1 \\ h_2  \end{array}\right).
\eea 
Here the rotation angle is defined as
\bea
	\tan2\alpha = \frac{2m_{h_1h_2}^2}{m_{h_1h_1}^2 - m_{h_2h_2}^2}, 
\eea
and the mass eigenvalues are
\bea
	m_{h,H}^2 = m_{h_1}^2 + m_{h_2}^2 \pm \sqrt{(m_{h_1}^2 - m_{h_2}^2)^2 - 4 m_{h_1h_2}^4}.
\eea
Here we identify $h$ as the SM Higgs boson.




\begin{thebibliography}{99}
	


\bibitem{Aad:2012tfa} 
  G.~Aad {\it et al.} [ATLAS Collaboration],
  Phys.\ Lett.\ B {\bf 716}, 1 (2012)
  doi:10.1016/j.physletb.2012.08.020
  [arXiv:1207.7214 [hep-ex]].
  
  
\bibitem{Chatrchyan:2012xdj} 
  S.~Chatrchyan {\it et al.} [CMS Collaboration],
  Phys.\ Lett.\ B {\bf 716}, 30 (2012)
  doi:10.1016/j.physletb.2012.08.021
  [arXiv:1207.7235 [hep-ex]].

\bibitem{Barbieri:2000gf} 
  R.~Barbieri and A.~Strumia,
  hep-ph/0007265.


\bibitem{Martin:1997ns} 
  S.~P.~Martin,
  Adv.\ Ser.\ Direct.\ High Energy Phys.\  {\bf 21}, 1 (2010)
  [Adv.\ Ser.\ Direct.\ High Energy Phys.\  {\bf 18}, 1 (1998)]
  doi:10.1142/9789812839657, 10.1142/9789814307505
  [hep-ph/9709356].

\bibitem{Kaplan:1983fs} 
  D.~B.~Kaplan and H.~Georgi,
  Phys.\ Lett.\ B {\bf 136}, 183 (1984).
  doi:10.1016/0370-2693(84)91177-8
  
\bibitem{Agashe:2004rs} 
  K.~Agashe, R.~Contino and A.~Pomarol,
  Nucl.\ Phys.\ B {\bf 719}, 165 (2005)
  doi:10.1016/j.nuclphysb.2005.04.035
  [hep-ph/0412089].


\bibitem{ArkaniHamed:2002qy} 
  N.~Arkani-Hamed, A.~G.~Cohen, E.~Katz and A.~E.~Nelson,
  JHEP {\bf 0207}, 034 (2002)
  doi:10.1088/1126-6708/2002/07/034
  [hep-ph/0206021].



%
\bibitem{Chacko:2005pe} 
  Z.~Chacko, H.~S.~Goh and R.~Harnik,
  Phys.\ Rev.\ Lett.\  {\bf 96}, 231802 (2006)
  doi:10.1103/PhysRevLett.96.231802
  [hep-ph/0506256]; JHEP {\bf 0601}, 108 (2006)
  doi:10.1088/1126-6708/2006/01/108
  [hep-ph/0512088].
  
  
\bibitem{Burdman:2006tz} 
  G.~Burdman, Z.~Chacko, H.~S.~Goh and R.~Harnik,
  JHEP {\bf 0702}, 009 (2007)
  doi:10.1088/1126-6708/2007/02/009
  [hep-ph/0609152].
  
\bibitem{Cai:2008au} 
  H.~Cai, H.~C.~Cheng and J.~Terning,
  JHEP {\bf 0905}, 045 (2009)
  doi:10.1088/1126-6708/2009/05/045
  [arXiv:0812.0843 [hep-ph]].



\bibitem{Chacko:2005un} 
  Z.~Chacko, H.~S.~Goh and R.~Harnik,
  JHEP {\bf 0601}, 108 (2006)
  doi:10.1088/1126-6708/2006/01/108
  [hep-ph/0512088].

  
\bibitem{Craig:2014aea} 
  N.~Craig, S.~Knapen and P.~Longhi,
  Phys.\ Rev.\ Lett.\  {\bf 114}, no. 6, 061803 (2015)
  doi:10.1103/PhysRevLett.114.061803
  [arXiv:1410.6808 [hep-ph]]; JHEP {\bf 1503}, 106 (2015)
  doi:10.1007/JHEP03(2015)106
  [arXiv:1411.7393 [hep-ph]].
  
 
  
\bibitem{Burdman:2014zta} 
  G.~Burdman, Z.~Chacko, R.~Harnik, L.~de Lima and C.~B.~Verhaaren,
  Phys.\ Rev.\ D {\bf 91}, no. 5, 055007 (2015)
  doi:10.1103/PhysRevD.91.055007
  [arXiv:1411.3310 [hep-ph]].
  
  
   
\bibitem{Craig:2015pha} 
  N.~Craig, A.~Katz, M.~Strassler and R.~Sundrum,
  JHEP {\bf 1507}, 105 (2015)
  doi:10.1007/JHEP07(2015)105
  [arXiv:1501.05310 [hep-ph]].
  
\bibitem{Barbieri:2005ri} 
 R.~Barbieri, T.~Gregoire and L.~J.~Hall,
 hep-ph/0509242.



\bibitem{Bellazzini:2014yua} 
  B.~Bellazzini, C.~Csáki and J.~Serra,
  Eur.\ Phys.\ J.\ C {\bf 74}, no. 5, 2766 (2014)
  doi:10.1140/epjc/s10052-014-2766-x
  [arXiv:1401.2457 [hep-ph]].



    
\bibitem{Chacko:2005vw} 
  Z.~Chacko, Y.~Nomura, M.~Papucci and G.~Perez,
  JHEP {\bf 0601}, 126 (2006)
  doi:10.1088/1126-6708/2006/01/126
  [hep-ph/0510273].
     
   
\bibitem{Goh:2006wj} 
  H.~S.~Goh and S.~Su,
  Phys.\ Rev.\ D {\bf 75}, 075010 (2007)
  doi:10.1103/PhysRevD.75.075010
  [hep-ph/0611015].


  
%



\bibitem{Chang:2006ra} 
  S.~Chang, L.~J.~Hall and N.~Weiner,
  Phys.\ Rev.\ D {\bf 75}, 035009 (2007)
  doi:10.1103/PhysRevD.75.035009
  [hep-ph/0604076].
  
  
\bibitem{Craig:2013fga} 
  N.~Craig and K.~Howe,
  JHEP {\bf 1403}, 140 (2014)
  doi:10.1007/JHEP03(2014)140
  [arXiv:1312.1341 [hep-ph]].


\bibitem{Batra:2008jy} 
  P.~Batra and Z.~Chacko,
  Phys.\ Rev.\ D {\bf 79}, 095012 (2009)
  doi:10.1103/PhysRevD.79.095012
  [arXiv:0811.0394 [hep-ph]].
  
\bibitem{Barbieri:2015lqa} 
  R.~Barbieri, D.~Greco, R.~Rattazzi and A.~Wulzer,
  JHEP {\bf 1508}, 161 (2015)
  doi:10.1007/JHEP08(2015)161
  [arXiv:1501.07803 [hep-ph]].
  
\bibitem{Low:2015nqa} 
  M.~Low, A.~Tesi and L.~T.~Wang,
  Phys.\ Rev.\ D {\bf 91}, 095012 (2015)
  doi:10.1103/PhysRevD.91.095012
  [arXiv:1501.07890 [hep-ph]].



\bibitem{Beauchesne:2015lva} 
  H.~Beauchesne, K.~Earl and T.~Grégoire,
  JHEP {\bf 1601}, 130 (2016)
  doi:10.1007/JHEP01(2016)130
  [arXiv:1510.06069 [hep-ph]].



\bibitem{Harnik:2016koz} 
  R.~Harnik, K.~Howe and J.~Kearney,
  arXiv:1603.03772 [hep-ph].
  

\bibitem{Yu:2016bku} 
  J.~H.~Yu,
  arXiv:1608.01314 [hep-ph].





 

\bibitem{Cheng:2015buv} 
  H.~C.~Cheng, S.~Jung, E.~Salvioni and Y.~Tsai,
  JHEP {\bf 1603}, 074 (2016)
  doi:10.1007/JHEP03(2016)074
  [arXiv:1512.02647 [hep-ph]].
  
  
\bibitem{Branco:2011iw} 
  G.~C.~Branco, P.~M.~Ferreira, L.~Lavoura, M.~N.~Rebelo, M.~Sher and J.~P.~Silva,
  Phys.\ Rept.\  {\bf 516}, 1 (2012)
  doi:10.1016/j.physrep.2012.02.002
  [arXiv:1106.0034 [hep-ph]].
  
  
 
    
\bibitem{Craig:2016kue} 
  N.~Craig, S.~Knapen, P.~Longhi and M.~Strassler,
  JHEP {\bf 1607}, 002 (2016)
  doi:10.1007/JHEP07(2016)002
  [arXiv:1601.07181 [hep-ph]].
  

  

\bibitem{Barbieri:2006dq} 
  R.~Barbieri, L.~J.~Hall and V.~S.~Rychkov,
  Phys.\ Rev.\ D {\bf 74}, 015007 (2006)
  doi:10.1103/PhysRevD.74.015007
  [hep-ph/0603188].
 

\bibitem{Craig:2015xla} 
  N.~Craig and A.~Katz,
  JCAP {\bf 1510}, no. 10, 054 (2015)
  doi:10.1088/1475-7516/2015/10/054
  [arXiv:1505.07113 [hep-ph]].
  
\bibitem{Garcia:2015loa} 
  I.~García García, R.~Lasenby and J.~March-Russell,
  Phys.\ Rev.\ D {\bf 92}, no. 5, 055034 (2015)
  doi:10.1103/PhysRevD.92.055034
  [arXiv:1505.07109 [hep-ph]];Phys.\ Rev.\ Lett.\  {\bf 115}, no. 12, 121801 (2015)
  doi:10.1103/PhysRevLett.115.121801
  [arXiv:1505.07410 [hep-ph]].


\bibitem{Farina:2015uea} 
  M.~Farina,
  JCAP {\bf 1511}, no. 11, 017 (2015)
  doi:10.1088/1475-7516/2015/11/017
  [arXiv:1506.03520 [hep-ph]].
  
  
  
   
  
\bibitem{Aad:2015gba} 
  G.~Aad {\it et al.} [ATLAS Collaboration],
  Eur.\ Phys.\ J.\ C {\bf 76}, no. 1, 6 (2016)
  doi:10.1140/epjc/s10052-015-3769-y
  [arXiv:1507.04548 [hep-ex]].
  
\bibitem{Khachatryan:2014jba} 
  V.~Khachatryan {\it et al.} [CMS Collaboration],
  Eur.\ Phys.\ J.\ C {\bf 75}, no. 5, 212 (2015)
  doi:10.1140/epjc/s10052-015-3351-7
  [arXiv:1412.8662 [hep-ex]].


\bibitem{Aad:2015pla} 
  G.~Aad {\it et al.} [ATLAS Collaboration],
  JHEP {\bf 1511}, 206 (2015)
  doi:10.1007/JHEP11(2015)206
  [arXiv:1509.00672 [hep-ex]].
  
\bibitem{Bernon:2015hsa} 
  J.~Bernon and B.~Dumont,
  Eur.\ Phys.\ J.\ C {\bf 75}, no. 9, 440 (2015)
  doi:10.1140/epjc/s10052-015-3645-9
  [arXiv:1502.04138 [hep-ph]].


\bibitem{atlas:HLLHC} 
 ATLAS Collaboration, ATL-PHYS-PUB-2013-014, CERN, Geneva, Oct, 2013. 


 


\bibitem{Baak:2014ora} 
  M.~Baak {\it et al.} [Gfitter Group Collaboration],
  Eur.\ Phys.\ J.\ C {\bf 74}, 3046 (2014)
  doi:10.1140/epjc/s10052-014-3046-5
  [arXiv:1407.3792 [hep-ph]].



\bibitem{Haber:2010bw} 
  H.~E.~Haber and D.~O'Neil,
  Phys.\ Rev.\ D {\bf 83}, 055017 (2011)
  doi:10.1103/PhysRevD.83.055017
  [arXiv:1011.6188 [hep-ph]].

\bibitem{Blinov:2015qva} 
  N.~Blinov, J.~Kozaczuk, D.~E.~Morrissey and A.~de la Puente,
  Phys.\ Rev.\ D {\bf 93}, no. 3, 035020 (2016)
  doi:10.1103/PhysRevD.93.035020
  [arXiv:1510.08069 [hep-ph]].
  


  

\end{thebibliography}
\end{document}